\documentclass[aps]{revtex4-1}
\usepackage[latin9]{luainputenc}
\usepackage{color}
\usepackage{verbatim}
\usepackage{bm}
\usepackage{amsmath}
\usepackage{graphicx}
\usepackage{nomencl}
\providecommand{\printnomenclature}{\printglossary}
\providecommand{\makenomenclature}{\makeglossary}
\makenomenclature

\makeatletter
\def\vecsign{\mathchar"017E}
\def\dvecsign{\smash{\stackon[-1.95pt]{\vecsign}{\rotatebox{180}{$\vecsign$}}}}
\def\dvec#1{\def\useanchorwidth{T}\stackon[-4.2pt]{#1}{\,\dvecsign}}
\usepackage{stackengine}
\stackMath

\usepackage{marginnote}

\makeatother

\begin{document}

\title{Mesoscopic simulations at the physics-chemistry-biology interface}

\author{Massimo Bernaschi}

\address{IAC-CNR, via dei Taurini 19, 00185, Rome, Italy}

\author{Simone Melchionna}

\address{ISC-CNR, c/o Dipartimento di Fisica, Università La Sapienza, P.le
A. Moro 5, 00185, Rome, Italy}

\address{IACS-SEAS, Harvard University, Oxford Street 29, 02138, Cambridge,
USA}

\author{Sauro Succi}

\address{Center for Life Nano Science@Sapienza, Italian Institute of Technology,
Viale Regina Elena, 295, I-00161 Rome, Italy}

\address{IAC-CNR, via dei Taurini 19, 00185, Rome, Italy}

\address{IACS-SEAS, Harvard University, Oxford Street 29, 02138, Cambridge,
USA}
\begin{abstract}
We discuss the Lattice Boltzmann-Particle Dynamics (LBPD) multiscale
paradigm for the simulation of complex states of flowing matter at
the interface between Physics, Chemistry and Biology. 

In particular, we describe current large-scale LBPD simulations of
biopolymer translocation across cellular membranes, molecular transport
in ion channels and amyloid aggregation in cells. We also provide
prospects for future LBPD explorations in the direction of cellular
organization, the direct simulation of full biological organelles,
all the way to up to physiological scales of potential relevance to
future precision-medicine applications, such as the accurate description
of homeostatic processes. 

It is argued that, with the advent of Exascale computing, the mesoscale
physics approach advocated in this paper, may come to age in the next
decade and open up new exciting perspectives for physics-based computational
medicine. 
\end{abstract}
\maketitle
\tableofcontents{}

\section{INTRODUCTION}

Thanks to the spectacular advances of computer technology (hardware
and software) on the one side, and mathematical modelling on the other,
in the last few decades modern science has come to the point of providing
a quantitative description of many biological systems, whose complexity
would have been regarded as mission-impossible until only recently.
In this Review, we shall illustrate the point through several concrete
examples. 

Notwithstanding such major advances, the challenge of modelling biological
and physiological systems remains formidable, as it actually amounts
to cover some ten decades in space (from molecules to the human body)
and easily twice as many in time. No mathematical/computational model
in the foreseeable future can take up such a challenge head-on, i.e.
by direct simulation of all the actual mechanisms, scales and levels
involved in the process. 

Coarse-grained methods come in many flavours and families, depending
on the range of scales and problems they are targeted to, but in this
review we shall focus on a specific mesoscale technique, known as
the Lattice Boltzmann method (LBM), namely a minimal lattice version
of the Boltzmann equation \cite{BOL,CERCI} which has witnessed a
burgeoning growth for the description of complex flow phenomena across
an impressively broad range of scales. 

The Lattice Boltzmann Equation (LBE) was developed as a computational
alternative to the discretization of the Navier-Stokes (NS) equations
of continuum fluid mechanics \cite{BSV,LB1}. 

Over the years, however, it has made proof of an amazing and largely
unanticipated versatility and ability to describe a broad variety
of phenomena involving complex states of flowing matter, beyond the
strict realm of continuum hydrodynamics, including non-trivial flows
at micro and nanoscale. Thanks to this versatility, and to the coupling
with various families of mesoscale particle methods, in the last decade
LBE has gained increased status for the simulation of many complex
flow problems at the interface between fluid dynamics, chemistry,
material science and biology. These include, for instance, multiphase
and multicomponent flows with complex interfaces, the motion of suspended
bodies under strong geometrical confinement, possibly with chemical
reactions \cite{LB1b,LB2}. After revisiting the main ideas behind
the Lattice Boltzmann (LB) theory, in this review we discuss current
and future prospects of multiscale/level Lattice Boltzmann-Particle
Dynamics (LBPD) simulations at the physics-chemistry-biology interface,
in an attempt to identify and portray outstanding problems of potential
relevance to clinical applications in a not-so-distant future, i.e.
\emph{computational medicine}.

Computer-assisted medicine is a consolidated branch of modern science,
which generally develops around two separate pillars; molecular biology
and macroscale physiology. The former is heavily leaning on bio-informatics
and data science tools, while the latter usually relies upon the methods
of continuum and fluid mechanics. In this Review, we portray a third,
alternative approach, based on \emph{mesoscale physics}, i.e. physics-informed
coarse-grained models of microscale biological processes, possibly
embedded within their physiological environment. This mesoscale approach
is grounded into the intermediate level of the description of matter,
namely kinetic theory, in both its versions, Boltzmann's kinetic theory
and Langevin stochastic particle dynamics. As a result, the main tools
of the mesoscale approach are fluids, particles and probability distribution
functions. The multiscale LBPD concept is thus apparent; Boltzmann
naturally connects upwards to continuum hydrodynamics, while Langevin
connects downwards to the molecular level. Combining the two in a
single computational harness, opens up a direct route from the continuum
to the atomistic word, and back. 

Several specific examples have already shown the potential of LBPD
simulations in areas straddling across Physics and Biology. A selected
set of applications will be described in detail in this Review, to
provide a taste for the breadth of applications that have been tackled
in the recent past. Readers keen on specific details are kindly directed
to the original literature. 

\begin{figure}
\begin{centering}
\includegraphics[scale=1.6]{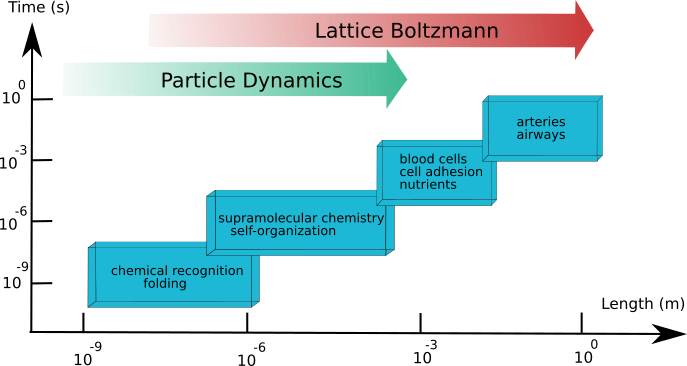}
\par\end{centering}
\caption{Scales related to biological and medical applications, whose stretch
can be covered by the LBPD approach. While the particle description
enables the representation of complex macromolecules or cellular organizations
by tracking the fate of each individual particle, the LB description
is based on the collective motion of solvent molecules. The boundary
between LB and PD is blurred and depends on the degree of microscopic
detail required by each single application (see also Fig. \ref{fig:quest}).\label{fig:BioScales}}
\end{figure}

Clearly, the PCB interface is enormously rich and varied, ranging
from nanometric macromolecular phenomena, to peptidic aggregation
and biopolymer translocation, to cellular motion and active matter,
as sketched in Fig. \ref{fig:BioScales}. In a few words, the topic
goes way beyond the scope of the present work. Nevertheless, it is
important for the reader to appreciate the great flexibility of the
method to cope with multiple scales of motion and, more importantly,
its ability to incorporate the desired degree of bio-chemical specificity
inherent to the problem at hand. 

The LBPD framework can be enriched in the direction of describing
complex flowing states of matter at micro and nanoscales. Although
not ``rigorous'', such variants prove capable of providing new physical
insights into highly complex states of flowing matter. Remarkably,
such extensions can be put in place without compromising the outstanding
parallel amenability to parallel processing of the method. On the
assumption that computing power will keep growing in the next few
decades, if only perhaps at sub-Moore's paces, this state of affairs
spawns tremendous opportunities to gain new insights into a series
of fundamental problems dealing with complex states of flowing matter
in general, and with special focus on those relevant to biology and
medicine. 

The paper is organised in three main parts. In the first one, we discuss
the basic aspects of Boltzmann's kinetic theory with special emphasis
on its lattice version for fluid dynamics and its extensions to soft-matter
and biological applications, including the coupling to particle dynamics
for the motion of suspended bodies.

\begin{figure}
\begin{centering}
\includegraphics{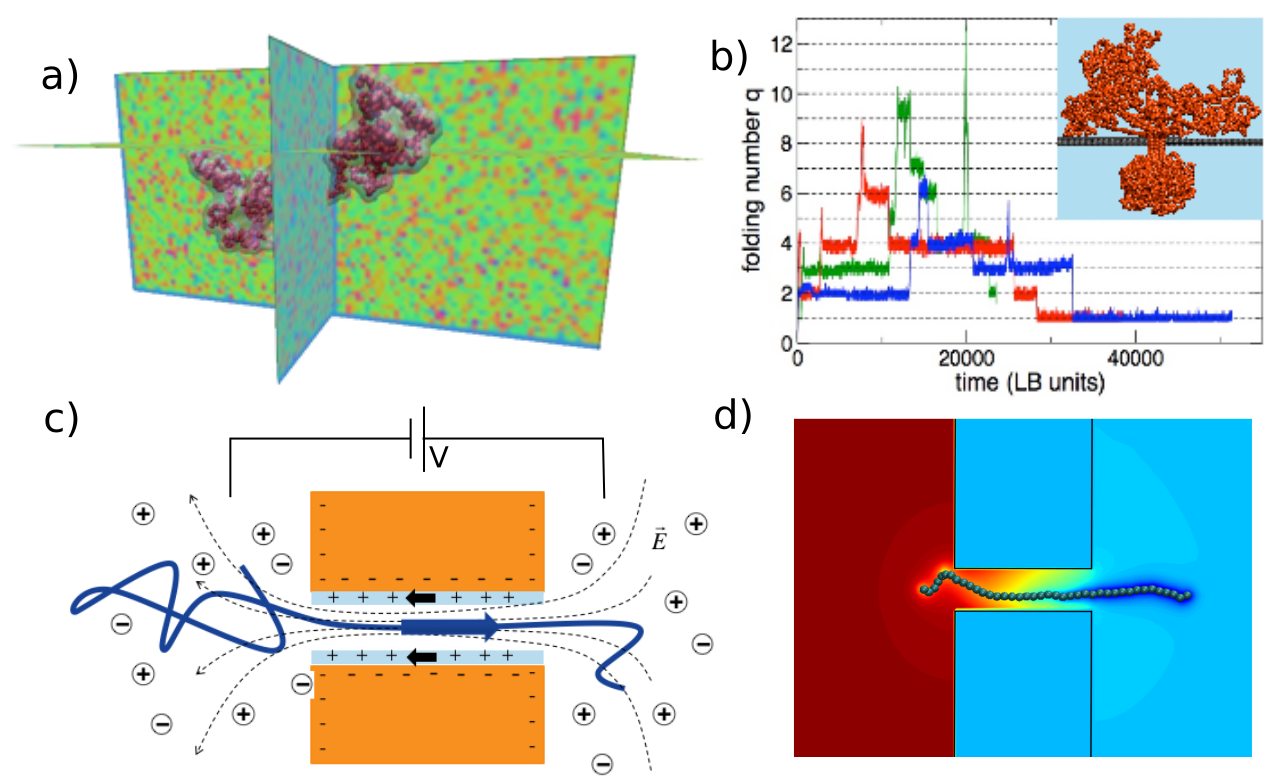} 
\par\end{centering}
\caption{Translocation of a biopolymer, as in the case of DNA, through a narrow
pore. Different representations can be used to study multiple levels
of detail during the translocation process. a,b) the biopolymer is
represented as a simple necklace of beads, by neglecting correlations
stemming from the local molecular rigidity or backbone charge. c,d)
at the next level, the macromolecule is charged and moves in an electrolyte
solution, whereby a neutral solvent and counterions and coions migrate
due to an externally applied electric field, giving rise to an electroosmotic
flow that ultimately causes the molecule to translocate \cite{DATAR2017}.\label{fig:Translocation}}
\end{figure}

\begin{figure}
\begin{centering}
\includegraphics[scale=1.0]{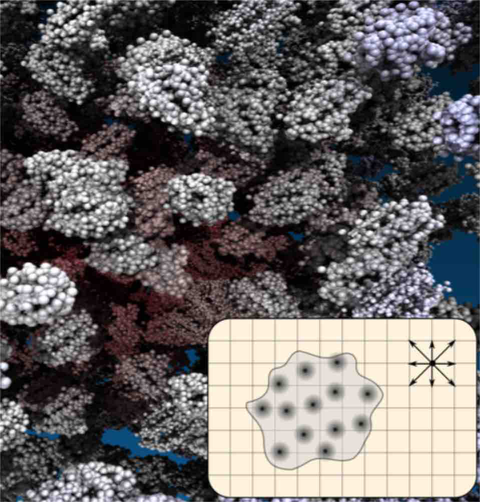} 
\par\end{centering}
\caption{Representation of the crowded interior of the cell as obtained from
simulations \cite{LBCROWDING}. The inset illustrates the embedding
of a protein in the LB mesh and each protein atom is represented via
the DPM particle-fluid exchange scheme. \label{fig:Crowding}}
\end{figure}

\begin{figure}
\begin{centering}
\includegraphics[scale=1.0]{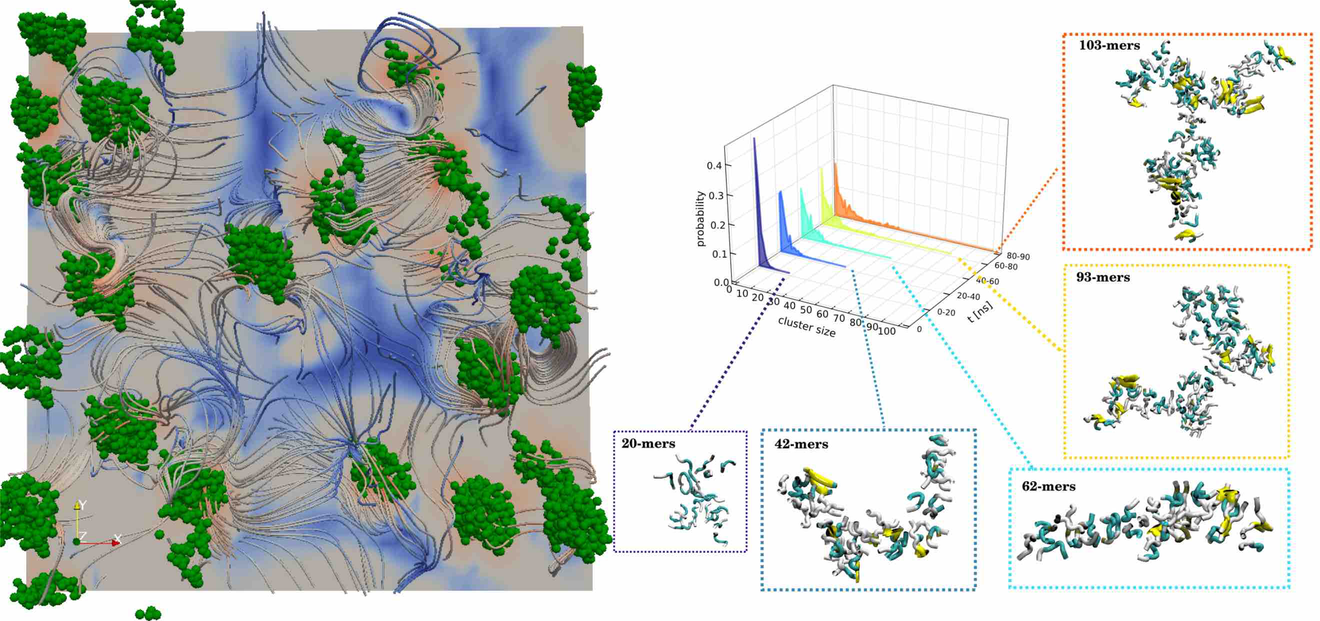} 
\par\end{centering}
\caption{Left: snapshot of a configuration of $1000$ amyloid peptides A$\beta_{16-22}$
simulated in a cubic box of size $30\times30\times30$ nm$^{3}$and
complex flow structure generated by their motion in the surrounding
solvent. Right: evolution of the size of the peptidic aggregates as
a function of time \cite{NASICA2015,LBMDbio2}.\label{fig:amyloids}}
\end{figure}

In the second part, we describe selected applications to biological
and physiological systems, such as biopolymer translocation, ion channels,
protein diffusion and amyloid aggregation in cellular environments.
For a quick visual appreciation, see figures \ref{fig:Translocation},
\ref{fig:Crowding}, \ref{fig:amyloids}, to be commented in detail
later in this manuscript.

In the third part, we provide a prospective view of a series of problems
at the physics-chemistry-biology interface, which may become accessible
once Exascale computers are with us. Special attention is paid on
their potential import for clinical applications, such as the direct
simulation of biological organelles and the quantitative description
of hemostatic processes.

\begin{figure}
\centering{} \includegraphics[scale=1.0]{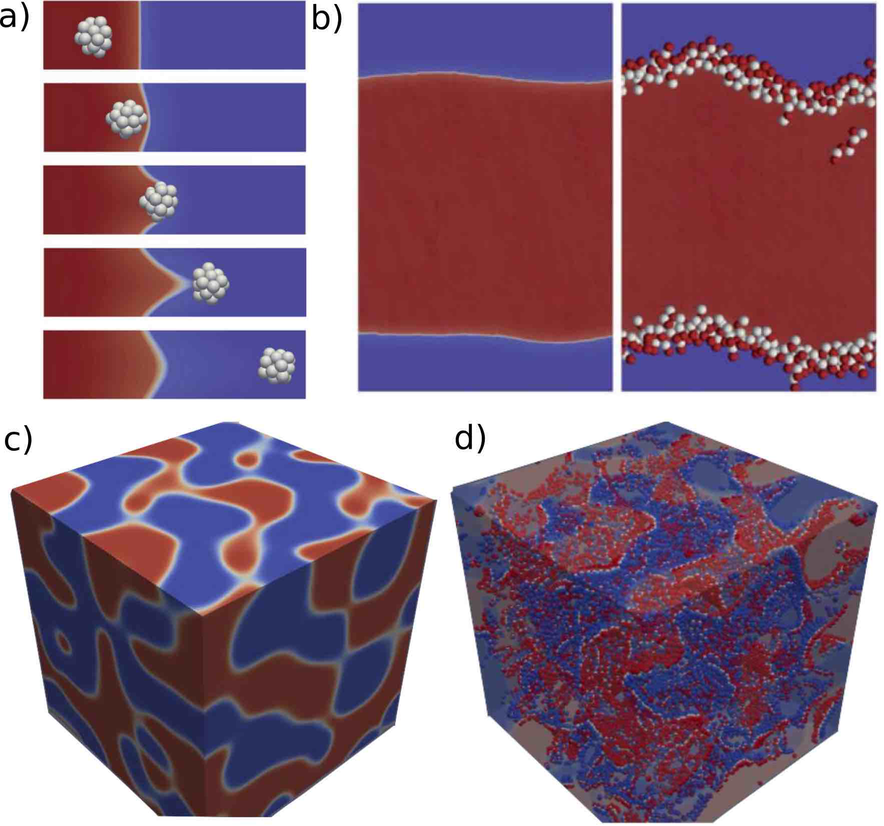}\caption{Snapshot of molecules and particles in a multiphase flow , either
a) dragging the fluid from one phase to another, b) sitting at the
fluid-fluid interface (\cite{SEGA} \textendash{} Reproduced by permission
of The Royal Society of Chemistry), in absence c) and in presence
of colloids d) (\cite{Tiribocchi19} \textendash{} Reproduced by permission
of The Royal Society of Chemistry) modifying the shape of the interface
to a corrugated one to reflect molecular correlations. \label{fig:Sega}}
\end{figure}

Finally, due to the crucial role played by high-performance computing
in this story, in the Appendix we provide an extended account of the
main issues involved with the implementation of the LBPD scheme on
high-end parallel computers in the Exascale range.

The main message we wish to convey in this Review is that a \emph{mesoscale
physics-based approach to computational medicine} may come to age
in the next decade. 

\section{BOLTZMANN KINETIC THEORY}

The Boltzmann equation (BE) is the core of Boltzmann's kinetic theory,
that, in turn, is the cornerstone of non-equilibrium statistical mechanics,
a pillar of theoretical physics at large \cite{BOL}. Besides its
paramount conceptual value as a bridge between the microscopic world
of atoms and molecules and the macroscopic world of thermo-hydrodynamic
fields, the BE also provides a concrete tool for the quantitative
investigation of a broad variety of practical non-equilibrium transport
problems~\cite{CERCI}. However, the BE is all but an easy piece
to work with: a non-linear integro-differential equation in $6+1$
(phase-space plus time) dimensions. This motivates a relentless search
for new methods to solve the BE either analytically or numerically,
the latter option usually covering a broader ground. Graeme Bird's
Direct Simulation Monte Carlo (DSMC) method has played a leading role
in this respect and continues to do so to the present days~\cite{BIRD}.
In principle, DSMC solves the BE directly and in full, i.e. accounting
for the specificity of molecular interactions, as well as strong non-equilibrium
effects, using a stochastic particle technique, whence the Monte Carlo
label. This comes at a major computational cost, which is why various
approximations have been developed and considerably refined over the
years~\cite{DVM1,DVM2}. Close to local equilibrium and away from
confining elements, however, molecular details become increasingly
irrelevant. Universality takes stage and more economical descriptions
can be devised. The basic idea is to relinquish the ``irrelevant''
details while still preserving the basic properties of macroscopic
physics, namely the symmetries and conservation laws which secure
the emergence of the NS equations from the underlying molecular dynamics.
Among others, a description which has gained major interest for the
last three decades is the LB method~\cite{BSV,LB1,LB1b,LB2}. LB
was devised with the specific intent of providing an alternative to
the discretization of the NS equations for the numerical solution
of continuum hydrodynamic problems. This still is its mainstay and,
for some authors, also the only place where it belongs. Indeed, the
use of LB for flows beyond NS was ruled out~\cite{LuoComment,KlarComment},
mostly on account of the lack of a rigorous asymptotic limit. The
above no-go has been proven largely over-restrictive and nowadays
applications beyond the strict realm of continuum fluid dynamics abound,
especially in the direction of soft matter. Since problems in biology
and medicine hardly involve fluid mechanics alone, such developments
are of direct relevance to computational explorations at the interface
between physics, chemistry and biology, the main scope of this Review.

\subsection{The Boltzmann Equation}

The Boltzmann equation (BE) of classic kinetic theory is basically
a continuity equation in six-dimensional \textit{phase-space}, namely
\cite{BOL}: 
\begin{equation}
\partial_{t}f+\vec{v}\cdot\nabla_{r}f+\vec{a}\cdot\nabla_{v}f=C(f,f)\label{BE}
\end{equation}
where $f\equiv f(\vec{r},\vec{v};t)$ is the probability density of
finding a molecule at position $\vec{r}$ in ordinary space, with
velocity $\vec{v}$ at time $t$. The left-hand side of the BE represents
the streaming of the molecules under the effect of a force field $\vec{F}=m\vec{a}$
and reflects the Newtonian mechanics $d\vec{r}/dt=\vec{v}$ and $d\vec{v}/dt=\vec{F}/m$
of a representative molecule, while the right-hand side describes
intermolecular interactions. For the case of a dilute gas, as originally
considered by Boltzmann, these interactions are typically taken in
the form of two-body local collisions, since higher order encounters
are much less frequent, hence negligible in the collision count. Even
under such major simplifications, solving the BE presents a very daunting
challenge on account of its high-dimensionality, six phase-space dimensions
plus time, as well as due to the non-linear (quadratic) integral character
of the collision operator~\cite{CERCI}.

Regardless of the complexity of the underlying microscopic interactions,
the collision operator must comply with the mass-momentum-energy conservation
laws, namely: 
\begin{eqnarray}
\int C(f,f)\{1,\vec{v},v^{2}\}d\vec{v}=0\label{CLAW}
\end{eqnarray}
In addition, it must also secure compliance with the Second Principle,
which amounts to supporting a so-called H-theorem, namely: 
\begin{equation}
-\frac{d}{dt}\;\int flogf\;d\vec{r}d\vec{v}\ge0\label{H}
\end{equation}
In other words, the dynamics of the distribution function must converge
to a universal global attractor, corresponding to the thermodynamic
equilibrium.

The macroscopic fluid variables are obtained by a linear and local
contraction of the Boltzmann distribution, namely: 
\begin{eqnarray}
\rho(\vec{r},t) & = & \int fd\vec{v}\label{MOM}\\
\rho\vec{u}(\vec{r},t) & = & \int f\vec{v}d\vec{v}\\
\rho k_{B}T(\vec{r},t) & = & \int fm(v-u)^{2}d\vec{v}=0
\end{eqnarray}
where $\rho$ is the mass density, $\vec{u}$ the flow speed, and
$T$ the flow temperature in $D$ spatial dimensions.

Central to the emergence of hydrodynamic behavior is the notion of
\textit{local equilibrium}. This is defined as the specific form attained
by the Boltzmann distribution once collisions come in complete balance,
i.e. 
\begin{equation}
C(f^{eq},f^{eq})=0\label{LEQ1}
\end{equation}

Inspection of the Boltzmann collision operator provides the following
universal Maxwell-Boltzmann (MB) local equilibrium distribution:

\begin{equation}
f^{eq}(\vec{v},\vec{r},t)=Z^{-1}\rho e^{-c^{2}/2}\label{MB}
\end{equation}
where $Z=(2\pi v_{T}^{2})^{D/2}$ is a normalization constant in $D$
spatial dimensions, $v_{T}=\sqrt{k_{B}T/m}$ is the thermal speed
and 
\[
\vec{c}=\frac{\vec{v}-\vec{u}}{v_{T}}
\]
is the peculiar speed, i.e., the molecular velocity relative to the
fluid one, in units of the thermal speed. The reader familiar with
statistical mechanics will readily recognize the canonical distribution
$e^{-E/k_{B}T}$ in the co-moving frame of the fluid, with the identification
$E=mc^{2}/2$ .

A few comments are in order.

First, the MB distribution depends on space and time only through
the hydrodynamic fields, $\lbrace n(\vec{r},t),\vec{u}(\vec{r},t),T(\vec{r},t)\rbrace$,
its dependence on the velocity variable being a universal Gaussian
distribution. This is a strict consequence of Eq. \ref{CLAW}, i.e.,
the microscopic conservation laws.

Such dependence is largely arbitrary, with the caveat that it should
be weak on the molecular scale. More precisely, the macrofields should
not show appreciable changes on the scale of the molecular \textit{mean-free
path}, that is 
\begin{equation}
Kn\equiv\lambda|\frac{\nabla_{r}M}{M}|\ll1
\end{equation}
where $M$ designates any macrofield and $\lambda$ is the molecular
mean free path. The above ratio, known as \textit{Knudsen number},
serves as the smallness parameter controlling the emergence of the
hydrodynamic limit from Boltzmann's kinetic equation. Ordinary fluids
dynamics holds in the range $Kn\sim0.01$ and below.

\subsection{From Boltzmann to Navier-Stokes hydrodynamics}

The conceptual path from BE to the NS equations of continuum fluids
is based on two fundamental steps: 
\begin{enumerate}
\item \textit{Projection of the Boltzmann equation upon a suitable basis
function in velocity space, typically Hermite polynomials in cartesian
coordinates}, 
\item \textit{Multiscale expansion using the Knudsen number as a smallness
parameter, on the assumption of weak departure from local equilibrum.
} 
\end{enumerate}
The projection generates a hierarchy of partial differential equations
for the kinetic moments: 
\begin{eqnarray}
\partial_{t}M_{0}+\nabla\cdot M_{1} & = & 0\\
\partial_{t}M_{1}+\nabla\cdot M_{2} & = & 0\\
\partial_{t}M_{2}+\nabla\cdot M_{3} & = & \frac{M_{2}^{eq}-M_{2}}{\tau}
\end{eqnarray}
where 
\begin{equation}
M_{n}\equiv M_{n}(\vec{r};t)=\int f(\vec{r},\vec{v};t)H_{n}(\vec{v})d\vec{v}
\end{equation}
and $H_{n}(v)$ denotes the $n$-th order tensor Hermite polynomial.
Note that $M_{n}$ is a tensor of rank $n$, namely $M_{0}$ (scalar)
is the fluid density, $M_{1}$ (vector) is the fluid current and $M_{2}$
(second order tensor) is the momentum-flux, whose trace delivers (twice)
the kinetic energy of the fluid and the triple tensor $M_{3}$ is
the flux of momentum flux.

The left hand side clearly shows that the moment hierarchy is open,
since the time derivative of $M_{n}$ is driven by the divergence
of $M_{n+1}$. This is the mechanism by which heterogeneities fuel
non-equilibrium. Also to be noted that the right-hand-side of the
first two equations is zero because collisions conserve mass and momentum.
However, they do \textit{not} conserve momentum-flux, which is why
the right-hand-side of the third equation is non-zero, expressing
the relaxation of the momentum-flux to its equilibrium expression.
Such relaxation takes place on a collisional timescale $\tau$, which
in turn fixes the \textit{kinematic viscosity} of the fluid. Instantaneous
relaxation ($\tau\to0$) denotes the infinitely strong collisional
regime whereby collisions do not leave any chance to non-equilibrium
to survive, formally corresponding to the idealized case of a \textit{perfect}
(zero dissipation) fluid.

Macroscopically, this corresponds to the inviscid Euler equations.

All real fluids, though, relax in a short but finite time (strictly
speaking this is also true for superfluids), and consequently the
moment equations present an open hierarchy which needs to be closed,
somehow. This is where the assumption of weak departure from local
equilibrium takes stage.

By that assumption, one formally expands the distribution function
\textit{and} space-time derivatives in powers of the Knudsen number,
replaces the expansion in the moment equations, and collects homologue
terms order by order in the Knudsen number. To zero order, the Euler
equations are obtained, whereas the first order delivers the NS equations
of dissipative fluids, namely

\begin{eqnarray}
\partial_{t}\rho+\nabla\cdot(\rho\vec{u})=0\label{NSE}\\
\partial_{t}(\rho\vec{u})+\nabla\cdot(\rho\vec{u}\vec{u})=-\nabla p+\nabla\cdot\dvec\sigma
\end{eqnarray}
where $\dvec\sigma$ is a second order tensor formed by spatial derivatives
of the flow field. Typically: 
\begin{eqnarray}
\dvec\sigma & = & 2\mu\dvec S+\lambda d\dvec I
\end{eqnarray}
where $\dvec S=1/2(\nabla\vec{u}+(\nabla\vec{u})^{T})$ is the symmetrized
gradient tensor, $d=\nabla\cdot\vec{u}$ is the divergence of the
flow and $\dvec I$ is the unit tensor. The first scalar $\mu$ is
the dynamic shear viscosity whereas $\lambda$ associates to the bulk
viscosity, $\eta=2\mu/3+\lambda$.

Despite their deceivingly simple physical content, essentially mass
and momentum conservation (Newton's law), as applied to a finite volume
of fluid, the NS equations prove exceedingly difficult to solve, as
they involve the non-linear evolution of a three-dimensional vector
field, often in a complex geometry set up. This sets a formidable
challenge to even the most advanced computational methods, whence
the ceaseless hunt for more efficient numerical methods. Those methods
include a broad array of techniques to discretize the NS equations,
using finite differences/elements/volume schemes.

Three decades ago, however, an entirely different route was devised,
which consists in attacking fluid dynamics ``from the bottom'',
i.e., appealing to a microscopic description of the fluid states matter,
namely a highly stylized version of molecular dynamics known as Lattice
Gas Cellular Automata \cite{LGCA,rivet2005lattice}. In a nutshell,
the idea is to introduce a boolean lattice fluid, consisting of a
set of boolean particles whose dynamics is confined to the lattice
sites. Boolean, here, means that the state of the system at a given
lattice site and instant in time is uniquely defined by a set of $b$
binary digits, coding for the absence/presence of a corresponding
particle moving with unit speed along one of the $b$ links connecting
each lattice site to its neighbours. By a suitable choice of the lattice
connectivity and interaction rules, such Boolean system can be shown
to reproduce the NS equations of continuum fluid dynamics. An associated
LB equation was also derived in the process of taking the boolean
automaton to NS, but its computational capabilities went under noticed.
Even though the LGCA did not make it into a competitive tool for computational
fluid dynamics, it nonetheless paved the way to the idea of computing
fluid flows by simulating fictitious particle dynamics instead of
discretising the NS equations. LB \cite{LBE} fully inscribes within
this line of thought.

\subsection{Boltzmann equation for Biology?}

The Boltzmann factor $e^{-E/k_{B}T}$ is a household name in biology,
as it governs the statistical behaviour of a broad class of equilibrium
and non-equilibrium (activated processes) phenomena of utmost relevance
to biological systems. But, how about the Boltzmann equation? At first
sight, Boltzmann kinetic theory, in its original form at least, offers
little scope for biological applications, since it formally applies
to dilute states of matter in which molecular collisions are rare,
the so-called weakly coupled regime, in which many-body interactions
can safely be neglected with respect to binary molecular encounters.
On the contrary, most biological phenomena are hosted primarily by
condensed and soft matter systems in which many-body effects play
a primary role.

Here, however, a change of perspective proves exceedingly fruitful.
Rather than the original Boltzmann equation for actual molecules,
we shall refer to \emph{model} Boltzmann's equations for \textcolor{black}{\emph{fluid-particles}},
the latter denoting the effective degrees of freedom describing the
behaviour of representative groups of molecules. Historically, the
distinctive feature of model Boltzmann equations is a dramatic simplification
of the collision operator, in an attempt to relinquish most mathematical
complexities while retaining the essential physics at hand. The most
popular Boltzmann model is the celebrated Bhatnagar-Gross-Krook equation,
in which the collision operator is replaced by a simple single-time
relaxation term \cite{BGK}:

\begin{equation}
C_{BGK}=-\omega(f-f^{eq})\label{eq:BGK}
\end{equation}
where $f^{eq}$ is the local equilibrium and $\omega$ is a relaxation
frequency controlling the relaxation to the local equilibrium on a
timescale $\tau=1/\omega$.

The key advantage of this change is neat: model equations are more
flexible by construction, hence they can be modified (extended) not
only to simplify the collision operator but also to describe a wide
host of physical effects not included in the original Boltzmann equation.

In particular, they can reinstate the effects of, $i)$ many-body
interactions, via effective one-body forces, in the spirit of Density
Functional Theory (Vlasov-Boltzmann Equation) \cite{hansen1990theory}
$ii)$ statistical fluctuations, through appropriate stochastic sources
(Fluctuating Boltzmann Equation) \cite{LADDFLUCT} $iii)$ far-from-equilibrium
inhomogeneities, via suitably extended collision-relaxation operators
in which the relaxation time is promoted to the status of a self-consistent
dynamic field \cite{CHEN2003,LBE,ECO2}.

The three extensions above inscribe to the general framework of \textcolor{black}{\emph{Reverse
Kinetic Theory}} (RKT), the strategy whereby the kinetic equation
is \textcolor{black}{\emph{designed}} top-down, based on prescriptions
securing compliance with macroscopic hydrodynamics in the first place,
and then adding ``molecular'' details ``on-demand'' by the specific
application under investigation.

RKT reverses the canonical bottom-up route, whereby kinetic equations
are \emph{derived} from the underlying microscopic models and proves
quite effective in bringing Boltzmann-like equations within the realm
of condensed and soft matter physics. However, care must be exercised
in securing compliance of the top-down approach with the basic principles
of statistical physics, namely symmetries/conservation laws as well
as evolutionary constraints (the Second Principle and its local form,
known as H-theorem).

This all said and done, a practical question still remains: although
simplified, the model BE's still leave in an unwieldy $6+1$ dimensional
space. Here, a time-honoured ally of any computational scientist,
the lattice, makes its glorious entry.

\subsection{Hydrodynamics for Biology}

Hydrodynamics and biology couple across an amazingly broad spectrum
of scales, ranging from the macromolecular level in the compartments
of living cells, all the way up to industrial bioreactors, rivers,
lakes and oceans. From macromolecular motion in crowded cellular spaces,
to the deformations of membranes, to the motion of cells and the self-propulsion
of bacteria, the common goal is to capture the effect of hydrodynamics
under both equilibrium and non-equilibrium conditions. The latter
may arise under the effect of an external flow, such as the transport
of proteins and cells in the blood stream, or via the consumption
of energy in cellular metabolic pathways. Indeed, the response of
biological matter to temperature, pH or mechanical forces, plays a
key role in most biological processes. For instance, biological macromolecules,
cells or tissues, are usually fragile and easily damaged by hydrodynamic
or shear forces, as they occur far from equilibrium and under strong
confinement. Following in the footsteps of hydrodynamic experimental
techniques, customarily used to measure basic molecular properties,
such as weight, size and shape, computer simulations can be used in
biophysical chemistry to explore and assess the mechanical and dynamical
properties of macromolecules far from equilibrium, hence much closer
to the \emph{in vivo} conditions relevant to medical purposes. The
benefits of including hydrodynamic forces is even more apparent whenever
thermodynamic forces, stemming from solute-solvent interactions, can
also be taken into account. A comprehensive computational framework
including both hydrodynamic and thermodynamic forces, stands therefore
as a most desirable target. Owing to its inherently intermediate nature,
between atomistic and continuum descriptions of matter, (lattice)
kinetic theory sits at a vantage point to meet this goal. 


\section{LATTICE BOLTZMANN FOR CONTINUUM HYDRODYNAMICS}


The basic idea of the LBM is to constrain the velocity degrees-of-freedom
to a discrete lattice with sufficient symmetry to protect the conservation
laws which secure the emergence of standard hydrodynamic behavior
in the macroscopic limit. LB is based on the idea of representing
fluid populations on a uniform, cartesian grid.

The standard LB scheme in single-relaxation time (BGK) form reads
as follows~\cite{LBE,LBGK}: 
\begin{equation}
f_{i}(\vec{r}+\vec{c}_{i};t+\Delta t)-f_{i}(\vec{r};t)=-\omega(f_{i}(\vec{r};t)-f_{i}^{eq}(\vec{r};t))+S_{i}(\vec{r};t),\;i=0,b\label{LBGK}
\end{equation}
where $f_{i}$ is the discrete Boltzmann distribution associated with
the discrete velocity $\vec{c}_{i}$, $i=0,b$ running over the discrete
lattice, to be detailed shortly. In the above, $\omega$ is a relaxation
parameter controlling the fluid viscosity and $f_{i}^{eq}$ is the
lattice local equilibrium, basically the local Maxwell-Boltzmann distribution
truncated to the second order in the Mach number. The truncation is
not a luxury, but a requirement dictated by Galilean invariance (GI). 

Let us remind that GI refers to the invariance of the  NS equations
under an arbitrary  change of  the local fluid velocity $\vec{u}\to\vec{u'}$;
upon such change, the NS equations stay  the same, provided  $\vec{u}$
is replaced by $\vec{u'}$. In continuum kinetic theory, Galilean
invariance is encoded within the dependence of the local Maxwell-Boltzmann
distribution on the relative velocity of the molecules with respect
to the fluid one, namely $\vec{v}-\vec{u}$.    As a result, an observer
in the comoving frame, that is a frame moving at the local fluid velocity,
experiences the same local equilibrium as if there were no fluid motion.
 To be noted that, in the above, $\vec{u}\equiv\vec{u}(\vec{r},t)$
is an arbitrary function of space and time, indicating that Galilean
invariance is a local and continuum symmetry, i.e., it holds even
if different regions of the fluid move  at different velocities, which
is of course the case in most fluids of practical interest. Galilean
invariance is reflected by the specific form taken by the moments
of the equilibrium distribution, and specifically by those explicitly
relevant to hydrodynamics, namely mass, momentum and the momentum
flux-tensor, that is:  
\[
\int f^{eq}\lbrace1,v_{a},v_{a}v_{b}\rbrace d\vec{v}=\lbrace\rho,\rho u_{a},\rho u_{a}u_{b}+p\delta_{ab}\rbrace
\]
where latin subscripts run over spatial coordinates $x,y,z$. The
above expression follows straight from the property of gaussian integrals
in velocity space.

One might naively expect that the same would be true in the lattice,
provided the local Maxwell-Boltzmann distribution  is retained  with
the plain replacement $\vec{v}=\vec{c}_{i}$. Straightforward algebra
shows that the situation is different not for a mere mathematical
accident, but as a consequence of the fact that a local and continuum
symmetry cannot remain unbroken in a discrete lattice. More precisely,
it cannot remain unbroken for any \emph{arbitrary} velocity field.
 It turns out, though, that this is possible whenever the fluid velocity
is much smaller than the sound speed, i.e., in the low-Mach number
limit.

Under such a perturbative approximation, replacing velocity integrals
with discrete sums returns exactly  the same moments as in the continuum,
namely: 
\begin{equation}
\sum_{i}f_{i}^{eq}\lbrace1,c_{ia}c_{ia}c_{ib}\rbrace=\lbrace\rho,\rho u_{a},\rho u_{a}u_{b}+p\delta_{ab}\rbrace\label{eq:MME}
\end{equation}
provided that lattice tensors up to fourth order are isotropic. Of
course, this by no means implies that GI is fully restored, but simply
that the GI-breaking terms are confined to kinetic moments higher
than order four.

In other words, the hydrodynamic constraints can be matched perturbatively,
by expanding the local Maxwellian to second order in the Mach number,
which is sufficient to recover the (isothermal) NS equations, since
those equations are quadratic in the fluid velocity. Full Galilean
invariance for an arbitrary flow field $u_{a}$ implies instead an
\emph{infinite} series in the Mach number, corresponding to the full
expansion of the local Maxwellian in Hermite polynomials. 

The actual expression of the discrete local equilibria reads as follows:
\begin{equation}
f_{i}^{eq}=w_{i}\rho(1+u_{i}+\frac{1}{2}q_{i})\label{eq:LEQ}
\end{equation}
where $u_{i}=c_{ia}u_{a}/c_{s}^{2}$ and $q_{i}=(c_{ia}c_{ib}-c_{s}^{2}\delta_{ab})u_{a}u_{b}/c_{s}^{4}$
represent the dipole and quadrupole contributions, respectively. Hereafter,
repeated indices are summed upon.  In the above, $w_{i}$ is a set
of lattice-dependent weights, normalized to unity, which represent
the lattice analogue of the global (no-flow) Maxwell-Boltzmann distribution.
Finally $c_{s}^{2}=\sum_{i}w_{i}c_{ia}^{2}$ is the lattice sound
speed.

We hasten to note that, at variance with its continuum counterpart,
the  expression (\ref{eq:LEQ}), being a polynomial truncation of
the Maxwell-Boltzmann equilibria,  is non-negative definite only in
a finite range of fluid velocities, typically of the order of $u/c_{s}\sim0.3$.
 This configures the LB method as an appropriate description of quasi-incompressible,
low Mach-number flows.

\begin{figure}
\centering{}\includegraphics[scale=0.6]{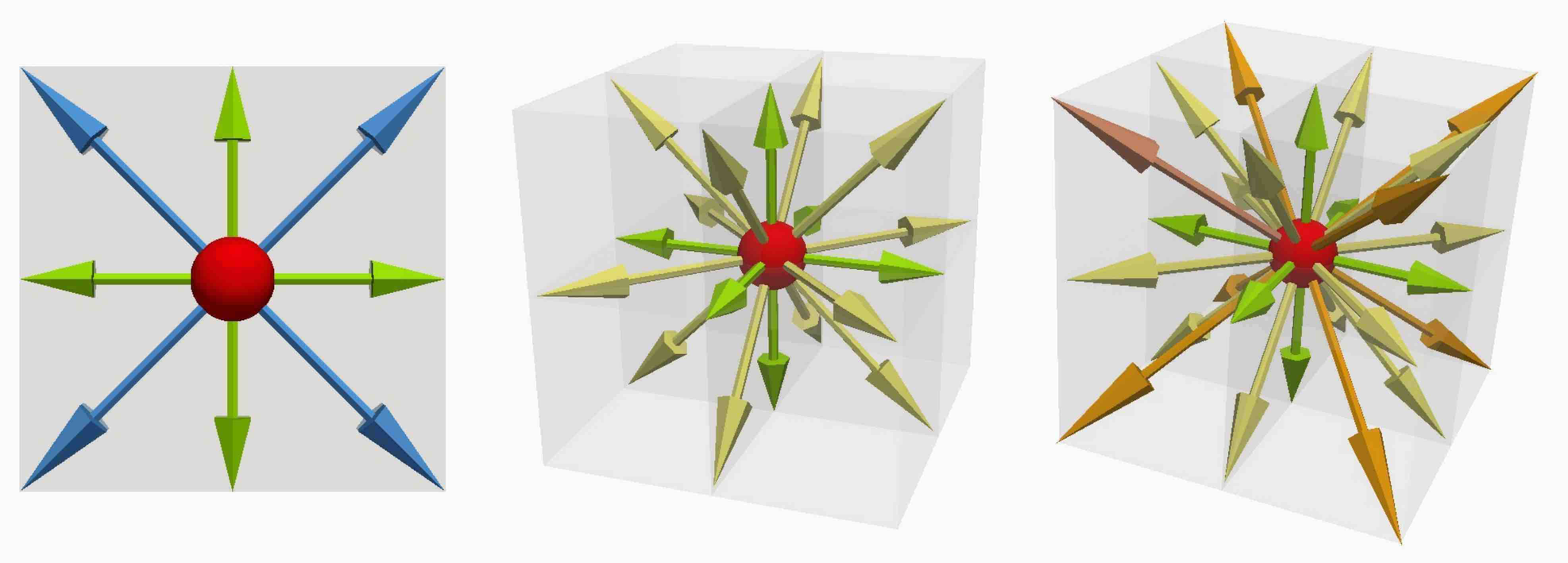}
\caption{Examples of standard 2d and 3d LB lattices, with $9$, $19$ and $27$
discrete speeds, typically denoted as D2Q9, D3Q19 and D3Q27, respectively.\label{fig:pops} }
\end{figure}

The discrete velocities matching isotropy constraints up to fourth
order can be shown to be subsets of the D3Q27 (D$n$Q$m$ is a widely
used notation to indicate a LB scheme in $n$ dimensions using a set
of $m$ velocities) mother lattice with $27$ discrete speeds in three
spatial dimensions. D2Q9 and D3Q27 are the direct tensor product of
the elementary one-dimensional D1Q3 stencil, $c_{ix}=\lbrace-1,0,+1\rbrace$,
in two and three spatial dimensions, respectively (see Fig. \ref{fig:pops}).

The expansion (\ref{eq:LEQ}) implies that hydrodynamic LB flows are
bound to be quasi-incompressible, i.e., Mach number well below unity.
Likewise, third order kinetic moments, describing energy and heat
flux, are not correctly reproduced, since these terms require sixth-order
isotropic lattice tensors. Such constraints can not be met by lattices
confined to the first Brillouin region described by D3Q27: higher
order lattices extending beyond the first Brillouin cell are necessary.
So much for the low Mach number approximation, which is specific to
the lattice. But what about the low Knudsen limit, which, on the contrary,
lies at the very roots of the convergence of Boltzmann to NS?

Since the Knudsen number controls the heterogeneity-driven departures
from local equilibrium, it is  intuitively clear that the low-Knudsen
hydrodynamic limit implies further constraints on the non-equilibrium
component of the momentum-flux tensor, which amounts to recovering
the continuum expression of the stress tensor.

Based on (\ref{eq:MME}) and (\ref{eq:LEQ}), it can be shown that
in order for such constraints to be met, isotropic tensors,  again
up to order four, need to be exactly reproduced in the lattice. It
may come as a surprise that non-equilibrium constraints can be matched
at the same order of isotropy of the equilibrium ones. The reason
is parity-invariance, namely the fact that each discrete lattice velocity
$\vec{c}_{i}$ comes with a mirror partner $\vec{c}_{i^{*}}=-\vec{c}_{i}$.
Indeed, non equilibrium constraints involve lattice corrections driven
by fifth-order lattice tensors, which are identically zero due to
parity invariance.  Readers interested in the straightforward but
laborious algebraic details, may look up the vaste literature on the
subject\cite{LGCA,rivet2005lattice,henon1987viscosity}.

Next, let us comment on the source term, $S_{i}$, at the right-hand
side of the LB equation (\ref{LBGK}). This term stands for a generic
source of mass/momentum/energy, describing the coupling of the fluid
to the surrounding environment. Mass sources are typically associated
with the presence of chemical reactions, turning one species into
another in multi-component versions of the LB for reactive flows.

In the case of inert flows, the source term typically encodes the
momentum exchange due to the coupling to external (or internal) fields,
such as gravity or more complex interactions, like self-consistent
forces reflecting potential energy interactions within the fluid,
as well as thermal fluctuations.  As we shall see, the two latter
cases are crucial for the extension of LB beyond NS hydrodynamics.

From the operational standpoint, the source term $S_{i}$ acts like
a bias promoting the populations which move along the local force
field and setting a penalty on those that move against it. It is therefore
clear that the strength of such term is subject to stringent stability
and positivity constraints. 

\subsection{From Lattice Boltzmann to continuum hydrodynamics}

The conceptual path taking LB to NS is exactly the same as in the
continuum theory, with the crucial caveat of turning around the (many)
catches associated with lattice discreteness. To make a long story
short, it amounts to securing the proper symmetries of the lattice
tensors entering the set of (lattice) moment equations. Like always
with lattice physics, the name of the game is to erase the lattice
dependence to the desired order. For the case of isothermal, incompressible
fluids, the desired order is the \textit{fourth} one. In equations,
and using coordinate notation for tensors \cite{henon1987viscosity}:
\begin{eqnarray}
\sum_{i}w_{i} & = & 1\\
\sum_{i}w_{i}c_{ia}c_{ib} & = & c_{s}^{2}\delta_{ab}\\
\sum_{i}w_{i}c_{ia}c_{ib}c_{ic}c_{id} & = & c_{s}^{4}(\delta_{ab}\delta_{cd}+\delta_{ac}\delta_{bd}+\delta_{ad}\delta_{bc})
\end{eqnarray}
where latin indices run over spatial coordinates. In the above, $w_{i}$
is a set of weights normalized to unity and odd-rank tensors are automatically
equal to zero by parity invariance, i.e., every discrete velocity
$c_{ia}$ comes with an equal and opposite partner, so that the sum
of the two gives a null. Finally $c_{s}^{2}$ is a lattice dependent
constant expressing the sound speed in the lattice, as per the expressions
above.

With the symmetries secured, everything proceeds like in the continuum
theory, with another important caveat though, namely the fact that
the \textit{effective} mean-free path of the lattice fluid is replaced
by the lattice spacing $\Delta x$ whenever the latter is larger than
the physical one, the typical case in most macroscopic LB simulations.

This is a very technical and somehow thorny issue, whose details are
out-of-the-scope of the present Review. Nevertheless, we wish to caution
the reader that the physics taking place at the scale of a few lattice
spacings should always be inspected with a big pinch of salt, because
it is constantly in danger of breaking the hydrodynamic assumptions.

Once these catches are disposed of, one ends up with a lattice fluid
obeying an ideal equation of state $p=\rho c_{s}^{2}$ and a kinematic
viscosity given by:

\begin{equation}
\nu=\left(\frac{1}{\omega}-\frac{1}{2}\right)\nu_{l}\label{eq:viscosity}
\end{equation}
where $\nu_{l}=c_{s}^{2}\Delta x^{2}/\Delta t$ is the natural lattice
viscosity. Note that the stability range of the discrete time-marching,
$0<\omega<2$, also secures the positivity of the kinematic viscosity.

Both ends of this range must be handled with care. In the low-viscosity
regime $\omega\to2$, typical of turbulence, strong gradients may
develop posing a serious threat to the numerical stability of the
scheme. A powerful variant of the basic LB includes a self-consistent
tuning of the relaxation parameter $\omega$ so as to ensure compliance
with local entropy growth (H-theorem) \cite{ELB1,ELB2}. That variant,
known as Entropic Lattice Boltzmann (ELB) is normally intended to
simulate high-Reynolds macroscopic flows, but lately is proving very
effective also to stabilize microscale LB simulations with sharp interfaces.

In the opposite high-viscous regime, $\omega\to0$, the one most relevant
to biological applications, the viscosity may formally diverge, signalling
a departure from hydrodynamic behaviour, due to the onset of ballistic
motion. As a result, whenever possible, LB simulations should be kept
away from both limits, say $\nu_{LB}\sim0.1$.

We shall return to this important point in the Section \char`\"{}Future
Challenges\char`\"{}.

\subsection{Boundary conditions}

\begin{figure}
\begin{centering}
\includegraphics[scale=0.4]{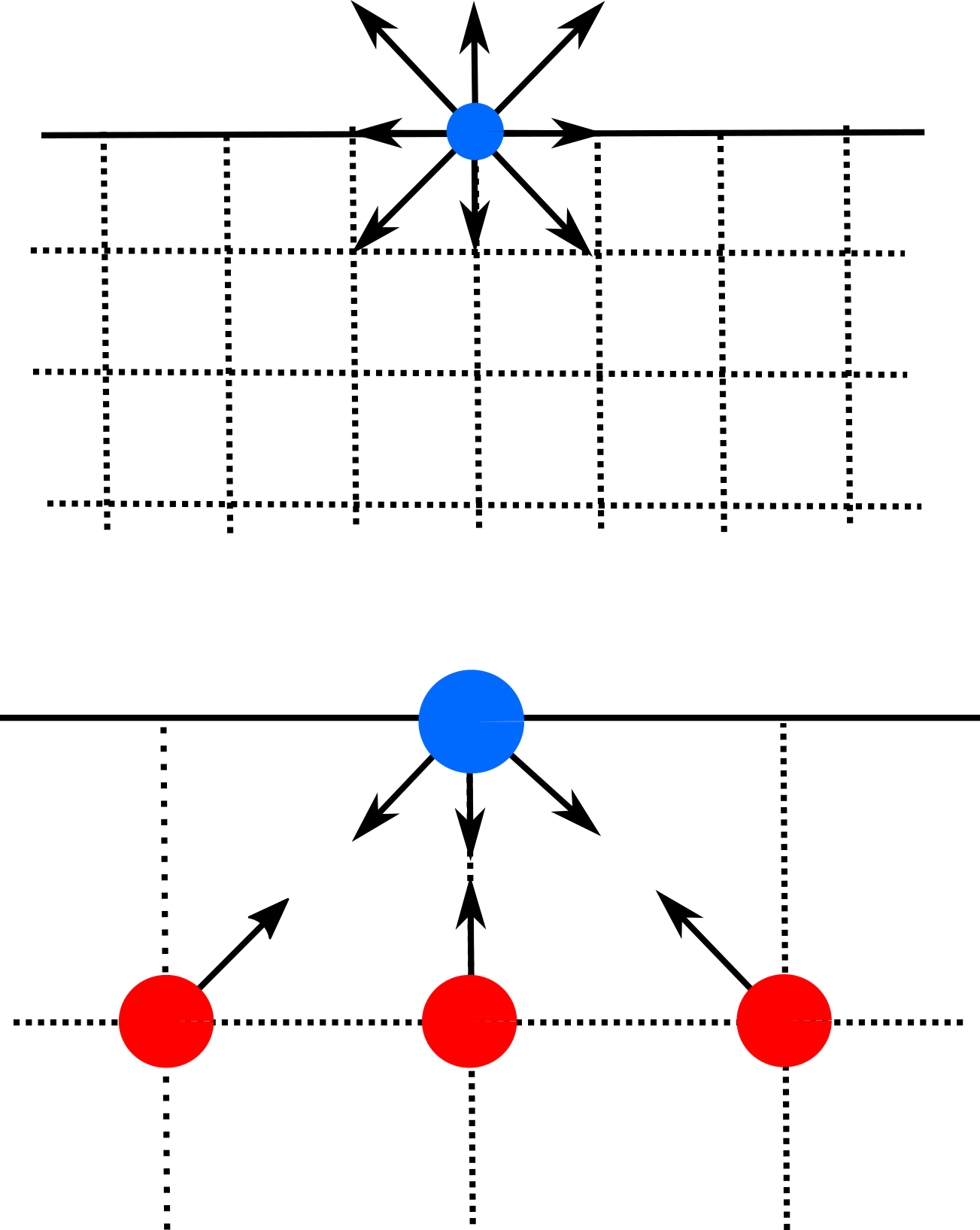} 
\par\end{centering}
\caption{The effect of wall boundaries on the LB populations according to the
Bounce Back scheme. Populations hitting a wall node are reflected
from the incoming fluid direction, giving rise to no-slip flow conditions.\label{fig:BounceBack}}
\end{figure}

In the early days, boundary conditions were hailed as one of the main
assets of LB, and, to a certain extent, they still are. As a matter
of fact, since information always travels along straight lines, even
complex geometries can be handled by comparatively straightforward
computational methods based on elementary mechanical operations. For
instance, no-slip on solid walls can be implemented through a simple
bounce-back between distributions propagating along opposite directions,
i.e., from the fluid to the wall and viceversa (See Fig. \ref{fig:BounceBack}).

For the sake of generality, let us consider a fluid flowing in a bounded
domain $\Omega$ confined by a surrounding boundary $\partial\Omega$.
The problem of formulating boundary conditions within the LBE formalism
consists in finding an appropriate relation expressing the \emph{incoming}
(unknown) populations $f_{i}^{<}$ as a function of the \emph{outgoing}
(known) ones $f_{i}^{>}$. Outgoing populations at a boundary site
$\vec{x}$ are defined by the condition 
\[
\vec{c}_{i}\cdot\vec{n}>0,
\]
where $\vec{n}$ is the outward normal to the boundary cell centered
in $\vec{x}$. Incoming populations are defined by the opposite sign
of the inequality. In mathematical terms, this relationship translates
into a linear integral equation 
\begin{equation}
f_{i}^{<}\left(\vec{x}\right)=\sum_{y}\sum_{j}B_{ij}\left(\vec{x}-\vec{y}\right)f_{j}^{>}\left(\vec{y}\right),\label{BCGEN}
\end{equation}
where the kernel $B_{ij}\left(\vec{x}-\vec{y}\right)$ of the boundary
operator generally extends over a finite range of values $\vec{y}$
inside the fluid domain. This boundary operator reflects the interaction
among the fluid molecules and the molecules in the solid wall. Consistently
with this molecular picture, boundary conditions can be viewed as
a special (sometimes even simplified) type of collisions between fluid
and solid molecules. Physical fidelity can make the boundary kernel
quite complicated, which is generally not the idea with LBE. Instead,
one usually looks for expressions minimizing the mathematical burden
without compromising the essential physics.

In particular, one seeks minimal kernels fulfilling the desired constraints
on the macroscopic variables (density, speed, temperature and possibly
the associated fluxes as well) at the boundary sites $\vec{x}$. This
may lead to a mathematically under-determined problem, more unknowns
than constraints\emph{, }\textcolor{black}{opening up an appealing
opportunity to accommodate more interface physics within the formulation
of the boundary conditions. However, it also calls for some caution
to guard against mathematical ill-posedness. }By now, a vast literature
which, however, goes beyond the scope of the present Review, is available
on this topic \cite{LB2,ZOUHE,chen1996boundary}.

Here, we just wish to point out that the main issue controlling the
complexity of the boundary problem is whether the collection of boundary
points lies on a surface aligned with the grid (so-called staircase
wall boundary) or it is given by a set of off-grid surface elements
(sometimes called surfels). The latter case is significantly more
complex and requires extra-care, the immediate benefit being that
the near-wall physics is second-order accurate as compared to the
first-order accuracy of the staircased approximation (for a detailed
account see \cite{LB1b,LB2}). Advanced applications of the off-lattice
boundary method prove capable of dealing with fairly complex geometries,
an important but highly technical topic which also goes beyond the
scope of the present Review. For full details see the recent books
\cite{LB2,LB1b} 

\subsection{The bright sides of Lattice Boltzmann}

Why does LB represent an appealing scheme for simulating complex states
of flowing matter ?

Several features merit highlight, but essentially they all track down
to the benefits of working with extra-dimensions opened up by the
six-dimensional phase-space inhabited by kinetic theory.

More in detail, the upshots are the following.

\emph{First}, the information always travels along straight lines,
regardless of the space-time complexity of the emergent hydrodynamics.

What we mean by this is that the discrete distributions move in sync
along the discrete light-cones, defined by: 
\[
\Delta\vec{r}_{i}=\vec{c}_{i}\Delta t,
\]
no matter the complexity of their space-time dependence.

Since the discrete velocities do not depend on space and time, the
streaming is \emph{exact}, no information is lost in hopping from
one lattice site to another, literally an error-free operation. This
stands in sharp contrast with the self-advection term $\vec{u}\cdot\nabla\vec{u}$
of macroscopic hydrodynamics, whereby the velocity fields, or any
other field for that matter, moves along space-time dependent material
lines defined by the fluid velocity itself, typically a complicated
function of space and time. 

Such independence explains the outstanding LB amenability to parallel
computing, a practical asset which can not be underestimated.

\emph{Second}, space and time always come on the same (first-order)
footing. In particular, this means that dissipation is not expressed
by second-order spatial derivatives (Laplace operator) but simply
as a local relaxation to a \emph{local} equilibrium, as described
earlier. This is a substantial advantage for \emph{confined} flows,
whose dynamics is largely dictated by the presence of solid boundaries,
where spatial gradients are usually at their largest. Taking space
derivatives near the boundaries is notoriously prone to numerical
inaccuracies, especially whenever the geometry of the domain is irregular,
as it is often the case for biological flows. LB equipped with suitable
formulations for curved boundaries can significantly mitigate such
difficulties.

\emph{Third}, LB carries pressure as just any other macroscopic field,
with no need of solving a (usually very expensive) Poisson problem
to compute the pressure field consistent with an incompressible flow.
This is because the discrete distributions carry the momentum-flux
tensor ``on their back'' and consequently the pressure obeys its
own dynamic equation. More specifically, the second order momentum-flux
tensor, $P_{ab}$, obeys the relaxation equation 
\begin{equation}
\partial_{t}P_{ab}+\partial_{c}Q_{abc}=-\omega(P_{ab}-P_{ab}^{eq})\label{eq:MOMFLUX}
\end{equation}
where $Q_{abc}=\sum_{i}f_{i}c_{ia}c_{ib}c_{ic}$ is the third-order
energy flux tensor, latin indices running over spatial dimensions.
In the limit where the $P_{ab}$ tensor is enslaved to its equilibrium
value, the time derivative can be dropped and the energy-flux tensor
can be approximated by its equilibrium expression. Under such conditions,
the above equation delivers $P_{ab}\sim P_{ab}^{eq}+\omega^{-1}\partial_{c}Q_{abc}^{eq}$.

Once the due lattice symmetries are fulfilled, the equilibrium component
delivers the advection and pressure terms of the NS equation, while
the third order non-equilibrium term delivers the dissipative term. 

\emph{Fourth}, coupling of the fluid to a broad variety of other physical
phenomena, is readily achieved by formulating suitable expressions
of the source term $S_{i}$.

Typically this reflects the action of mesoscale forces describing
the effective interactions between fluid-fluid and fluid-solid molecules.
The positive side effect is that an entire new world of complexity
may be supported \emph{simply by inserting a few tens of lines of
additional code}.

This feature is crucial to the extension of LB beyond the traditional
realm of dilute gases and particularly to biological flows, as we
shall detail in the sequel.

Once again, this rosy picture only refers to the conceptual fresco,
actual implementations requiring great care in dribbling lattice artefacts.

\begin{figure}
\begin{centering}
\includegraphics[scale=1.2]{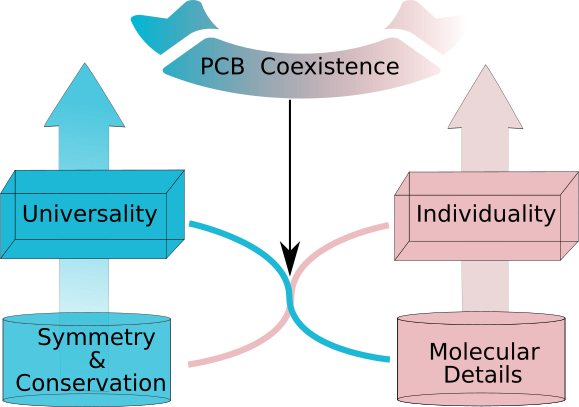} 
\par\end{centering}
\caption{The description of Physical-Chemical-Biological (PCB) systems requires
a suitable integration of microscopic details (Individuality) within
the universal harness dictated by symmetry principles and ensuing
conservation laws which govern the macroscopic behaviour (Universality).
Kinetic theory is expected to strike an optimal (problem-dependent)
balance between Universality and Individuality.\label{fig:quest}}
\end{figure}

\subsection{The dark side of the LB moon }

Following a witty line by Daan Frenkel, it is often more instructive
to analyse what can go wrong in a computer simulation, rather than
indulging in the glorious description of its success stories. To abide
by this wise attitude, in the following we mention things that can
still go wrong with LB simulations, the dark side of the LB moon.

LB draws much of its computational simplicity to the features of uniform
and regular lattices, that, at times, recalls an ideal Legoland. Realistically
complex geometries may sometimes challenge this setup, and call for
more flexible and elaborate formulations, such as interpolation procedures
for curved boundaries, local grid-refinement, mergers with finite-volumes
techniques, to name just a few of some popular options. It is only
fair to acknowledge that implementing such extensions maybe laborious
and possibly tax the computational simplicity\textcolor{blue}{{} }as
well.

Another limitation is the finite compressibility. While buying major
computational savings through dispensing with the Poisson problem
for pressure, a finite amount of compressibility must be tolerated
in return. Such effects are usually negligible for low-Reynolds flows,
but they need nevertheless to be watched carefully in open flows,
especially at outlets, where spurious density waves may eventually
back-propagate into the fluid and undermine the accuracy and sometimes
even the stability of the simulation.

A similar observation applies to flows with strong thermal transport.
For one, it should be appreciated that LB is essentially \emph{athermal},
in the sense  that the discrete Boltzmann distribution is represented
by a linear superposition of mono-chromatic beams with no dispersion
in velocity space, hence, zero temperature in a strict equilibrium
thermodynamic sense. Nevertheless, a kinetic temperature can still
be defined as a measure of the dispersion around the  mean flow velocity,
namely, in $D$ spatial dimensions, $Dk_{B}T=\sum_{i}f_{i}(\vec{c_{i}}-\vec{u})^{2}$.
The inclusion of strong thermal effects typically requires higher
order lattices, accommodating sixth order tensors  describing the
flux of energy \cite{LB2,nie2008thermal}. Several alternatives are
also possible which go beyond the scope of the present work (for details
see chapter 22 in \cite{LB1b}).  Despite major progress, it is fair
to say that thermal LB schemes still lag behind  their athermal counterparts
in terms of numerical robustness. Consequently, LB is often coupled
to  independent thermal solvers typically based on finite-difference
or finite-volume methods.

These limitations off the breast, we believe it is fair to surmise
that the appealing features of the LB method largely overweigh its
weaknesses.


\section{\label{sec:GeneralizedLB}LATTICE BOLTZMANN FOR GENERALIZED HYDRODYNAMICS}


The ideas and methods presented so far deal with flows of ``simple''
fluids that can be described in terms of the NS equations (note that
simple fluids can give rise to highly \emph{complex} flows, turbulence
being a prime example in point). 

Modern high-tech applications, nano-engineering and biology in the
first place, set a pressing demand of quantitative understanding of
more complex states of flowing matter, in which fluids interact with
external or self-consistent fields, undergo chemical reactions, phase-transitions,
or interact with a variety of suspended bodies, such as colloids or
biological molecules.

This emerging sector of modern science, often referred to as ``complex
fluids'', or more trendily, ``soft matter physics'', portrays a
multidisciplinary scenario whereby fluid dynamics makes contact with
other disciplines, primarily chemistry, material sciences and biology
as well.

There is a growing evidence that LB, and extensions thereof, holds
a vantage point as a computational framework for the simulation of
complex states of flowing matter. Ideally, LB would fill the gap between
fluid dynamics and molecular dynamics, namely the huge and all-important
region where fluid dynamics breaks down and molecular dynamics is
not yet ready to take over for lack of efficient algorithms and computing
power (see Fig. \ref{fig:quest}) \cite{boon1991molecular}.

LB is a natural candidate to fill that gap because of its mesoscopic
ability to incorporate microscopic details into the kinetic theory
formalism via suitable external fields and/or equivalent generalizations
of local hydrodynamic equilibria: a microscope for fluid mechanics,
a telescope for molecular dynamics \cite{LB1b}.

The extension of LB to generalised flows is based on a number of major
upgrades of the basic LB theory. In the sequel, we focus on the following
selection:
\begin{itemize}
\item \textit{Reactive systems} 
\item \textit{Charged flows} 
\item \textit{Flows far from equilibrium} 
\item \emph{Fluctuating hydrodynamics} 
\item \textit{Non-ideal fluids } 
\item \textit{Flows with suspended bodies } 
\end{itemize}
We now proceed to a more detailed discussion of the items above.

\subsection{Advection-Diffusion-Reaction systems}

Chemical reactivity is an essential element to deal with when facing
the task of simulating systems at the PCB interface \cite{boon1996,LB1b,coveney2016bridging,alowayyed2017multiscale}. 

Biochemical reactions lie at the heart of most biological phenomena,
controlling species interconversion. They involve the breaking and
making of covalent bonds, often catalyzed by enzymes, ubiquitous in
all metabolic and synthetic reactions within the cell and in the body.
Reactions occur in bulk conditions in single-phase (homogeneous) or
multi-phase (heterogeneous) environment, typically at the interface
between phases or in a porous-like environment. Biological reactions
take place under a wide host of different conditions. The antigen-antibody
binding in solution or the binding of small molecules to plasma proteins
in blood, clotting reactions on the surface of blood vessels, the
hydrolysis of adenosine triphosphate (ATP\nomenclature[ADR]{ATP}{adenosine triphosphate})
on the surface of endothelium, ion transport, and the release of nitric
oxide are examples of homogeneous reactions. Heterogeneous reactions
within tissues include aerobic metabolism, glucose consumption, and
receptor-mediated endocytosis.

In all cases, reactions can occur either in no-flow or under flow
conditions, as for example the enzyme reactions on the surface of
the blood vessels. Diffusion and convection affect the rates of homogeneous
and heterogeneous reactions. Because reactants must diffuse or convect
to the surface where they react, the mass transfer mechanisms occur
in sequence with the reaction process. In many cases, the effects
of reaction and diffusion on the reaction rates cannot be easily separated
and numerical methods provide the only viable route to study the process.

LB is a powerful framework to capture diffusion and reaction since
its mathematical apparatus is by no means confined to fluid equations.

As a matter of fact, by suitable tuning of the local equilibria, relaxation
matrix and external forces, it permits to generate a very broad variety
of linear and non-linear partial differential equations, including
those describing relativistic and non-relativistic quantum phenomena
\cite{SUCCI2002,MENDOZA2010}.

Of relevance are advection-diffusion-reaction (ADR) \nomenclature[ADR]{ADR}{Advection-Diffusion-Reaction}
equations of the general form: 
\begin{equation}
\partial_{t}C+\vec{u}\cdot\nabla C=D\Delta C+R(C)\label{ADR}
\end{equation}
where $C$ is a species concentration, say molecules per unit mass
or volume. The left-hand side describes the passive transport along
the fluid material lines, whereas the right-hand-side describes diffusion
plus the effects of chemical reactions occurring in the fluid moving
at the barycentric velocity $\vec{u}$.

Although thermodynamically favorable, many reactions are limited by
the energy barrier which needs to be crossed in order to form the
activated state, so that, in the absence of the catalyst, reactions
simply do not occur. In its presence, the rate of reaction can increase
dramatically, although the change in energy between reactants and
products is not affected. In fact, the enzymes affect the rate of
a reaction, not its equilibrium. For many reactions involving a single
substrate, the rate of consumption of the substrate follows the Michaelis-Menten
equation 
\[
R(C)=\frac{RC}{K+C}
\]
where $R$ is the magnitude of the rate of disappearance of the substrate
(or reactant) and $K$ is the Michaelis constant, the concentration
of the substrate at which the reaction attains half of its maximum
rate.

Another reaction term particularly relevant to populations biology
is the logistic expression 
\[
R(C)=aC-bC^{2}
\]
where $a$ describes the (Malthusian) growth rate and $b$ characterizes
the non-linear decay of the species due to competition for, say, space
and/or food. The ADR class falls naturally within the formalism, by
simply defining a LB distribution $C_{i}$ for the concentration,
such that $\sum_{i}C_{i}=C$ and specifying the following local equilibria:
\[
C_{i}^{eq}=w_{i}C\left(1+\frac{\vec{u}\cdot\vec{c_{i}}}{c_{s}^{2}}\right)
\]
where both fluid and lattice speeds are normalised by the sound speed
$c_{s}$. To be noted that in the above $\vec{u}$ is the fluid speed,
which does \emph{not} match the current $C\vec{u}\ne\sum_{i}C_{i}\vec{c}_{i}$,
because ADR's conserve mass but not momentum. The LB is then particularly
well suited to solve the ADR equations in complex geometries, such
as those occurring in morphogenesis \cite{PONCE1993,AYODELE2011},
heterogeneous catalysis \cite{FALCUCCI2016} and related phenomena.

\subsection{Charged Fluids}

Electrostatics plays a vital role in biological processes and requires
handling electrolytic solutions and charged solutes typically in flow
conditions, the realm of the so-called electrokinetics. Solutes range
from strongly charged molecules, such as DNA, to weakly charged macromolecules,
such as proteins, where partially screened mobile charges and charged
surfaces are ubiquitous conditions of biological settings. Prominent
examples include \textcolor{black}{\emph{i)}} the stability of proteins
as a function of pH and ionic strength, such as salting-in and salting-out
effects due to the interplay of protein charges with the aqueous/saline
environment; \textcolor{black}{\emph{ii)}} protein-ligand association
processes, including enzymatic allostery modulated by salt-bridges,
salt-bridges in virus assemblies and thermal stability; \textcolor{black}{\emph{iii)}}
membrane proteins and their specific electrostatic properties, whereby
the collective charge of the intracellular residues tend to be more
positive as compared to the extracellular ones (the so-called ``positive
inside'' rule), and so on. The associated biological flows are driven
by electro-osmosis and electrophoresis. Well-known examples of those
processes include \textcolor{black}{\emph{i)}} ion channels and the
gating of ions across the cell membrane that regulate electrical signalling
in secretory and epithelial cells, as much as the cell volume; \textcolor{black}{\emph{ii)}}
the electrokinetic flow due to the glycocalyx layer covering cells,
a polyelectrolyte exhibiting a surface charge and responsible for
an electrochemical gradient that regulates mechanotransduction; \textcolor{black}{\emph{iii)}}
the mitochondria and the chloraplast, where proton gradients generate
a chemi-osmotic potential, also known as a proton motive force, for
the synthesis of ATP.

Although the general principles of electrokinetics are fairly well
understood \cite{ELECTROKINETICS}, methods that translate these principles
into accurate numerical predictions are still in their infancy. In
fact, deriving the interactions between charged solutes and the solvent
requires computing the interactions of a large number of molecules
and the averaging of these over many solvent configurations. Such
daunting computational requirements are partly overcome through the
continuum mesoscopic approach, whereby solvent and ions are described
in the continuum and in a pre-averaged sense \cite{CAPUANI2004,CAPUANI2006,WANG2010}.

The LB framework solves complex electrokinetic problems through an
efficient formulation of the electrolytic solution as a multi-species
problem: one species for the neutral solvent, water, and two for the
positively and negatively charged ionic components. The Poisson equation
provides the solution for electrostatics and the self-consistent forces
for the transport of the fluid species, in the so-called Vlasov-Poisson
approximation.

Let us briefly survey the LB method for charged multi-component systems.
The fluid mixture is composed by three sets of populations labelled
by index $\alpha=0,1,2$, two ionic components with charges $z_{\alpha}e$,
$e$ being the proton charge, density $n^{\alpha}$ and velocity $\vec{u}^{\alpha}$.
Given the barycentric velocity $\vec{u}=\frac{\sum_{\alpha}n^{\alpha}\vec{u}^{\alpha}}{\sum_{\alpha}n^{\alpha}}$,
the species relative velocity is $\delta\vec{u}^{\alpha}=(\vec{u}^{\alpha}-\vec{u})$.
The aqueous medium accommodates solute biomolecules, with the i-th
particle having position $\vec{r}_{i}$ and valence $z_{i}$. The
electrostatic potential is obtained by solving the Poisson equation,
\begin{equation}
\nabla^{2}\psi=-\frac{1}{\epsilon}e[n^{+}-n^{-}+\sum_{n}z_{n}\delta(\vec{r}_{n}-\vec{r})]\label{eq:POISSON}
\end{equation}
where the medium has dielectric permittivity $\epsilon$, duly complemented
by boundary conditions of the Dirichlet or Neumann kind. For insulating
confining walls, where the surface has local charge density $\Sigma$,
this reads $-\nabla\phi\cdot\hat{n}=\Sigma/\epsilon$, where $\hat{n}$
is the unit vector normal to the surface.

The dynamics of each species follows the evolution equation:

\begin{equation}
f_{i}^{\alpha}(r+c_{i},t+1)=f_{i}^{\alpha}(r,t)+\omega(f_{i}^{\alpha,eq}-f_{i}^{\alpha})+S_{i}^{\alpha}\label{eq:LBMCHARGE}
\end{equation}
where the Maxwellian equilibrium for mixtures is given by \cite{MARCONI2011,MARCONI2011b},

\begin{equation}
f_{i}^{\alpha,eq}=w_{i}n^{\alpha}\left[1+\frac{\delta\vec{u}^{\alpha}\cdot\vec{c}_{i}}{v_{T}^{2}}+\frac{(\delta\vec{u}^{\alpha}\cdot\vec{c}_{i})^{2}-v_{T}^{2}(\delta u^{\alpha})^{2})}{2v_{T}^{4}}\right]\label{eq:LEQCHARGE}
\end{equation}
and the force term 
\[
S_{i}^{\alpha}=w_{i}n^{\alpha}\left[\frac{\vec{F}^{\alpha}\cdot\vec{c}_{i}}{v_{T}^{2}}+\frac{(\vec{c}_{i}\cdot\vec{u})(\vec{c}_{i}\cdot\vec{F}^{\alpha})-v_{T}^{2}\vec{F}^{\alpha}\cdot\vec{u})}{v_{T}^{4}}\right]
\]
The local self-consistent forces are $F^{\alpha}=F^{\alpha,drag}-ez^{\alpha}\mathbf{\nabla}\psi$,
where $F^{\alpha,drag}=-\omega_{drag}^{\alpha}\sum_{\beta}\frac{n^{\beta}}{n}(u^{\alpha}-u^{\beta})$
is the drag force exerted on species $\alpha$, resulting in a cross-diffusion
coefficient $D^{\alpha}=\frac{v_{T}^{2}}{\omega_{drag}^{\alpha}}$,
where $\omega_{drag}^{\alpha}$ is a relaxation frequency.

So much for the governing equations. However, a hallmark of charged
systems is that local electrostatic forces can be dramatically intense,
often exhibiting rapid spatial modulations of the electrolytic densities,
as in the presence of double layers.

Such occurrence can lead to severe numerical instabilities in methods
based on the direct solution of NS equation, as a consequence of local
violations of the Courant-Friedrich-Levy stability condition \cite{COURANT1928}.
Thanks to its inherently small time step (in macroscale units), LB
can often handle stiff forces without loosing stability. However,
strong polyelectrolytes are problematic to handle and cannot be treated
directly, thus some sort of charge rescaling is required \cite{DATAR2017}.
Despite these liabilities, LB simulations of electrolytic systems
show stable behaviour over a wide range of solute charges and molarities,
in particular in the sub-molar range that covers a large portion of
biological conditions.

\subsection{Flows far from equilibrium}

Most biological systems operate far from equilibrium, i.e., they draw
energy from their  environment and dissipate heat back to it, thereby
lowering their own entropy content  at the expense of the environment.
This is the operating principle of the so called ``dissipative structures'',
a cornerstone of  non-equilibrium thermodynamics \cite{prigogine2017non}.
In the process of dumping entropy to the environment, they manage
to migrate across a sequence of different non-equilibrium steady-states
(NESS), in which they deliver different functions.

Examples of NESS in biology include molecular machines, cells in motion,
metabolic pathways and many others.  For instance, the kinesin protein
walks along the microtubule, carrying cargos from one part of the
cell to another, by absorbing energy from ATP hydrolysis and converting
chemical energy into mechanical work, of which $\sim60\%$ is used
for motion and the rest is dissipated to the surroundings. Off-equilibrium
conditions imply that biological agents exchange mass, momentum, energy
and entropy with the environment through transient or steady currents
and fluxes. Clearly, such complex space-time dependent network of
currents and fluxes, typically operating on a broad spectrum of concurrent
space and time scales, needs to be captured by  the numerical approach
\cite{TAKAHASHI2005}. Whence a major need of multiscale methods.

Indeed, these currents and fluxes typically occur through thin (atomic
scale) interfaces, which resolve the tension between  competing mechanisms,
say chemical reactions and molecular diffusion, through a sudden spatial
 transition between distinct bulk phases, say the liquid and vapour
in a multiphase flow.

For the case of a diffusion-reaction system, the width of the transition
region can be estimated as 

\[
w\sim\sqrt{D\tau_{ch}}
\]
 where $D$ is the diffusion coefficient and $\tau_{ch}$ is a typical
chemical timescale. By expressing $D\sim\lambda^{2}/\tau_{c}$, with
$\lambda$ being the mean-free path and $\tau_{c}$ the collisional
timescale, we obtain $w\sim\lambda(\tau_{ch}/\tau_{c})^{1/2}$.  This
shows that, unless chemical reactions are much slower than inert collisions
(slow-chemistry), the interface  width is comparable with the molecular
mean free path, or shorter (fast-chemistry).

Another example in point are foams and emulsions, i.e. droplets (bubbles)
of liquid (vapor) dispersed  in a continuum liquid phase, typically
water. In these multiphase systems, the transition between the dense
and light phases is controlled  by the surface tension which is, in
turn, dictated by molecular interactions, notably the strength of
the  potential and its spatial range.  That leads to interface widths
of the order of the spatial range of such  interactions, typically
nanometers or below. A typical estimate of the interface width $w$
is as follows: 

\[
w\sim\sqrt{\frac{k_{B}T}{\sigma}}
\]
where $\sigma$ denotes the surface tension. Typical values, in MKS
units are $k_{B}T\sim4\times10^{-21}$ and $\sigma\sim7\times10^{-2}$,
deliver $w\sim4\times10^{-10}$, i.e a fraction of nanometer. 

\begin{figure}
\centering{}\includegraphics{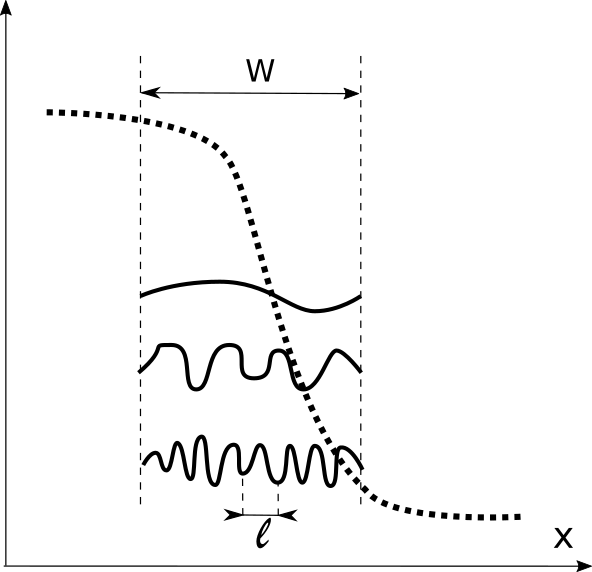}
\caption{A sketch of a high-density/low-density diffuse interface of width
$w$, as typical for many biosystems (dotted line), for different
values of the mean free path, of length $\ell$. Each solid line shows
a different profile of the interface, depending on its intrinsic structure.
Here we assume that the wavelength of the density profile coincides
with $\ell$. In the limit $\ell/w\ll1$ (solid line, bottom), non-equilibrium
effects are negligible and hydrodynamics holds. On the other hand,
when $\ell$ becomes comparable to the interface width $w$ (solid
line, top), non-equilibrium effects can no longer be neglected and
higher order kinetic moments must be accounted for. \label{fig:MFP} }
\end{figure}

As shown in Fig. \ref{fig:MFP}, such nanometric interfaces challenge
the low-Knudsen assumption which lays at the foundations of the hydrodynamic
description. In fact, given that the interface thickness is comparable
with the molecular mean free path, the result are local Knudsen numbers
of the order unity. In the last decade, a number of technical extensions
of the original LB method have been developed with the aim of gaining
insights into these complex non-equilibrium interfacial phenomena
(and yet, much remains to be done).

Here, we limit ourselves to summarize the main upgrades, namely: \textcolor{black}{\emph{i)}}
\emph{Higher-Order lattices}, \textcolor{black}{\emph{ii)}} \emph{Kinetic
Boundary Conditions}, \textcolor{black}{\emph{iii)}} \emph{Regularization}
techniques.

\subsubsection{Higher-Order Lattices}

As discussed in the initial part of this paper, hydrodynamics represents
the ``infrared'' limit of kinetic theory, whereby all macroscopic
heterogeneities live on much longer scales than the molecular ones
(hydrodynamic transport regime). From the formal viewpoint, this means
that the Boltzmann distribution is accurately described by its lower-order
moments, typically density (order $0$), current (order $1$) and
momentum-flux (order $2$).

In the hydrodynamic transport regime, all higher order moments (non-equilibrium
excitations) are directly enslaved to the low order ones, hence they
have no independent dynamics of their own.

Far from equilibrium, where strong inhomogeneities may persist down
to near-molecular scales, such low-order picture breaks down, and
more kinetic moments concur to define the motion of matter beyond
the hydrodynamic regime.

Without entering details, it is intuitively clear that this far from
equilibrium regime requires the use of higher order lattices, securing
the proper recovery of the correspondingly higher order moments. A
formal theory of LB beyond NS, based upon higher order lattices, was
laid down by X. Shan \textcolor{black}{\emph{et al}}~\cite{ShanJFM}.
Those authors developed the discrete analogue of Grad's expansion
in Hermite polynomials, with full details on its specific implementation
on a series of higher order lattices associated with different numerical
quadrature rules. The first observation as compared to Grad's 13 moment
formulation is that the scheme provides a larger set of degrees of
freedom~\cite{GRAD13}. The discrete speeds of the corresponding
lattices are typically of order $40$ or more, hence many more than
the $13$ Grad's moments~\cite{MENGZHANG,ZhangShanChen}.

\subsubsection{Kinetic Boundary Conditions}

Higher-order lattices offer room for extra-moments, but this is not
sufficient per-se, unless suitable boundary conditions are formulated
near solid boundaries. This is where the lattice formulation makes
a distinct contribution: while it appears very hard to devise well-posed
boundary conditions for the kinetic moments, in fact complicated nonlinear
tensors, lattice formulations lend themselves to conceptually transparent
and mathematically well-posed formulations. The reason is always the
same: the information moves along straight lines and boundary conditions
can be formulated in terms of mechanical relations between the outgoing
(fluid-to-wall) and the incoming (wall-to-fluid) discrete distributions.
Non-equilibrium flows exchange momentum with the solid walls along
both tangential and normal directions: describing this exchange is
the mandate of Kinetic Boundary Conditions. Empirical forms adapting
this constraint to the lattice environment have been formulated in
terms of accommodation coefficients~\cite{SSLIP}. Following Maxwell,
the idea is that molecules impinging on the wall loose track of their
incoming speed~\cite{MAX}. Consequently, they are re-injected into
the fluid along a random direction, with a velocity drawn from a Maxwellian
at the local wall speed and temperature. Albeit handy, these accommodation
schemes remain empirical in nature. A more satisfactory formulation
was developed by Ansumali and Karlin, who basically expressed the
accommodation coefficients in terms of the outgoing (fluid-to-wall)
distribution functions and their equilibrium version, thus providing
a closed and consistent recipe~\cite{AKBC}. The Ansumali-Karlin
boundary conditions exhibit a number of appealing properties. First,
they preserve the positivity of the distribution at boundary nodes,
i.e., if the incident distribution is positive, the reflected one
is guaranteed to be positive too. Second, they readily extend to general
wall scattering kernels, such as those allowing a blend of slip and
reflection. At the time of this writing, these kinetic boundary conditions
are the tool of the trade for finite-Knudsen LB simulations.

\subsubsection{Regularization}

Regularization is a general and powerful idea across many fields of
theoretical physics, to remove various forms of divergences and singularities
which arise whenever a given description/theory fails to capture the
physics in point.

Kinetic theory is no exception. Indeed, it is known since long that
post-hydrodynamic equations beyond the NS level, suffer a number of
problems, primarily short-scale linear instabilities\textasciitilde{}\cite{BOBYL}.
Many regularization schemes have been proposed ever since to tame
such instabilities~\cite{REG1,REG2}. Regularization procedures have
been (re)-discovered only recently by the LB community~\cite{CHOPAREG,MONTEREG}
and it is not yet clear how they relate to the corresponding counterparts
in continuum kinetic theory. LB regularization consists of filtering
out the contribution of non-hydrodynamic moments (ghosts) to the hydrodynamic
ones: mass, momentum and momentum-flux.

Let us dig a little bit deeper into the subject.

The standard LB scheme in BGK form reads (time-step made unit for
simplicity): 
\begin{equation}
f_{i}(\vec{r}+\vec{c}_{i};t+1)=(1-\omega)f_{i}(\vec{r};t)+\omega f_{i}^{eq}(\vec{r};t)
\end{equation}
The actual distribution can be split as follows: 
\[
f_{i}=h_{i}+g_{i}
\]
where the hydro-component $h_{i}$ collects terms up to third order
Hermite polynomials associated with mass, momentum, momentum-flux
and energy flux, while $g_{i}$ collects all higher order terms, transport
plus genuinely kinetic fields with no immediate macroscopic interpretation
(ghosts in LB jargon). Formally, one defines a regularisation operator
$\mathcal{R}$, projecting the actual distribution onto the hydrodynamic
subspace, i.e., $h_{i}=\mathcal{R}f_{i}$, that is $\mathcal{R}g_{i}=0$.

Both hydro and ghost terms further split into equilibrium and non-equilibrium
components, the ghost equilibrium being identically null by construction.
Thus, what remains to be filtered out is just the ghost non-equilibrium,
which is constantly revived at each free-streaming step, the non-equilibrium
engine.

By applying the regularisation projector to the right-hand-side of
the BGK equation, an operation corresponding to filtering out ghost
components after streaming, we obtain the following Regularized-LB:
\begin{equation}
f_{i}(\vec{r}+\vec{c}_{i};t+1)=(1-\omega)h_{i}(\vec{r},t)+\omega f_{i}^{eq}(\vec{r},t)
\end{equation}

The Regularized-LB has recently made proof of providing significant
benefits in terms of improving the stability of the LB scheme under
finite-Knudsen conditions.

The ``post-hydro'' LB literature is vast and growing, but not conclusive
yet. In particular, it is not clear how the three main ingredients
mentioned above should be combined in order to obtain correct finite-Knudsen
behavior.

A very valuable step in the comprehension of LB post-hydro capabilities
was provided by Ansumali, Karlin and coworkers, who discovered exact
solutions to the hierarchy of nonlinear LB kinetic equations for stationary
planar Couette flow at non-vanishing Knudsen numbers~\cite{AKexact}.
By using a 16-speed two-dimensional lattice and kinetic boundary conditions,
these authors have derived closed-form solutions for all higher-order
moments, and solved them analytically.

The results indicate that the LB hierarchy with larger velocity sets
does indeed approximate kinetic theory beyond the NS level. If only
for a simple set-up, those exact solutions indicate that LB equipped
with kinetic boundary conditions is able to carry quantitative non-hydrodynamic
information, hence it can be regarded as a kinetic closure in its
own right.

It thus appears that the extension of the LB formalism to higher order
lattices can result into an effective tool to probe deeper into non-equilibrium
regimes beyond the hydrodynamic description \cite{AIGUO}.

\subsection{Fluctuating Lattice Boltzmann}

For the case of nanoscale flows, reproducing thermal behavior implies
that the LB must incorporate the effects of statistical fluctuations.

To achieve this goal, following in the footsteps of Landau-Lifshitz
fluctuating hydrodynamics, Ladd added a source of random fluctuations
of the momentum-flux tensor \cite{LADDFLUCT}, namely: 
\begin{equation}
\tilde{S}{}_{i}=\tilde{A}k_{B}T\;S_{ab}(c_{ia}c_{ib}-c_{s}^{2}\delta_{ab})
\end{equation}
where $S_{ab}=(\partial_{a}J_{b}+\partial_{b}J_{a})/2$ is the shear
tensor and $\tilde{A}$ is the amplitude of the fluctuations. 

The latter must be tuned so as to comply with the Fluctuation-Dissipation-Theorem.
The resulting fluctuating NS equation reads 
\begin{equation}
\rho\left(\frac{\partial}{\partial t}\vec{u}+\vec{u}\cdot\nabla\vec{u}\right)=\nabla\cdot(\dvec P+\dvec S)+\eta\nabla^{2}\vec{u}+\vec{G}
\end{equation}
where $\dvec P$ is the fluctation-free momentum-flux tensor and $\vec{G}$
is the body force acting on the fluid.

\begin{figure}
\centering{}\includegraphics{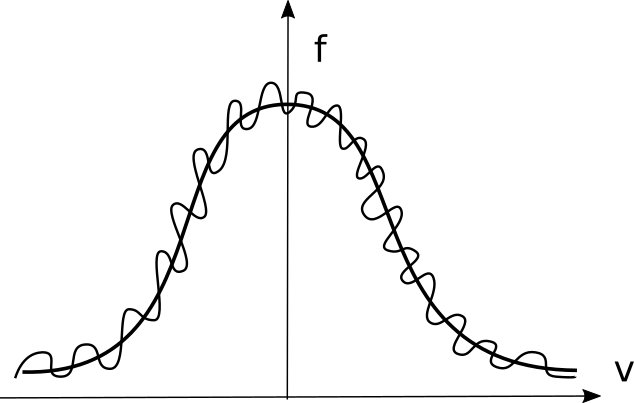}
\caption{A sketch of the fluctuating distribution, $\tilde{f}$ (wiggly line),
featuring rapid oscillations with respect to the non-fluctuating one,
$f$ (smooth line). Fluctuating hydrodynamics emerges as a consequence
of the stochastic component of the distribution function. \label{fig:FLUCTLB} }
\end{figure}

The LB scheme is modified accordingly, by adding a stochastic source
to the right hand side:

\begin{equation}
\tilde{f}_{i}(\vec{r}+\vec{c}_{i};t+1)=(1-\omega)\tilde{f}_{i}(\vec{r};t)+\omega\tilde{f}_{i}^{eq}(\vec{r};t)+\tilde{S}_{i}
\end{equation}
as sketched in Fig. \ref{fig:FLUCTLB}. The source term $\tilde{S}_{i}$
is local in space and time and acts at the level of the stress tensor
and non-hydrodynamic modes, with no effect on mass and momentum conservation. 

cd 

In actual practice, the source term $\tilde{S}_{i}$  is constructed
via a set of the lattice eigenvectors $\{\chi_{k}\}$ with $k=0,Q-1$
orthonormal according to the scalar product $\sum_{m=0}^{Q-1}w_{m}\chi_{km}\chi_{lm}=\delta_{kl}$.
In the D3Q19 scheme, the eigenvectors correspond to the kinetic moments:
$k=0$ is relative to density, $k=1:3$ to current, $k=4:9$ to the
momentum flux tensor, the remaining $k=10:(Q-1)$ eigenvectors to
non-hydrodynamic modes. The stochastic forcing reads as follows: 
\begin{equation}
\tilde{S}_{i}=\sqrt{\frac{\rho k_{B}T\omega(2-\omega)}{c_{s}^{2}}}\sum_{k=4}^{Q-1}w_{i}\chi_{ip}{\cal N}_{k}\label{eq:NOISE}
\end{equation}
where ${\cal N}_{k}$ is a set of $15$ random numbers with zero mean
and unit variance. Given the fact that the thermal velocity is fixed
and equals the underlying lattice speed $c$, the thermal mass is
chosen in such a way as to obtain the thermal fluctuations according
to $k_{B}T=mv_{T}^{2}$.

The stochastic forcing has been subsequently improved so as to produce
consistent fluctuations at all spatial scales, in particular at short
distances where the effect on the translocating molecule is critical
\cite{DUNWEGLADD,ADHIKARI}. The fluctuating LB passes a number of
litmus tests, particularly the compliance of velocity-velocity and
force-force autocorrelation functions with the principle of stochastic
kinetic theory, in particular the fluctuation-dissipation theorem.

\subsection{Lattice Boltzmann for non-ideal fluids}

Boltzmann originally derived his equation under the assumption of
diluteness, whereby molecular collisions take place as zero-ranged,
instantaneous events, leading to large scattering angles. This rules
out long-range, soft interactions giving rise to small-angle deflections.
Such interactions, however, are of utmost importance for biological
applications, since soft interlayers based on membranes and biopolymers
define the spatial boundaries between different phases in biological
systems. In aqueous media containing various ions, interactions are
governed by a complex interplay of generic and specific interfacial
interactions, typically controlled by dispersion and electrostatic
forces, and locally shaped as hydrogen bond networks, disulphide bridges,
hydrophobic, entropy-induced interactions and so on.

Soft-core interactions can be included in the kinetic equation in
the form of effective one-body forces, resulting from the collective
interaction of a representative particle with the self-consistent
environment (bath) due to all other particles. Formally, such effective
one-body force takes the following form 
\begin{equation}
\vec{F}_{1}=\int\nabla_{r_{1}}V(\vec{r}_{1}-\vec{r}_{2})f_{12}\;d\vec{v}_{2}d\vec{r}_{2}\label{F1}
\end{equation}
where $f_{12}\equiv f(\vec{r}_{1},\vec{r}_{2},\vec{v}_{1},\vec{v}_{2})\;d\vec{v}_{2}d\vec{r}_{2}$
is the (unknown) two-body distribution and $V_{12}\equiv V(\vec{r}_{1}-\vec{r}_{2})$
is the two-body atomistic potential. Following a customary practice,
one writes $f_{12}=f_{1}f_{2}g_{12}$, where $g_{12}$, the two-body
correlation function, collects the two-body physics. The exact form
of the correlation function is known exactly only in a few precious
cases; yet one can introduce several useful ansatzs which turn the
formal expression~\ref{F1} into an operational one. A typical ansatz,
which has proven very fruitful for LB modelling of non-ideal fluids,
looks like follows: 
\begin{equation}
\vec{F}(\vec{r}_{1})=\psi(\vec{r_{1}})\int G(\vec{r}_{1},\vec{r}_{2})\psi(\vec{r}_{2})(\vec{r}_{2}-\vec{r}_{1})d\vec{r}_{2}\label{G1}
\end{equation}
where $G(\vec{r}_{1},\vec{r}_{2})$ is a model Green function and
$\psi[\vec{r}]\equiv\psi[\rho(\vec{r})]$ is a local functional of
the fluid density $\rho(\vec{r}_{1},t)$. The above expression is
quite general and allows a wide degree of latitude in \textit{modeling}
non-ideal fluid interactions.

\subsection{Pseudo-potential models}

The most popular lattice transcription of the above, due to Shan and
Chen~\cite{SC1,SC2}, reads as follows:

\begin{equation}
\vec{F}(\vec{r})=\psi(\vec{r})\sum_{i}w_{i}G_{i}\psi(\vec{r}+\vec{c}_{i})\vec{c_{i}}\label{SC}
\end{equation}

\begin{figure}
\centering{}\includegraphics{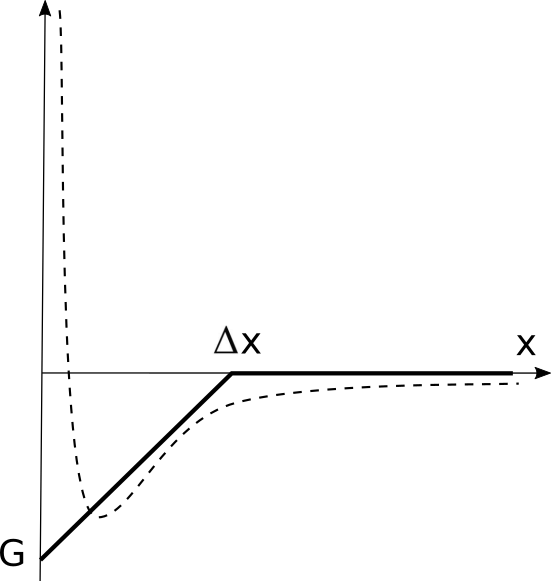}
\caption{Sketch of the Shan-Chen piece-wise linear potential, extending over
the first Brillouin lattice cell at a distance $\Delta x$. The potential
mimics the attractive tail of a van der Waals potential (dashed line)
while the repulsive one is quenched to zero. Notwithstanding the absence
of hard-core repulsion, the Shan-Chen potential does not cause any
unstable density pileup because, in the high density limit, $\rho\gg1$,
the generalised density $\psi(\rho)=1-e^{-\rho}$ flat-tops to a constant
$1$, yielding a zero gradient, hence zero force \label{fig:SHANCHEN}.}
\end{figure}

The sum runs over the prescribed set of neighbors, typically the first
Brillouin region in the original version, and subsequently extended
to the second or even the third one. Typically, all discrete speeds
in the same Brillouin region share the same $G_{i}$, so that in the
original Shan-Chen (SC) formulation, there is just one coupling strength
$G$ (see Fig.\ref{fig:SHANCHEN}). This is nonetheless sufficient
to generate the main ingredients of non-ideal fluids, namely a non-ideal
equations of state supporting phase-transitions, as well as surface
tension. Moreover, this limitation can be readily lifted by extending
the Shan-Chen interaction beyond the first Brillouin region, thereby
explicitly accounting for both repulsive and attractive interactions. 

The Shan-Chen expression delivers a non-ideal equation of state of
the form 
\begin{equation}
p=\rho c_{s}^{2}+\frac{G}{2}c_{s}^{2}\psi^{2}[\rho]\label{SCEOS}
\end{equation}
By choosing the generalized density in the form: 
\[
\psi[\rho]=1-e^{-\rho}
\]
it is readily checked that the Shan-Chen fluid becomes critical for
$G<G_{crit}=-4$ at a critical density $\rho_{c}=\ln2$. Note that
even though the interaction is purely attractive ($G<0$), no pile-up
instabilities take place because the force becomes increasingly faint
as the density increases. This is the reason for using the generalized
density $\psi[\rho]$ instead of the physical one $\rho$. This is
a very expedient trick to trigger phase-transitions without any repulsive
force. The translation from the force to the LB source $S_{i}$ proceeds
through a systematic expansion in lattice Hermite polynomials. To
leading order $S_{i}\sim\vec{F}\cdot\vec{c}_{i}$, but higher order
terms, at least quadrupole ones, are definitely needed to obtain accurate
results.

More generally, thick interfaces lead to large values of the Cahn
number, namely the ratio of the interface width to a typical mesoscopic
scale, say the droplet (bubble) diameter: 

\[
Cn=\frac{w}{D}.
\]

For most applications $D$ is in the range of tens, up to one hundred,
microns, leading to very small Cahn numbers, $Cn\sim10^{-5}$. Replicating
this number in LB simulations is totally unviable, for it would amount
to placing $O(10^{5})$ lattice spacings across the particle diameter.
LB simulations are forced to operate at much higher Cahn numbers of
the order of $0.01\div0.1$, which means that potential inaccuracies
due to finite Cahn number effects need to be carefully monitored. 

Lack of sufficient symmetry also leads to the appearance of spurious
currents,  which may seriously affect the physical results whenever
the density ratio between the light and dense phases is above 30 \textendash{}
50, a problem which is further exacerbated in the case of fast-moving
interfaces. Finally, the model is not thermodynamically consistent,
as it is not derived from a mean-field free-energy functional. That
limitation is, however, less serious than it seems, since the method
is, actually, equipped with a quasi-free energy \cite{sbragaglia2009continuum}.
As a matter of fact, more disturbing, instead, is the fact that the
Shan-Chen state equation features a sound speed smaller in the liquid
 with respect to the vapour phase, with significant consequences on
the stability of the light phase across the interface.

Most of these limitations have been significantly mitigated, if perhaps
not fully resolved, by subsequent  developments. Among others, particularly
noteworthy for the simulation of soft-glassy materials, is  the so-called
multi-range pseudo potential method.

The main idea is to augment the original Shan-Chen short-range attraction
in the first Brillouin region (belt, in LB jargon), say the D3Q27
lattice in three dimensions, with a repulsive interaction  acting
on the second Brillouin region, i.e., including discrete velocities
up to $\sqrt{5}$. The main advantage of this formulation is that
a proper tuning between the first-belt attraction  and the second-belt
repulsion, makes possible to achieve smaller values of the surface
tension, thereby permitting to sustain long-lived, multi-droplet configurations
with a highly complex interfacial dynamics. Despite its many limitations,
the SC method remains the most popular version of LB for non-ideal
fluids, mostly on account of its simplicity and transparency.

However, for applications leading to complex interfacial dynamics,
more advanced schemes are required. 

\subsubsection{Free-Energy models}

Another successful route to LB schemes for non-ideal fluids is (lattice)
Density Functional Theory (DFT), whereby the non-ideal physics is
encoded within a free-energy functional of the form: 
\begin{equation}
\mathcal{F}[\rho]=\int\phi(\rho,\nabla\rho)\;d\vec{r}\label{eq:FREEENERGY}
\end{equation}
where $\phi$ is a local functional of the fluid density and its gradients
\cite{SWIFT}. 

The former encodes the non-ideal equation of state, while the latter
describes the effect of surface tension. 

\textcolor{black}{Variational minimization over density changes delivers
the equations of motion of the non-ideal fluids, governed by the Korteweg
pressure tensor: 
\begin{equation}
P_{ab}(\vec{r})=p(\vec{r})\delta_{ab}+\kappa\partial_{a}\rho\partial_{b}\rho\label{KORTE}
\end{equation}
where
\[
p(\vec{r})=p_{0}(\rho)-\kappa[\rho\Delta\rho-\frac{1}{2}(\partial_{a}\rho)^{2}]
\]
is the non-local pressure, consisting of the bulk contribution $p_{0}(\rho)$,
fixing the equation of state, plus an interface contribution associated
to surface tension. In the above, $\partial_{a}$ stands for the space
derivative along direction $a=x,y,z$ and $\kappa$ is a tunable coefficient
controlling the surface tension. }

\textcolor{black}{From the practical standpoint, the free-energy formulation
leads to non-local equilibria, which involve second order derivatives
of the density field, thus taxing the simplicity of the method, and
sometimes its stability as well, as compared to the pseudo-potential
method. Nevertheless, the free-energy method has found broad use for
many applications such as microflows over geometrically or chemically
patterned surfaces and various types of droplet motions \cite{PANCAKE}. }

Several variants have been developed over the years, which have considerably
improved over the original versions, especially in terms of reaching
higher density ratios without compromising numerical stability. Among
others, high-order finite-difference schemes \textcolor{red}{\cite{LEE}}
and mergers with flux-limiting methods \textcolor{red}{\cite{LAMURA}}.
These methods are currently used to simulate a variety of multiphase
and multi-component fluid applications, such as Rayleigh-Taylor instabilities,
droplet collisions, cavitation and free-surface flows. For a review,
see the recent books \textcolor{red}{\cite{LB1b,MFBook}} and references
therein.

\subsubsection{Chromodynamic models}

\textcolor{black}{Another class of LB methods for non-ideal fluids
which has been revamped in the recent past  is the so-called Color-Gradient
model. Here, the two phases/components, call them Red and Blue for
convenience, segregate  based on a top-down ``colouring'' rule,
which sends the particles along the color gradient, namely Red towards
Red and Blue towards Blue, thereby  triggering an instability leading
to interface formation \cite{gunstensen1991lattice}.}

\textcolor{black}{The recoloring amounts to correcting the the Red
and Blue populations as follows: }

\textcolor{black}{
\begin{equation}
f_{i}^{R,B}=\phi^{R,B}f_{i}^{*}\pm\beta\phi^{R}\phi^{B}\mu_{i}f_{i}^{eq,0}\label{RECOLOR}
\end{equation}
where $\phi^{R,B}=\rho^{R,B}/(\rho^{R}+\rho^{B})$ is the mass density
fraction, $\mu_{i}$ is the cosine of the angle between the color
gradient $\vec{G}=\nabla(\rho_{B}-\rho_{R})$ and the $i$-th discrete
velocity. In the above $f_{i}^{*}=f_{i}^{R,*}+f_{i}^{B,*}$, where
star indicates the population after the application of the non-ideal
force stemming from surface tension, and $f_{i}^{(eq,0)}=f_{i}^{(eq,0),R}+f_{i}^{(eq,0),B}$,
where superscript $^{(eq,0)}$ denotes the equilibria at zero flow.
Finally, $\beta$ is a free parameter controlling the strength of
the recoloring procedure, to all effects and purposes an anti-diffusive
operator. The crucial term is the second one at the right-hand side
of eq. (\ref{RECOLOR}), which, by construction, is active only at
the interface between the two components. For more details see \cite{Leclaire2017}
and \cite{MLS2019}.}

The interface is then stabilised by means of a chromo-dynamic force
proportional to the amplitude of the color gradient, but opposite
to it, so as to level out the deficit of one species over the other
(color gradient). Although essentially rule-driven, modern variants
of this scheme have proved capable of accessing parameter regimes
which appear to be off-limits for Shan-Chen schemes and extensions
thereof, as well as for Free-Energy methods.\} For instance, such
methods have been recently applied to the design of microfluidic devices
for droplet generation, such as flow-focusers and step-emulsifiers
\textbackslash{}cite\{AM2018a,AM2018b\}. 

The interface is then stabilised by means of a chromo-dynamic force
proportional to the amplitude  of the color gradient, but opposite
to it, so as to level out the deficit of one species over the other
(color gradient). Although essentially rule-driven, modern variants
of this scheme have proved capable of accessing  parameter regimes
which appear to be off-limits for Shan-Chen schemes and extensions
thereof, as well  as for Free-Energy methods.

\textcolor{black}{The recoloring amounts to correcting the Red and
Blue populations as follows: where $\phi^{R,B}=\rho^{R,B}/(\rho^{R}+\rho^{B})$
is the mass density fraction, $\mu_{i}$ is the cosine of the angle
between the color gradient $\vec{G}=\nabla(\rho_{B}-\rho_{R})$ and
the $i$-th discrete velocity. In the above $f_{i}^{*}=f_{i}^{R,*}+f_{i}^{B,*}$,
where the star indicates the population after the application of the
non-ideal force stemming from surface tension, and $f_{i}^{(eq,0)}=f_{i}^{(eq,0),R}+f_{i}^{(eq,0),B}$,
where superscript $^{(eq,0)}$ denotes the equilibria at zero flow.
Finally, $\beta$ is a free parameter controlling the strength of
the recoloring procedure, to all effects and purposes an anti-diffusive
operator. The crucial term is the second one at the right-hand side
of eq. (\ref{RECOLOR}), which, by construction, is active only at
the interface between the two components. For more details see \cite{Leclaire2017,MLS2019}.}

\textcolor{black}{The interface is then stabilised by means of a chromo-dynamic
force proportional to the amplitude of the color gradient, but opposite
to it, so as to level out the deficit of one species over the other
(color gradient). Although essentially rule-driven, modern variants
of this scheme have proved capable of accessing parameter regimes
which appear to be off-limits for Shan-Chen schemes and extensions
thereof, as well as for Free-Energy methods. For instance, such methods
have been recently applied to the design of microfluidic devices for
droplet generation, such as flow-focusers and step-emulsifiers \cite{AM2018a,AM2018b}. }

\subsubsection{Entropic models for multiphase flows}

The lattice Boltzmann method can also be formulated by minimizing
a suitable lattice H-function of the form (Kullback-Leibler entropy):
\[
H[f]=\sum_{i}f_{i}log(f_{i}/w_{i})
\]
 where $w_{i}$ are suitable lattice weights. The resulting scheme
takes the usual form of a standard LB, with the crucial twist that
the relaxation time is adaptively adjusted in such a way as to secure
compliance with the second principle of thermodyamics, namely: 
\[
\frac{dS}{dt}=\int H[f]d\vec{r}d\vec{v}\;\ge0
\]

Leaving a detailed description to the original literature, here we
just mention that compliance with the second principle translates
into a significant enhancement of numerical stability \cite{ELB}.
For this reason, ELB has found profitable use for the simulation of
low-viscous flows typical of fluid turbulence.

In a recent time, the ELB has been extended to the case of multiphase
flows, and shown to provide stability benefits also in the viscous
regime of relevance to many biological applications \cite{ELBMU,ELBAM}.
Although these developments are too recent to permit a solid prononciation,
the results appear very encouraging and leave hope that the entropic
version of LB may become a major player in the field for the years
to come. 

\section{FLUIDS AND PARTICLES: THE LATTICE BOLTZMANN - PARTICLE DYNAMICS SCHEME}

The multiphase LB schemes discussed in the previous Section have generated
a mainstream of applications in soft matter research, since they permit
to deal with flows of major dynamic and morphological  complexity,
such as foams and emulsions. However, they are unsuited to handle
rigid bodies suspended in the continuum phase, nor can they describe
in detail mechanical properties of deformable objects, say membranes,
vesicles, cells and other biological bodies. To this purpose, LB needs
to be explicitly coupled to particle methods, tracking the dynamics
of the biological bodies immersed in the flow.

Indeed, most flows of biological interest consist of biological bodies
of assorted nature: cells, polymers, proteins, floating in a fluid
solvent, typically water. Such flows often operate at low, often near-zero,
Reynolds number \cite{PURCELL1977}, but this does not mean that hydrodynamic
interactions (HI) can be neglected. To the contrary, HI have repeatedly
been shown to accelerate a variety of nanoscale biological transport
processes, such as biopolymer translocation across nanopores, amyloid
aggregation of proteins in the cell and other related phenomena. This
explains why the combination of LB with particle dynamics has made
the object of extensive research and the development of multiple simulation
schemes.

\subsection{The extended particle model (EPM)}

Flows with suspended objects have been modelled since the early days
of LB research, starting with the trailblazing work of A. Ladd \cite{LADD1,LADD2}.
Ladd's original method consists of tracking the motion of rigid spherical
bodies under the impact of the surrounding solvent hitting the surface
of the body. The scales pertaining to body and solvent are fairly
separate, since the former is much larger and heavier than the latter,
hence the mass ratio $m/M$ serves as a suitable scale separator.

\begin{figure}
\begin{centering}
\includegraphics[scale=0.5]{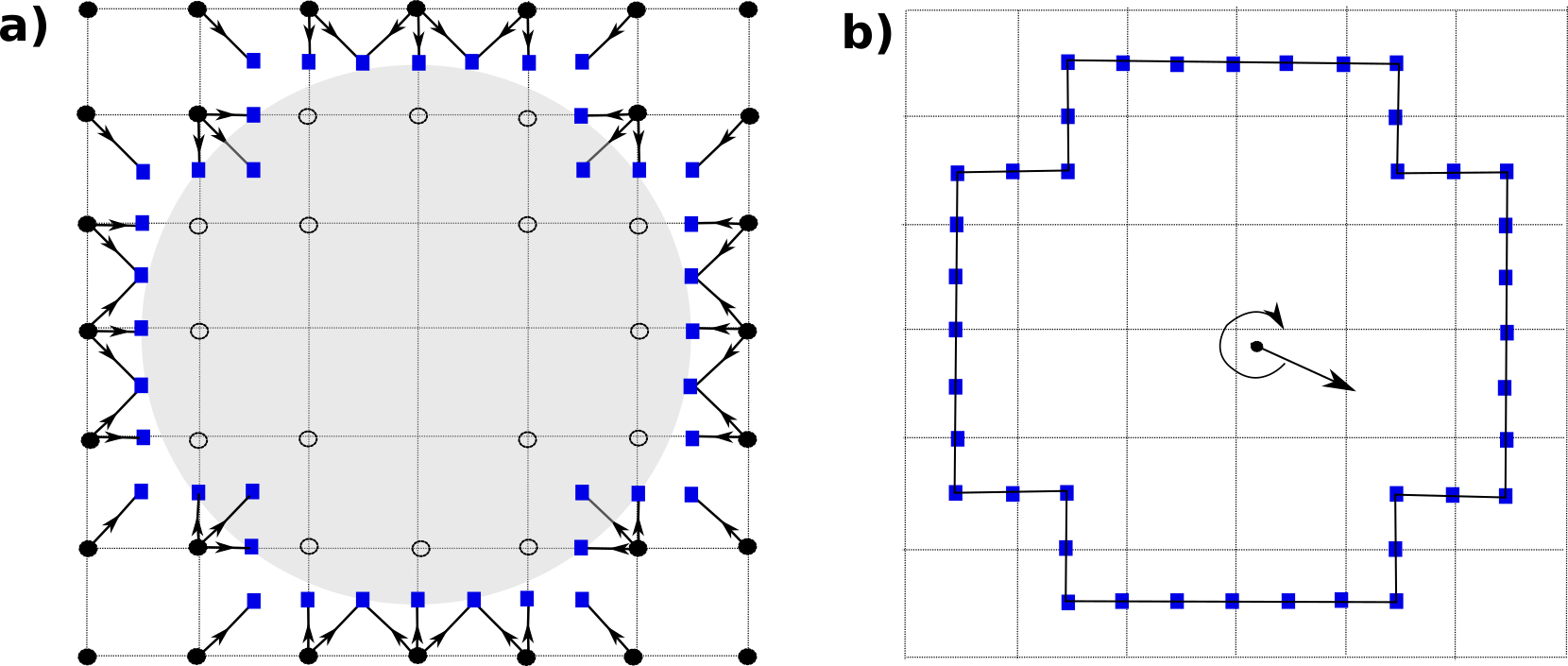} 
\par\end{centering}
\caption{Ladd's method for a suspended (spherical) body: a) the body-fluid
interface is tracked by searching the mesh points inside and outside
the particle \cite{LADD1}. Here a modified bounce-back scheme is
applied to account for the local exchange of momentum. The mid-point
of the connecting links identifies the stair-cased surface that corresponds
to the no-slip condition; b) the particle center moves in the continuum
according to the momentum exchange, and the rotational motion is accounted
for by computing the total torque. \label{fig:FIGLADD}}
\end{figure}

In the EPM, the exchange of momentum between particles and LB fluid
is a boundary-collision method whereby the suspended body interact
with each other only through the mediation of local collisions with
the solvent molecules (LB). The solvent-body collision is conservative,
the momentum lost (gained) by the solvent to the body is gained (lost)
by the body on the solvent. The method therefore is based on the local
exchange of momentum by computing the amount of momentum that every
population hitting the body surface exchanges with the latter. Being
the total force $\vec{F}$ and torque $\vec{T}$ the sum of fluid-body
momentum exchange (or drag interactions), direct particle-particle
mechanical interactions or, in case of microscopic conditions, stochastic
forces due to the Brownian motion, the body linear momentum $\vec{p}$
and angular momentum $\vec{l}$ obey the classical equations of motion
\begin{eqnarray}
\frac{d\vec{p}_{i}}{dt} & = & \vec{F}_{i}\nonumber \\
\frac{d\vec{l}_{i}}{dt} & = & \vec{T}_{i}\;\;\;\;\;\;\;\;\;\;\;\;\;\;\;i=1,N_{p}\label{eq:rigidbodymotion}
\end{eqnarray}
It is worth mentioning that the fluid-particle coupling is hydrokinetic
in nature because it is treated at collisional level rather than at
hydrodynamic level, so that, hydrodynamic forces such as long-time
tails, naturally emerge at times larger than the LB marching one.
The approach takes naturally into account the solvent fluctuations,
if present, since the latter are transmitted across the body surface
analogously to the drag forces (see Fig.\ref{fig:FIGLADD}).

Since the solvent-body collisions are conservative, no extra stochastic
source is needed on the body side.

Ladd's method finds its best use in simulating colloidal suspensions
of rigid particles and in this sense has found a comparatively limited
application in the biological context. However, it provides a first
and reliable example of coupling between LB and suspended bodies.
At the same time, the method shows some inaccuracy due to the lattice
discreteness, especially for near-contact colloid-colloid interactions,
occurring below the grid scale where the LB method can not resolve
the lubrication forces. Some variants have been developed in the literature,
making use of grid refinement and/or dynamic interpolations. Unsurprisingly,
they only add to the computational complexity of the method, which
is comparatively laborious on its own, as it demands full knowledge
of the local fluid-solid connectivity, namely the dynamic list of
fluid nodes interacting with the solid ones.

\subsection{The point-particle model (PPM) }

A qualitatively different strategy has been proposed by Ahlrichs and
Dünweg \cite{AHLRICHSDUENWEG}. This is a minimal way to embed point-like
particles in the LB fluid which, as opposed to Ladd's method, is entirely
\emph{force-driven}, hence metric instead of topologic. In essence,
particles carry a phenomenological friction coefficient $\gamma_{T}$
(also known as the bare friction coefficient) and the fluid exerts
a drag force based on the instantaneous difference between the particle
and fluid velocities, reading 
\begin{equation}
\vec{F}^{drag}=-\gamma_{T}\left[\vec{V}-\vec{u}\delta(\vec{r}-\vec{R})\right]\label{eq:DRAG}
\end{equation}
The addition of the point force into the fluid equations introduces
a singularity into the flow field, as represented by Dirac's delta
function. This is automatically smoothed out by interpolation procedures,
but requires nonetheless due care in the numerical treatment.

\begin{figure}
\begin{centering}
\includegraphics[scale=0.8]{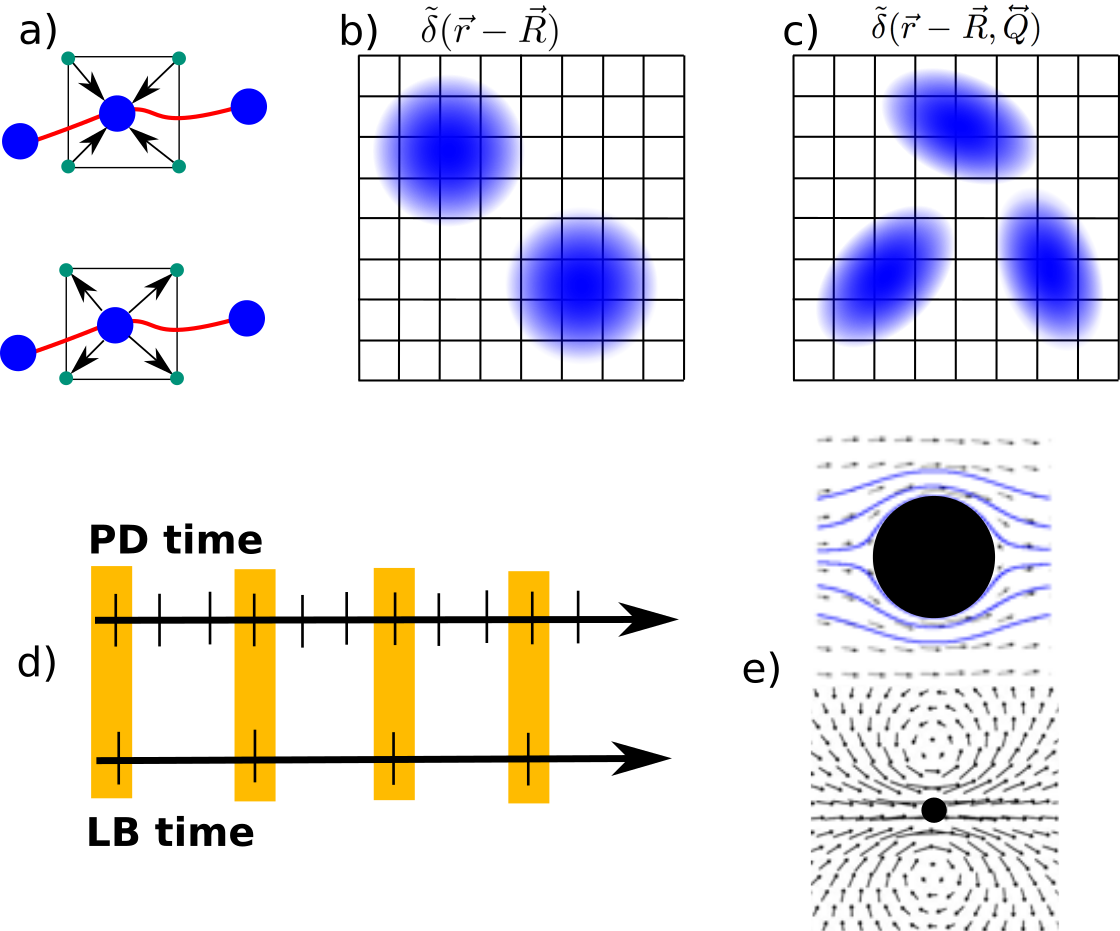} 
\par\end{centering}
\caption{Examples of immersed bodies in the LBPD approach. Panels a)-c) show
the different coupling methods between particles and fluid, as detailed
in the text: a) PPM with point-like particles exchanging information
with the nearest grid node, b) DPM for diffused isotropic particles,
c) DPM for diffused anisotropic particles. Panels d) shows the timestepping
method between particle dynamics (PD) and fluid dynamics (LB). The
timestepping scheme of the LB and PD individual components can be
either synchronous or asynchronous. Typically, the PD component ticks
more frequently than the LB component, allowing for more efficient
simulations. Panel e) illustrates a typical configuration of the fluid
around a moving particle. Hydrodynamic response manifests itself as
Stokes flow field (upper part) and long-range flow structure around
a spherical particle or its far-field flow pattern (lower part).\label{fig:FIGPD}}
\end{figure}

On the other hand, the flow field around a finite-sized particle is
generated by a distributed force located around the particle position,
as shown in Fig. \ref{fig:FIGPD}.

In the scheme, the flow field is treated as everywhere finite and
the force density acts onto the LB fluid at position ${\bf r}$ as
\begin{equation}
\vec{G}=-\vec{F}^{drag}I(\vec{r},\vec{R})\label{eq:DRAGFORCE}
\end{equation}
where $I(\vec{r},\vec{R})$ is a function interpolating between the
particle position $\vec{R}$ and the surrounding mesh node $\vec{r}$.
In its original formulation, the method replaced $u\vec{}\delta(\vec{r}-\vec{R})$
and $I(\vec{r},\vec{R})$ by the values of these fields at the mesh
node nearest to the particle position. 

A refined version resorts to a simple linear interpolation scheme
using the mesh points on the elementary lattice cell containing the
particle. Denoting the relative position of the particle in this cell
by $(d_{x},d_{y},d_{z})$ with the origin being at the lower left
front edge, one defines $\delta_{0,0,0}=(1-d_{x})(1-d_{y})(1-d_{z})$,
$\delta_{1,0,0}=d_{x}(1-d_{y})(1-d_{z})$ etc., the interpolation
weights. The interpolated velocity then reads $\vec{u}\delta(\vec{r}-\vec{R})\simeq\sum_{r\in cell}\delta_{r}\vec{u}(\vec{r})$,
where the sum is over the mesh points on the considered elementary
lattice cell. The force exerted back to the fluid can then be chosen
with the same weight coefficients or by spreading the force equally
on the edges of the cell, the corresponding rule to preserve linear
momentum being easily derived.

Due to the regularization of the point-force, one should expect that
the mobility of the suspended particle is not simply given by the
bare friction coefficient $\gamma_{T}$, but is somehow renormalized.
Indeed, the mobility is given by the sum of the bare mobility and
a Stokes-type contribution due to the lattice discretization. The
effective friction coefficient relates to the bare one via $1/\gamma_{T}^{eff}=1/\gamma+1/g\eta\Delta x$,
where $g\simeq25$ is a geometrical correction coefficient \cite{AHLRICHSDUENWEG}.

The PPM finds application in simulating microscopic systems, with
a stochastic term added on each particle in addition to the fluctuating
LB bath. In this way, the particle does not leak energy on average
and the fluctuation-dissipation balance maintains a well-defined temperature
of the fluid-particle system. Again, to balance and preserve momentum,
such stochastic force is also restituted to the fluid with an opposite
sign. The presence of long-time tails, that is, the inherent hydrodynamically-sustained
motion of the moving particle generating a long-time decay of velocity,
has also been observed \cite{AHLRICHSDUENWEG}. On the downside, the
PPM is permeable to fluid momentum, as the local force is unable to
create enough resistance to the incoming flow, and the classical Stokes-like
picture of the streamlines does not apply.

Owing to its inherent simplicity, the PPM was first recognized to
be useful to simulate topologically connected particles, such as polymers,
in the presence of hydrodynamic interations. Whenever the embedded
particles represent micro/mesoscopic objects, such as atoms or the
beads of a polymer, mass diffusion plays a role comparable to mechanical
and hydrodynamic forces. Another interesting extension involves coupling
PPM with the Shan\textendash Chen multicomponent LB to simulate efficiently
complex fluid\textendash fluid interfaces. The idea is to introduce
a solvation free-energy for the particle\textendash fluid interaction
proportional to the fluid density gradients \cite{SEGA} and reads
as follows 
\begin{equation}
\vec{F}^{solv}=-\zeta\nabla\rho\label{eq:SOLVATION}
\end{equation}
Such force drives particles towards maximal or minimal density gradients,
depending on the sign of the coupling coefficient $\zeta$. The approach
is particularly appealing when used in conjunction with multiphase
conditions. As a matter of fact, it makes possible to treat different
multiphase fluids in the presence of suspended molecules, such as
amphiphilic molecules as surfactants or, by adding another level of
detail with electrostatics, polyelectrolytes in bicomponent fluids.

\subsection{The diffused-particle model (DPM) }

An extension of PPM represents particles with finite-size extension,
still relying on a force-based mechanism. Given the finite extension,
it is well suited for anisotropic particles, providing a very handy
computational flexibility for the description of biological suspensions,
cellular compartments or even entire cells (recalling that the interior
of the cell is anisotropic due to intracellular organelles). The strategy
is to consider the particle roto-translational response as originating
from the coupling of the finite-size extension of the particle with
the fluid momentum and vorticity. The particle is an effective diffused
body, with no need to track its boundary to control its coupling with
the environment.

The particle shape is described as an ellipsoid having three major
radii along the three principal directions, $\xi_{\alpha}$ with $\alpha=1,2,3$.
To fit in the discrete nature of the lattice, $\xi_{\alpha}$ are
taken as three integers, one for each cartesian component (such requirement
can be lifted but a few interesting properties described below, would
be lost). The roto-translation is governed by rigid body dynamics
of eqs. \ref{eq:rigidbodymotion}. The particle rotational state is
encoded by the matrix $\dvec Q$ whose rows are three orthogonal unit
vectors aligned along the principle axis of the particle, that is,
the basis to transform between the laboratory and the moving frames.
The roto-translational state is given by the tensorial product

\begin{equation}
\tilde{\delta}(\vec{r},\dvec Q)=\prod_{\alpha=x,y,z}\tilde{\delta}[(\dvec Q\vec{r})_{\alpha}]\label{eq:IBM}
\end{equation}
where 
\[
\tilde{\delta}(y_{\alpha})=\begin{cases}
\frac{1}{8}\left(5-4|y_{\alpha}|/\xi_{\alpha}-\sqrt{1+8|y_{\alpha}|/\xi_{\alpha}-16y_{\alpha}^{2}/\xi_{\alpha}^{2}}\right) & |y_{\alpha}/\xi_{\alpha}|\leq0.5\\
\frac{1}{8}\left(3-4|y_{\alpha}|/\xi_{\alpha}-\sqrt{-7+24|y_{\alpha}|/\xi_{\alpha}-16y_{\alpha}^{2}/\xi_{\alpha}^{2}}\right) & 0.5<|y_{\alpha}/\xi_{\alpha}|\leq1\\
0 & |y_{\alpha}/\xi_{\alpha}|>1
\end{cases}
\]
is a shape function having compact support and the normalization property
$\sum_{r}\tilde{\ensuremath{\delta}}(\vec{r})=1$. Typically $\xi_{x}=\xi_{y}=\xi_{z}=2$
corresponds to a spherically symmetric diffused particle with a support
extending over 64 mesh points. The translational response of the suspended
body is designed according to the fluid-particle exchange kernel 
\[
\vec{\phi}(\vec{r},\vec{R},\vec{V})=-\gamma_{T}\tilde{\delta}\left(\vec{V}-\vec{u}\right)
\]
where $\tilde{\delta}\equiv\tilde{\delta}(\vec{r}-\vec{R},\dvec Q)$
and the hydrodynamic DPM force is obtained via integration over the
particle spatial extension. It reads as follows: 
\[
\vec{F}^{drag}=\sum_{r}\vec{\phi}(\vec{r},\vec{R},\vec{V})=-\gamma_{T}\left(\vec{V}-\tilde{\vec{u}}\right)
\]
where $\tilde{\vec{u}}\equiv\sum_{r}\tilde{\delta}\vec{u}$.

The coupling between the body motion and the fluid vorticity is represented
by a rotational kernel 
\[
\bm{{\bf \tau}}(\vec{r},\vec{R},\bm{\omega})=-\gamma_{R}\tilde{\delta}\left(\vec{\Omega}-\vec{\omega}\right)
\]
with $\gamma_{R}$ a rotational coupling coefficient. The corresponding
drag torque is 
\[
\vec{T}^{drag}=\sum_{{\bf r}}\vec{\tau}(\vec{r},\vec{R},\vec{\omega})=-\gamma_{R}\left(\vec{\Omega}-\tilde{\vec{\omega}}\right)
\]
where $\tilde{\vec{\omega}}\equiv\sum_{{\bf r}}\tilde{\delta}\vec{\omega}$.

In the DPM, the smooth function $\tilde{\delta}$ is a very natural
interpolating device, and moving information back and forth between
the grid and the particles bears a significant influence on the effective
hydrodynamic size of the body. In particular, spherical particles
acquire an effective radius given by: 
\[
\frac{R_{eff}}{R}=1+a\frac{\Delta x}{R}-b\frac{R}{L}
\]

In the above, $a$ and $b$ are positive numerical prefactors taking
into account the range of the interpolators and finite-size effects.
The former make the effective radius larger, hence larger dissipation,
due to defective autocorrelations triggered by grid discreteness.
The latter have the opposite sign because the lack of large-scale
modes above the box size results in lesser dissipation. 

These corrections are crucial to the purpose of matching the diffusion
coefficient of the simulated bodies to the physical one \cite{AHLRICHSDUENWEG,DUNWEGLADD}.
\}

\subsection{Immersed Boundary Methods}

The hydrodynamic force and torque acting on the DPM are obtained via
integration over the particle spatial extension (volume force). A
different approach represents the finite extension of particles by
tracking only the particle boundary degrees of freedom, the so-called
Immersed Boundary Method (IBM), developed by C. Peskin decades ago,
to deal with immersed moving boundaries within fluid flows~\cite{PESKIN}.
Borrowing from Ladd's approach, on one hand, and Ahlrichs and Dünweg,
on the other, the LB-IBM procedure is both boundary-driven and force-driven.
The major benefit is that the body surface is treated like a deformable
membrane, emanating an elastic force field towards the outside fluid.
In equations: 
\begin{equation}
\vec{F}_{f}(\vec{r};t)=\int_{\mathcal{M}}\vec{F}_{m}(\vec{R},t))\delta[\vec{r}-\vec{R}(t)]d\vec{R}\label{M2F}
\end{equation}
where $\vec{r}$ is the generic fluid location, $\vec{R}$ runs over
the two-dimensional membrane and $\vec{F}_{m}$ is the force acting
on the membrane at location $\vec{R}$. The latter is usually computed
as the divergence of the elastic stress tensor of the membrane. The
membrane equation of motion is given by the Lagrangian condition 
\begin{equation}
\frac{d\vec{R}}{dt}=\vec{u_{f}}(\vec{R},t)\label{MEOM}
\end{equation}
where $\vec{u}_{f}$ is the fluid velocity extrapolated to the membrane
location, i.e. 
\begin{equation}
\vec{u}_{f}(\vec{R},t)=\int_{\mathcal{M}}\delta(\vec{r}-\vec{R})\vec{u}(\vec{r})\;d\vec{r}\label{eq: MEOM1}
\end{equation}
The fluid equation of motion is the standard (Eulerian) LB, with the
force given by the numerical version of the integral in eq.~(\ref{M2F}).

To be noted that the accuracy and efficiency of the LB-IBM scheme
is highly dependent on the discrete representation of the Dirac's
delta. This is usually replaced by piecewise polynomials extending
over a few lattice sites, typically $4$ for cubic splines, sometimes
also known as ``smoothed particles''. The LB-IBM is a fully coupled
non-linear and non-local Eulerian-Lagrangian scheme, hence it presents
a very demanding computational task. However, it provides the major
benefit of smoothness, due to the integral nature of the convolutions
which control the exchange of information between the LB fluid and
the IBM membrane. The LB-IBM scheme is gaining increasing popularity
for soft-matter applications involving the interaction of micro-nanoscale
fluids with deformable suspended bodies.

\subsection{Chemical specificity and Coarse-Graining}

Realistic microscopic biological simulations set a key quest for chemical
specificity. As summarised in the announcement of the 2013 Nobel prize
in Chemistry (http://www.nobelprize.org/nobel\_prizes/chemistry/laureates/2013/,
2013), a tremendous amount of information has been gained during the
last forty years by using Molecular Dynamics to simulate proteins.
That is, by reproducing molecular forces to high accuracy and the
complex motion of biological settings based on Newton's equations
of motion.

The hallmark of protein simulations has been the discovery that complex
macromolecules can be represented by force fields that include intra
and inter-aminoacidic interactions via comparatively compact potential
functions. Several force fields, such as the well-known Charmm and
Amber force-fields \cite{PONDER2003,GUVENCH2008,LINDORFF2012}, have
been developed in the all-atoms context, that is, by taking into consideration
all possible atoms stemming from the macromolecule and the aqueous
solvent. The all-atom strategy was dictated by the high level of heterogeneity
of biological interactions, ranging from the local hydrogen bonds,
to dispersion forces, to the long-range electrostatic interactions.

Computing such diverse forces at high-performance rates, mandates
sophisticated parallel algorithms and dedicated hardware. In order
to unveil macromolecular dynamics and, most notably, folding within
the characteristic \textit{funneled} free-energy landscapes \cite{levitt1975computer,Frauenfelder1991,WOLYN},
to large stretches of time, up to milliseconds, few outstanding simulations
have been carried out to date, by exploiting custom/specialized hardware
\cite{SHAW2008,FEIG2017}. While providing extraordinary results,
such efforts are still isolated and, more importantly, cannot deal
with large size systems, due to the huge memory requirements required
in practice, and the large hardware and power costs of extreme-scale
simulations.

For these reasons, more simplified versions of force fields have emerged
in recent years, the so-called Coarse-Grained Force Field (CGFF) with
the aim to tackle large systems while retaining the desired degree
of chemical specificity and accuracy in place \cite{warshel:2011}.
A few examples of CGFF are currently available (MARTINI, OPEP) \cite{LBMDbio2,LBMDbio3}
and they are being deployed to study many different macromolecular
systems, ranging from polypeptides to polynucleotides. A key point
of the CGFF approach is the mesoscale nature of the molecular representation,
in fact multiple atoms are replaced by effective particles. Such particles
are then connected by bonding potentials that enforce local backbone
connectivity and structure. Other non-bonding interactions are used
to enforce dispersion and hydrogen bond forces packaged into a lumped
form.

Importantly, the CGFF was developed to remove the explicit need to
represent the molecular details of the solvent and as such, solvent-mediated
interactions, as much as electrostatics, are effectively recasted
in terms of the CGFF potential form. It is true, though, that coming
up with a consistent form of the solvent-mediated interactions is
not an easy task and some force fields, such as the MARTINI one, use
some form of particle-based method to represent a minimalistic type
of solvent. Setting aside details, CGFF provide a major step towards
bio-simulations. In future years, we are most likely to witness more
force fields, with the goal of dispensing with the explicit representation
of the solvent. 

\begin{figure}
\begin{centering}
\includegraphics[scale=0.7]{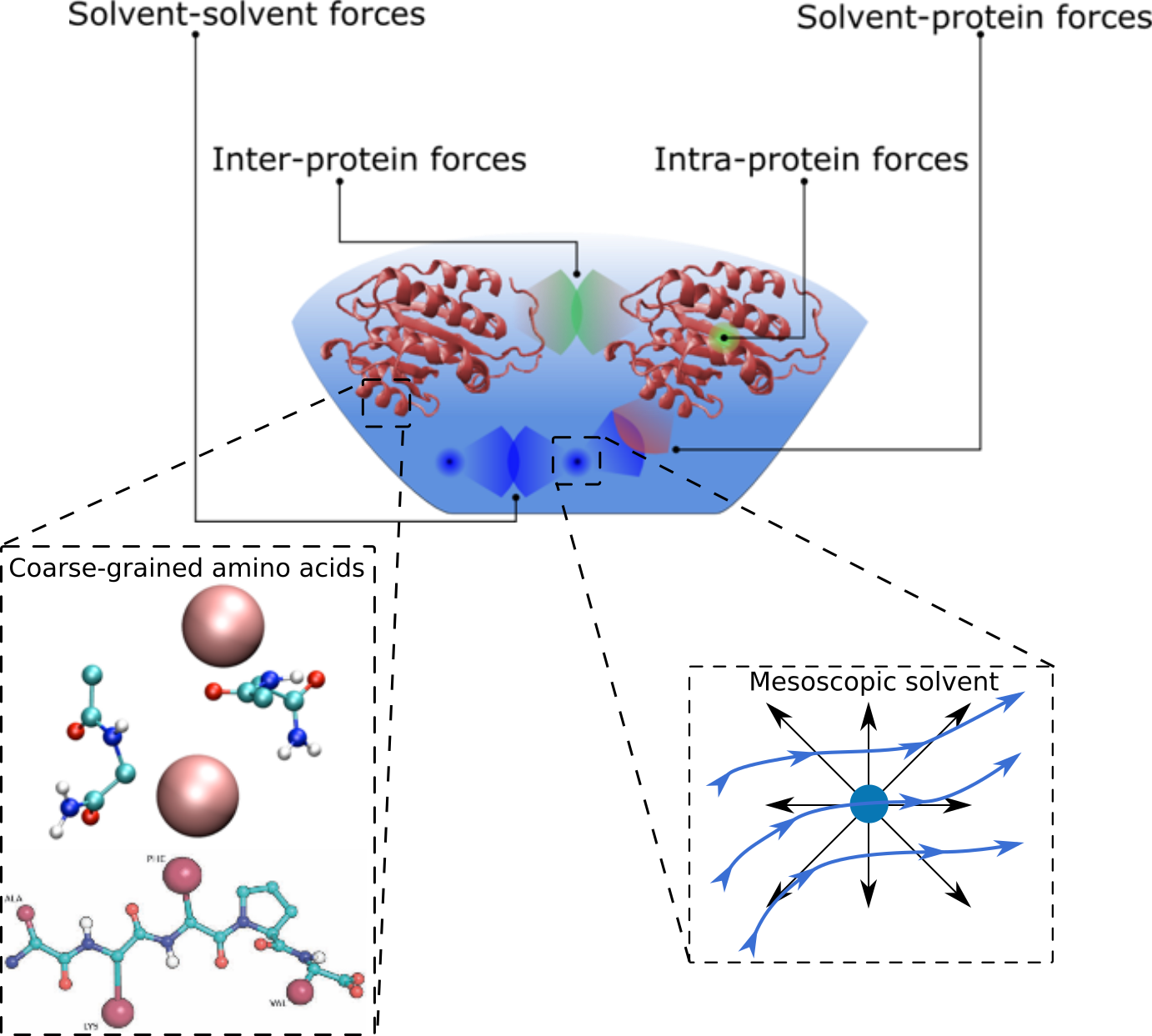} 
\par\end{centering}
\caption{Proteins immersed in a LB solvent are treated as mesoscopic objects.
Consequently, their representation is coarse-grained, that is, a level
of detail based on a handful of pseudo-atoms per amino acid (five
pseudo-particles in the current case from the OPEP model \cite{chebaro2012}).
The pseudo-particles are needed to encode the details of the peptidic
backbone and effective inter-particle interactions. \label{fig:FIGPROTEIN}}
\end{figure}

The most relevant aspect in the present context, is that the CGFF
approach sits well with the mesoscale description for the embedding
solvent and LB provides an ideal partner to simulate large protein
assemblies. Since thermal fluctuations are key at this scale, the
simulations require Langevin terms as well as thermal fluctuations
on the LB side (see Fig. \ref{fig:FIGPROTEIN}).

As the basic form of LB does contain only ideal thermodynamics, the
CGFF does not need to be modified with respect to the original version
used in absence of HI, this holds true because from the statistical
mechanics viewpoint, effective forces under equilibrium conditions
do not depend on hydrodynamic, velocity-dependent interactions\cite{hansen1990theory,noid2008multiscale}.
Conversely, under non-equilibrium conditions, say subject to external
macroscopic flow, the CGFF requires changes. As the need arises in
the direction of reintroducing solvent-mediated interactions stemming
from non-ideal thermodynamics or hydrogen bond interactions, one may
think of reformulating the CGFF to account for an explicit non-ideal
thermodynamics of the solvent. Many research works and applications
ahead are awaiting for a full exploration.

As for any simulation of proteins in solution, a word of caution is
in order about the usage of mesoscale simulations to describe the
aqueous solvent as a continuum. The LB mesh spacing $\Delta x$ is
defined as a coarse-grained representation of the collective kinetic
behaviour of a group of solvent molecules. In order to observe hydrodynamic
behavior down to the mesh spacing distances, the mean free path is
usually expected not to exceed $\Delta x$. In the liquid state, the
molecular mean free path extends over just a few Angstroms, thus commanding
subnanometric lattice spacings to allow the emergence of hydrodynamic
behaviour at scales larger than $\Delta x$. 

As shown by Horbach and Succi \cite{HORBACH}, this strategy is effective
for simulating nanofluidic coherent patterns in close agreement with
those obtained by Molecular Dynamics simulations. 

\subsubsection{Life at low Mach and Reynolds numbers: numerical caveats}

In this section we touch briefy at some caveats that need to be taken
into account in the LBPD simulations at vanishingly small values of
the Reynolds and Mach numbers. 

The three main dimensionless groups characterising incompressible
flows are the Reynolds number $Re=UL/\nu$, the Mach number $Ma=U/C_{s}$
and the Knudsen number $Kn=\lambda/L$, where $U$ is the flow speed,
$C_{s}$ the (physical) sound speed, $L$ a typical length scale and
$\lambda$ the molecular mean free path. Elementary manipulations
show that the three groups are related through the so-called von Karman
relation: 
\[
Re\;Kn=Ma.
\]
In order for an ensemble of molecules to behave collectively like
a fluid, the Knudsen number must be significantly below unity, $Kn\ll1$,
which means that low Reynolds numbers imply even smaller Mach numbers.
Since LB explicitly tracks sound waves, this implies extremely small
timesteps. A few numbers help clarifying the point. Consider a nanoscale
LB simulation of water with a lattice spacing $\Delta x=10^{-9}$
m. Given that the kinematic viscosity of water is $\nu=10^{-6}m^{2}/s$,
using $\nu_{LB}=0.1$ implies a timestep $\Delta t=\frac{\nu}{\nu_{LB}}\;\Delta x^{2}=10^{-13}\;s$.
This shows that LB ticks ``just'' two orders of magnitude above
molecular dynamics. While appropriate for nanoscale simulations, it
is clear that such tiny timesteps do not allow to cover macroscopic
time-spans. At a given spatial resolution, the only way to increase
the time step is to increase the lattice viscosity $\nu_{LB}$, but,
as explained earlier on, taking $\nu_{LB}$ above unity runs against
the hydrodynamic constraint $Kn<1$. Hence LB simulations in the Stokes
limit $Re\to0$ are restricted to very small time steps. One way of
short-circuiting the problem is to artificially enhance the Mach,
hence the Reynolds number, in the hope that the physics in point be
rather insensitive to the specific value of these numbers, as long
as they are both well below unity. Under such benevolent conditions,
Reynolds number, say $10^{-6}$, could be safely inflated to, say
$Re=10^{-2}$, without incurring any major disruption of the physics
at hand. Needless to say, this \char`\"{}inflationary stratagem\char`\"{}
only works as long as the physics exhibits analytic and non-singular
dependencies on the Reynolds number. Although characterized by the
onset of non-local, long-range interactions, the limit $Re\to0$ is,
in principle, a non-singular one, thus leaving some chances to the
inflationary strategy discussed above. 

\subsection{LB vs PD resolution}

The LBPD scheme relies on two essential timescales for both the fluid
solvent dynamics, controlled by kinematic viscosity $\nu$, and the
molecular dynamics, controlled by the friction, or drag coefficient,
$\gamma_{T}$. The two are related via Stokes law: 
\[
\gamma_{T}=3\pi\eta{\cal D}/M
\]
where $\eta=\rho\nu$ is the solvent dynamic viscosity, ${\cal D}$
is the particle (protein) equivalent diameter and $M$ its mass. In
the LBPD scheme, however, fluid and particles are handled as numerically
independent quantities, so as to achieve the desired level of coarse-graining
between fluid and particles degrees of freedom. By obeying the fluctuation
dissipation principle $D=kT/\gamma_{T}$, at given level of thermal
fluctuations, mass diffusivity and coupling parameter are changed
as a single parameter. Another important quantity is the ratio of
solvent viscous and solute mass diffusivities, the so-called Schmidt
number, 
\[
Sc=\frac{\nu}{D}
\]
In an ideal gas $Sc=1$, but in a liquid it typically exceeds $100$. 

When simulating proteins within the LBPD scheme, a conservative approach
is to represent proteins at coarse-grained level by taking a mesh
spacing $\Delta x=2$ Angstrom and a timestep of $\Delta t=1$ fs.
The latter guarantees that the protein internal motion is duly resolved
by the PD solver, whereas the mesh spacing guarantees that local hydrodynamic
signal is captured by the LB solver. Since proteins and biological
aggregates range in size from tens to thousands nm in diameter, such
mesh spacing fully resolves the mesoscale structure. One may reasonably
wonder whether at the sub-nanometric scales the notion of Boltzmann
distribution functions still makes sense, due to the dominance of
statistical fluctuations. This is indeed possible by introducing the
notion of effective thermal mass, to be detailed in the sequel.

The starting setting is such that a protein particle sits typically
inside a mesh voxel, a comfortable situation for numerical production
purposes. On the other hand, it is often desirable to increase $\Delta x$
to minimize the computational burden. Under such conditions, the numerical
effort involved in handling the LB solver dominates the MD component,
typically by a factor of $3-5$. However, it is well known that, in
principle, resolving the sub-nanometric hydrodynamics is unnecessary,
since hydrodynamic interactions typically set in above the $3-5$
Angstrom scale.

Under operating conditions, it is convenient to fix the time step
$\Delta t$, as this is a rather strict requirement from the PD side.
Thus, changing $\Delta x$ at fixed physical viscosity or equivalently
Reynolds number, implies altering the numerical viscosity. If we utilize
the method of matching advection, we require that the characteristic
velocity remains unchanged, $u=\frac{\Delta x}{\Delta t}u_{lb}$,
implying that numerical viscosity changes as $\nu_{lb}=\frac{\Delta t}{\Delta x^{2}}\nu=\frac{u_{lb}\nu}{u}\frac{1}{\Delta x}$.

However, this simple scheme is inconvenient if the numerical mass
diffusivity of solutes must kept fixed at varying $\Delta x$. In
fact, mass diffusivity is primarily a function of the interatomic
distances, their mutual interactions and the solvent viscosity, so
that the numerical counterpart, expressed as $D_{pd}=D\Delta t/d^{2}$,
where $d$ is the interparticle distance, should be left unchanged
as much as $d$. In summary, we are left with a varying $\Delta x$,
with fixed $Sc$,$\Delta t$ and $d$. 

To analyze the implications on the solvent viscosity, let us recast
the Schmidt number as follows: 
\[
Sc=\frac{\nu}{D}=\frac{\nu_{lb}}{D_{pd}}\frac{\Delta x^{2}}{d^{2}},
\]
From this expression, we see that in order to keep $Sc$, $\Delta t$
and $d$ unchanged at changing $\Delta x$, the LB viscosity must
scale as $\nu_{LB}\propto1/\Delta x^{2}$. 

This shows that $\Delta x$ cannot grow too large, for pain of undermining
numerical stability. Typically, $\nu_{LB}$ should stay above $0.005$,
although resort to entropic version of LB may loosen this constraint.

Overall, this strategy can achieve a significant speed-up of the simulation,
reaching a factor three in solvent coarsening and a sizeable speed-up
of $27$ for the LB solver, without affecting protein diffusivity
and related phenomena, such as the kinetics of peptidic aggregation. 

\subsubsection{The Boltzmann number}

As discussed earlier on in this paper, the fluctuating LB (FLBE) is
meant to account for the statistical fluctuations about the equilibrium
state being dictated by the number of particles in the discrete states,
as it is the case for actual molecules.

The relative intensity of the fluctuations, proportional to $1/\sqrt{\Delta N}$,
is controlled by the so-called \emph{Boltzmann number}, defined as
\cite{DUNLADD}: 
\[
Bo=(\frac{\theta}{n\Delta x^{3}})^{1/2}
\]
For most fluids $\theta\equiv v_{T}^{2}/c_{s}^{2}\sim O(1)$, $\theta=1$
corresponding to the ideal gas limit. The denominator is the number
of molecules represented by a single LB particle, once we stipulate
a number density $n_{lb}=1$ in lattice units, which is always possible
in an incompressible fluid.

The (squared) Boltzmann number can also be recast in the alternative
form: 
\[
Bo^{2}=\theta\;(\frac{d}{r_{0}})^{3}\;(\frac{r_{0}}{\Delta x})^{3}
\]
where $d$ is the interparticle distance and $r_{0}$ is the range
of the intermolecular potential.

The second term at the right-hand side is the granularity parameter,
controlling the degree of non-diluteness, while the third one is a
direct measure of the spatial coarse-graining.

By definition, the standard non-fluctuating LB corresponds to the
macroscopic limit 
\[
Bo\to0
\]
To be noted that in a liquid, where $d/r_{0}\sim O(1)$, and even
more so in a dilute gas, $d/r_{0}\gg1$, the smallness of the Boltzmann
number is entirely in charge of spatial coarse-graining, i.e., $r_{0}/\Delta x\ll1$.

In a nanofluid however, the latter condition is no longer guaranteed.

For macroscopic flows, the Boltzmann number is pretty small indeed,
typically of the order of the inverse square root of the Avogadro
number.

For a millimeter cube of water, about $30$ molecules per cubic nanometer:
with $\Delta x=10^{-3}$, $d/r_{0}\sim1$, and $\theta\sim1$, we
obtain $Bo\sim10^{-10}$.

In microfluidics, the Boltzmann number gets larger: with $\Delta x=10^{-6}$
m, we obtain $Bo\sim10^{-5}$, still much smaller than $1$, no point
for FLBE.

Nanofluidics, though, tells another story; with $\Delta x=10^{-9}$
m, $Bo\sim0.2$. By pushing LB even further down, to atomistic scales,
$\Delta x=0.1$ nm, see \cite{HORBACH}, the Boltzmann number may
even get larger than 1: the LB particles become ``quarky'', i.e.,
they represent \emph{a fraction} of a real molecule!

Clearly, this is a strongly fluctuating regime, for which the notion
of FLBE as a weak perturbation on top of LBE goes under heavy question.
In fact, at such sub-nanometric scales, the very notion of distribution
function becomes shaky because of many-body density correlations.
The question is: \emph{how can we possibly use the fluctuating LBE
altogether at such small scales}?

Here, mass and temperature come to some rescue, as detailed below. 

FLBE simulations work at $v_{T}^{2}\sim10^{-4}\div10^{-3}$, much
smaller than $c_{s}^{2}\sim O(1)$, both in lattice units, for otherwise
numerical instabilities arise due to the stochastic source being too
strong.

This constraint is conveniently analyzed in terms of the ``thermal
mass'' introduced by Duenweg and Ladd \cite{DUNLADD}, namely: 
\[
m_{T}\equiv\frac{k_{B}T}{c_{s}^{2}}=m\theta
\]
where $m$ is the mass of the solvent molecules, i.e. $m/m_{i}\ll1$.

Given that the inertial mass is typically set to $m=1$ in LB units,
if the thermal mass were to coincide with the inertial one, i.e. $m_{T}=m=1$,
one would indeed obtain $v_{T}^{2}/c_{s}^{2}\sim O(1)$ in lattice
units.

Since in actual practice the above ratio is of the order $10^{-3}$
at most, the thermal mass is much smaller than $1$, thus ensuring
the condition $Bo\ll1$ even when fluctuations in particle number
are of order $O(1)$.

In practical terms, it is like replacing a particle of mass $m$ with
$m/m_{T}\gg1$ particles of mass $m_{T}$, a procedure with many similarities
with variance reduction techniques popular in Monte-Carlo simulations.

\section{SIMULATIONS AT THE PHYSICS-CHEMISTRY-BIOLOGY INTERFACE}

In order to appreciate the specificity of the LBPD approach to tackle
diverse problems at the PCB interface, we consider the computational
complexity of the distinct LB and PD components of the methodology.
The number of floating point operations needed to simulate a given
problem over its characteristic dynamical evolution depends on the
representation adopted for its constituents. In addition, $k=\#Flops/(\Delta x^{3}\Delta t)$
is the computational density, being given by the ratio of the number
of floating point operations needed to update the degrees of freedom
contained in an elementary cell of volume $V=L^{3}$ and for a time
$T$, both in lattice units. The complexity (Flops) is thus expressed
as 
\begin{equation}
\mathcal{C}=(k_{LB}+pk_{PD})L^{3}T\label{eq:CPU}
\end{equation}
where $p$ is the fraction of particles contained in the elementary
lattice volume $\Delta x^{3}$. Note that the cost of LB-PD cross-coupling
has been empirically absorbed by the prefactors $k_{LB}$ and $k_{PD}$. 

For diffusion-dominated applications, the typical case in biological
processes, $T\sim L^{2}$ thus $\mathcal{C}\sim L^{5}$, while for
ballistic or convective dynamics, $T\sim L$ thus $\mathcal{C}\sim L^{4}$.
Granted that for quantitative purposes, $\mathcal{C}$ needs to be
evaluated for the specific application at hand, as a first estimate,
we assume $k_{LB}\sim pk_{PD}\sim10^{4}$.

A similar argument goes for the memory demand, which can be written
as 

\begin{equation}
\mathcal{M}=8(n_{LB}+pn_{PD})L^{3}\mbox{{Bytes}}\label{eq:MEM0}
\end{equation}
where $n_{LB}$ is the number of discrete LB populations per lattice
cell and $n_{PD}$ is the number of degrees of freedom per discrete
particle. For a standard single-species LB scheme $n_{LB}\simeq20$,
while for pointlike particles $n_{PD}=6$ (position and momentum),
for rigid bodies $n_{PD}=12$, (position, momentum, angles and torque)
and for extended deformable bodies $n_{PD}$ can reach up to several
hundreds. For the applications to be discussed in the sequel, we shall
take simply:

\begin{equation}
\mathcal{M}\simeq10^{3}L^{3}\mbox{{Bytes}}\label{eq:MEM}
\end{equation}
which is an overestimate for dilute rigid particles and a likely underestimate
for dense deformable ones. 

\subsection{Biopolymer Translocation}

The translocation of biopolymers, in particular DNA or RNA strands,
in nanometric pores provides a showcase of the synergistic hydrodynamic
effects assisting or interfering with the translocation process. The
translocation mimics a genuinely biological one, whereby viral penetration
takes place via the injection of viral genetic material into the host
cell's cytoplasm. At technological level, understanding how the physics
of nanopores controls translocation inspires new paths to fast DNA
sequencing~\cite{FytaReview}. Nanopores-based technologies ultimately
aims at translocating polynucleotidic chains through a nanoconfined
environment, where the genetic information can be decoded by optical
mapping, ionic or electronic detection. The challenge is to control
the process and the random, squiggly forms that the polynucleotide
takes in solution, eventually designing nanofluidic devices according
to stringent photolithographic requirements.

Computer simulations have the ability to access the fine details of
the translocation process, both for technological innovations and
for a better understanding of the biological processes involving the
migration of small biopolymers. Consequently, various applications
of LBPD have also appeared in recent years \cite{DEPABLO2007,HAMMACK2011,YEOMANS2012,HICKEY2014,ALFAHANI2015}. 

Biopolymer translocation has been analysed in different set-ups and
modeling details for the translocating biopolymer, starting by a single
necklace and neutral polymer threading between two chambers driven
by a localized force acting only within the pore region (bead pulling)
\cite{DNA1,DNA2,DNA2b}. This set-up mimicked real experiments where,
given the presence of two electrodes at large distance from the pore,
the electric field is overly intense where resistance is higher and
mostly constant inside the pore region. In addition, electrostatic
interactions stemming from the charged polymer are modelled in terms
of effective beads of the chain. Simulations of translocation in small
pores have thus focused on the dependence of the translocation time
on the polymer length, thereby showing that the effect of hydrodynamic
interactions is best seen on the translocation time vs chain length
$T\propto N_{b}^{\alpha}$, with the characteristic exponent being
$1.27$ for short and $1.32$ for long polymer chains, a result that
is explained in terms of scaling analysis and energetic considerations,
that are peculiar to such hydrodynamic-assisted process. Translocation
in large pores showed a even richer phenomenology, with the appearance
of several different configurations of the polymer folds, the consequence
of fast translocation events that create discrete states that reflect
on quantized current blockades on the measurable ionic currents \cite{DNA2b,DNA3}.
Even more central is the role of electrokinetic forces on the process,
the physical ingredient that can be included only by the full solution
of the charged polymer whose translocation is driven by the self-consistent
electric field. Even by including the double helix structure of polynucleotides,
the complexity of the numerical apparatus can be optimally handled
within the coherent LBPD framework, complemented by the solution of
the Poisson equation for electrostatics. The result is a detailed
description of translocation and the measure of the ionic currents,
locally modulated by the threading polymer and being the result of
the concurrent effects of excluded volume, drag and electrostatic
forces \cite{DATAR2017}. Importantly, the development of new coarse-grained
potentials for DNA \cite{hsu2012ab}and RNA \cite{cruz2018coarse}
paves the way to reveal the effect of the strong charging of the nucleic
backbone that could not be elicited by using more aggressive coarse-grained
models \cite{miocchi2018mesoscale}.

For the translocation process, the associated threading time is proportional
to the number of beads of the polymer $N_{b}$ and in lattice units,
and in lattice units it can be estimated as
\[
T\approx10^{2}\times N_{b}^{1.27}
\]
where the characteristic exponent is a direct signature of hydrodynamic
interactions assisting translocation \cite{storm2005fast,FytaReviewBOH}.
Most of the computational time goes into solving hydrodynamics via
the LB component, while the time to compute the mechanical forces
and evolve the PD component is negligible. Consequently, for a typical
size $L=10^{2}$, the problem requires 

\[
\mathcal{C}\simeq10^{13}\mbox{Flops}=10\mbox{ TeraFlops}
\]
to translocate a chain of about $100$ beads. This is well within
the capabilities of present day computers. In fact, scaling up the
size $L$ by a factor $10$ would take to the order of Exaflops, still
feasible on present-day leading-edge Petaflops/s computers. 

As to memory requirements, based on eq. (\ref{eq:MEM}), one estimates$\mathcal{M}\simeq10^{9}$,
a rather modest Gigabyte. 

These figures reflects the fact of working with highly stylized polymers,
without internal structure and chemical specificity (Physics-Biology
instead of PCB). 

The set-up of translocation consists of two large chambers, a \emph{cis}
and a \emph{trans} chamber, containing the pre-translocating and post-translocated
portions of the DNA strand. The chambers are typically much larger
than the nanopore characteristic size. Translocating a long DNA or
RNA chain into extremely narrow pores results in large entropy loss
caused by the confinement and the need to stretch the macromolecule.
The associated free-energy barrier reduces the biopolymer capture
rates and causes clogging at the nanochannel/pore entrance. On the
other hand, solvent-assisted interactions lubricate the process. It
is key to understand that the hydrodynamics of the translocating biopolymer
in such fluidic device, being modulated by competing forces acting
in the chambers and in the pore, give rise to a genuine multiscale
scenario~\cite{DNA1,DNA2,DNA3,DNA4,DNA5,LBMD1,LBMD2}. When facing
such complex set-up, all-atoms Molecular Dynamics methods, or even
coarse-grained representations of the translocating biopolymer, neglect
the explicit representation of the solvent, thus imposing severe limitations
to the overall accuracy. Resorting to strategies based on a direct
solution of the NS equations, or using other mesoscopic numerical
methods (Lagrangian or Eulerian based) is challenging in terms of
generating consistent fluctuations under confinement and achieving
a stable numerical method. In this respect, the LBPD method is attractive
because it allows generating the thermal fluctuations in a natural
way and guarantees numerical stability over a wide range of translocation
rates. In addition, one can analyze biomolecules of different size
and initial configurations, in situations where the biomolecule is
approaching the pore or is already in a docked configuration. LBPD
has been utilized to analyze multiple scenarios \cite{DNA4,FytaReview,DEPABLO2007}
when the biopolymer has lateral size comparable or smaller than the
pore diameter, conditions giving rise to single-file or multi-file
translocation configurations, as illustrated in Fig. \ref{fig:Translocation}.

When accounting for the simultaneous presence of hydrodynamic and
frictional forces, one can initially rely on the assumption of charge
neutrality for the biopolymer and saline solution, a simplification
justified by the need to reduce the computational effort. However,
electrostatics is essential to guide the ionic currents and the current
blockades caused by the impeding DNA molecule. A direct understanding
of the ionic current blockades provides a stringent comparison with
experimental measurements. The situation is even more complicated
under flow conditions, whereby the interplay between electrostatics
and flow does not allow to utilize simplified solutions based on the
assumption of global or local equilibrium. The inclusion of electrokinetics,
that is, the representation of the multi-component saline solution
that flows together with DNA from chamber to chamber, provides a direct
access to the electrohydrodynamic process \cite{FytaReview}.

\subsection{Ion Channels}

As anticipated, electrohydrodynamics is a fundamental aspect of the
biological function, in particular as regarding to ion channels, the
prototypical example of nanoscopic pores that subtend to the passage
of ions in and out of the cell and regulate its volume. Ion channels
are found within the membrane of most cells and are basically proteins
that form the pore connecting the inner and outer parts of the cell.
They look as narrow, water-filled pores that allow ions of certain
types to pass through via selective permeability, privileging specific
species, typically sodium or potassium. The transport of monovalent
or divalent species depends crucially on the morphological properties
of the confining elements that decorate the pore, notably charged
peptidic groups that form the inner scaffold of the channel \cite{DOYLE1998,HILLE2001}.

Knowledge of the way that ionic transfer takes place, unveils the
biological functioning, but simulating a large biological aggregate
composed of a membrane, ion channel and the inner and outer side of
the cell comprises a number of degrees of freedom, easily in excess
of millions. Due to the large spread of relevant timescales, often
unaccessible to today's computers. In principle, an alternative route
is to leverage the statistical-mechanical approach such that the atomistic
representation of the pore proteins is substituted by higher-level,
coarse-grained descriptions. Another pillar of kinetic modeling, the
Nernst-Planck equation, makes drastic simplifications by neglecting
hydrodynamics altogether, but provides the fluxes of ionic species
as a function of the concentration and applied voltage. Such drastic
simplification misses the fine details of ionic transport and the
imperfect screening occurring inside the narrow cavities of the ion
channel. From the operational standpoint, studying ionic transport
requires matching the atom-based with the continuum-based description,
a computationally unviable route due to the huge space / time gap
separating the two levels \cite{MARCONI2013}. 

As to translocating DNA, an optimal strategy is to proceed along the
tandem LBPD path, whereby any feature that takes place at the fine
atomic scale can surface up at the largest available scale, with its
full content of long-range and unscreened electrokinetics. The numerical
approach grants access to the characteristic ionic response by combining
the fluid dynamics of multiple species in solution, the interplay
of electrostatics and viscous forces, together with chemical specificity
for the confining protein. The latter is particularly effective in
determining the fine features within the pore lumen and vestibules
that are responsible for ionic selectivity. 

Another intriguing aspect of ion channels functioning is the fact
that transport takes place under strictly microscopic confinement,
whereby the competition among diffusive, stochastic and migration
forces together with the channel walls act as an effective thermalizing
bath for the moving ions. 

Ion channels have provided a stringent benchmark to quantify the mechanisms
by which local details arising from the channel geometry and the surface
charge, the salinity of the electrolytic solution and the physical
scale under study, affect ionic transport and the ensuing biological
function. Electrokinetic forces have been shown to be highly modulated
by geometrical details and by the channel surface charge \cite{marini2012charge,melchionna2011electro}.
The presence of internal vestibules of the biological channel, for
example, are easily modelled in the simulation and provide a direct
inspection on the way that the electric field focuses along the channel
axis, thereby modifying its activity. The role of axial asymmetries
can be probed directly. Assimilating the channel shape to a conical
one  revealed the peculiar characteristic curves where currents are
highly rectified by rather modest shape asymmetries. Similarly, the
presence of boundary effects at the channel inlet, are crucial to
capture ions from the bulk and convey them under confinement by lowering
\emph{de facto} the involved energy barriers \cite{chinappi2014modulation}.
The effect of millimolar concentrations of electrolytes has been studied
in terms of the double layer theory and unvealed the role of screening
on confined transport. Finally, and possibly most importantly, the
role of nanoscale forces stemming from excluded volume interactions,
acting among solvent molecules and ions, provide the critical ingredient
to understand transport under strong confinement \cite{MARCONI2013}.

Within the LBPD framework, let us estimate the computational effort
by considering a typical current of $1$ pA , traversing a ion channel
corresponding to a flux of $10^{7}$ ions per second. Consequently,
to study the passage of a single ion in a simulation box of edge length
$L=10^{2}$ (accomodating a membrane of thickness $4$ nm) with a
timestep of $10^{-12}$ s, required to cover $10^{-7}$ s, delivers
a computational complexity: 
\[
\mathcal{C}\simeq10^{4}\times10^{6}\times10^{5}\mbox{ Flops}=1\mbox{ PetaFlops}
\]
This is well within reach of standard computational resources, a Teraflop
computer would deliver in less than a hour.

With $L=10^{2}$, the memory requirements position around the tens
of Gigabytes.

However, such effort may not be needed to understand the ionic currents
semi-quantitatively. In fact, under confinement, hydrodynamics and
long-range coherent motion of the aqueous solution dissipate away,
due to the channel wall. Therefore, the continuum picture of fluid
flow is not the most effective way to represent the ion channel or
the entire embedding membrane in 3D. Alternatively, the Fokker-Planck
equation describes well the action of the thermalizing channel wall.
The mandate is then to cast the Fokker-Planck equation within the
LB framework, a task that has been succesfully accomplished a decade
ago, \cite{MORONI2006,MELCHIONNAFP2006}. Although less popular than
it fluid dynamic ounterpart, the lattice Fokker-Planck methodology
shows the same levels of accuracy, robustness and scalability. 

The LB can be comfortably extended to a broad variety of kinetic equations
and one more proof comes from handling excluded volume interactions.
The atomic correlations stemming from both electrostatics and excluded
volume interactions are particularly intense under the channels operating
conditions. Modeling correlations is a crucial element that other
methodologies, such as the Nernst-Planck or Dynamical Density Functional
Theory, cannot provide in conjunction with the solution of the NS
dynamics. 

The LB applied to ion channels has found various applications, although
most of them neglect the role of excluded volume and local specificity.
Excluded volume forces acting between molecules can be determined
starting from the Enskog collisional kernels, a revised version of
kinetic theory of gases, by resolving the ballistics of hard core
collisions \cite{LBMFluid1,MARCONI2009,MARCONI2011,MARCONI2011b}.
The LB scheme accommodates this new collisional kernel in a natural
way, another example of the versatility of the LB framework.

An upsurge of more chemical-specific LB schemes is awaited for and
expected in the forthcoming years. 

\subsection{Protein Diffusion and Amyloid Aggregation}

Mesoscopic simulations of macromolecules in aqueous solvent not only
allow to account for nanometric-scale hydrodynamics, but also for
macromolecular interactions, that are of paramount importance to avoid
misfolding \cite{dobson2015} and molecular recognition. An important
question is the extent to which molecular details are sufficient to
reach the required level of biological realism. The answer is definitely
problem-specific: representing a protein, a DNA chain or a lipidic
chain, may require different degrees of chemical specificity, depend
on the research target in point. 

It is also legitimate, however, to utilize coarse-grained force fields
in a rather flexible way, as long as the mesoscopic properties, fixed
at the nanometer/nanosecond scale, are reproduced. 

Following these lines, simulations of $18,000$ proteins have been
performed to demonstrate the capabilities of the computational method,
together with the parallel scalability on the Titan hardware platform
composed of $18,000$ GPUs\cite{LBCROWDING}, a multi-protein configuration
being sketched in Fig. \ref{fig:Crowding}. Such simulation allowed
for the first time to observe the diffusional properties of the solution
under realistic crowding conditions. Another valuable application
is provided by the joint usage of the particle-based approach with
a multiphase LB scheme \cite{SEGA}. A direct illustration of the
non trivial fluid-particle interplay in the formation and modulation
of membranes driven by the action of drag and solvophilic forces,
is shown in Fig. \ref{fig:Sega}. 

To answer the question about the optimal scale to represent a given
biological solutions, this is where kinetic modeling, particularly
for the liquid state solution, and the force fields match in accuracy.

At larger scales, micrometers and above, such level of detail may
become irrelevant, therefore a fair representation for thermodynamics,
possibly via an equation of state, and an accurate representation
of fluid mechanics, may fulfill most practical needs. The scenario
should also cope with the possible action of long-range forces, especially
of electrostatic origin. Fortunately, cellular conditions are such
that in bulk conditions and away from the compartment boundaries,
screening acts as a powerful localizer of interactions, that die off
at distances above few nanometers. 

In order to integrate the protein force-fields with the physico-chemical
features of solvation, the LB framework should also be enriched with
water-like features, inclusive of directional interactions, and hydrogen-bond
features, having deep implications on the macromolecular structures
\cite{papoian2004water}. Preliminary efforts in this direction have
been made in the past, but their thorough validation remains entirely
open \cite{LB-WATER}. 

Many applications of a more water-specific methodology naturally suggest
themselves: thermal stability of proteins, the onset of neurodegenerative
diseases due to peptidic aggregation, diffusion of proteins and trafficking
in cellular crowding, being just some examples in point. 

Besides initial foot-in-the-door applications, this plan requires
a massive amount of implementation and validation work, one that possibly
suggests the need for coordinated community efforts. 

The aggregation of misfolded soluble proteins into fibrils is the
precursor of several neurodegenerative diseases, such as the Alzheimer,
Parkinson, and Huntington ones. In particular, Alzheimer's disease
is marked by atrophy of cerebral cortex showing accumulation of amyloid
plaques and numerous neurofibrillary tangles made of filaments of
the phosphorylated tau proteins. The major constituents of plaques
are made of the amyloid $\beta$ peptides made of 40 and 42 amino
acids \cite{BUCKNER2005}.

The fibrillogenesis of amyloid $\beta$ peptides is a complex process
whereby fibrils extend up to hundred of nanometers, and the time scale
of full growth exceeds hours in vitro. The details of the emergence
of amyloid protofilaments are still debated but it has been observed
that the formation of ordered arrays of hydrogen bonds drives the
formation of $\beta$-sheets within aggregates that form early under
the effect of hydrophobic forces\cite{auer2008}. Understanding the
mechanisms of amyloid aggregation is key to the design of drugs able
to prevent fibril formation and toxicity in the brain and computer
simulation is an essential tool to explore the aggregation process.
First and foremost, describing the kinetics of amyloid formation via
conventional nucleation theory lacks information on the structure
and size of the primary nucleus.

Mimicking amyloid aggregation can not be done by implicit solvent
models, since the lack of solvent interactions does not include the
treatment of solvation thermodynamics and neglects altogether the
action of solvent-mediated correlations. Self-assembly initiates via
a hydrophobic collapse and the formation of molten oligomers, with
the common feature of fibrils being the inter-digitation of the side-chains,
the so-called steric zipper. Since fibril formation is under kinetic
and not thermodynamic control, this is a great showcase for the LBPD
strategy \cite{NASICA2015}. 

Large-scale aggregates would only form with the correct kinetics and
showing up the correct intermediate and metastable states by using
the highest level of physical fidelity. If a simplifying assumption
on the dynamics, such those provided by Langevin level, is used, spurious
intermediate states would eventually kick-in, as long as the final
aggregate is kinetically driven.

By including hydrodynamic interactions and by employing the OPEP force
field, the LBPD methodology has shown that the solvent mediated interactions
have a key role in regulating amyloid aggregation, see Fig. \ref{fig:amyloids}
\cite{LBMDbio2,LBMDbio3}. As a matter of fact, hydrodynamics enhances
peptidic mobility, thus facilitating mutual encounters and collapse
to the aggregated structure. This should be appreciated in view of
the non-trivial computational effort required to simulate the aggregation
process, which is not only driven by diffusion but also featuring
a slow-down due to the energetic barriers involved in the process.
For a dense number of peptides, $p_{PD}\simeq1$ and a typical box
edge of length $L=10^{2}$, the complexity is about two orders of
magnitude larger than a purely diffusive process, resulting in a total
of 
\[
\mathcal{C}\simeq10^{2}\mbox{ TeraFlops}
\]
With current hardware facilities, the aggregation process can be observed
during its full course. This is an important, possibly first real-life
application of the methodology, in a situation where macromolecular
realism and solvent-mediated interactions change drastically the conventional
picture of amyloid aggregation as compared to assuming negligible
hydrodynamic forces \cite{sterpone2014}. 

This becomes even more interesting when looking at the effect of shear
flow onto the kinetics of aggregation. In a Couette flow, fibril formation
can accelerate from one month down to a few hours \cite{DUNSTAN2009,BEKARD2011}. 

A possible mechanism for the effect of shear is the alignment of aggregates,
which in turn facilitates their assembly. Even changing the specific
nature of the shear flow can enhance the formation of protofibrils
and the growth of fibrils. Clearly, there is still a long way to go
towards a full characterization of the aggregation process, but at
this point, it is clear that the LBPD strategy is highly apt at coping
cope with this complex scenarios.

A side observation regards the computational efficiency of LBPD to
simulate pepditic solutions. Clearly, inter-particle interactions
are a major bottleneck, due the burden of computing a large number
of non-bonding and bonding forces, with a significant share of computing
a large number of non-bonding and bonding forces, with a significant
portion of computing time spent in searching interacting pairs and
bookkeeping them. In addition, as seen earlier, chemical realism requires
to account for rather stiff forces and therefore a consequent small
timestep imposed on the particle solver. If the system is dense in
particles, handling inter-particle forces is going to be the slowest
segment of the simulation, while the optimal setting is when particles
are in diluted or semi-diluted conditions. This is precisely the operating
conditions pertaining to the aggregation of amyloid $\beta$ peptides.

The effect of hydrodynamics on protein diffusion has been studied
for a solution made of $18,000$ Rat1 proteins in a bulk simulation
at 40\% volume concentration \cite{LBCROWDING,sterpone2014}. The
study showed that in such crowding conditions, protein diffusion proceeds
according to the experimental values measured by quasielastic neutron
scattering and pertaining to the $3.5$ ns \textendash{} $5$ ns temporal
range, exceeding the hydrodynamic timescale arising from the propagation
of vorticity over the protein linear size, and ultimately slowing
down the protein self-diffusion. A drop of the diffusion coefficient
at volume fraction between 10 and 30 \% marks the onset of caging
mechanisms, whereas at larger volume fractions the diffusivity dangerously
approaches a jamming transition, while protein motion is still in
action. Amyoid aggregation represents an examplar instance of hydrodynamic
forces impacting the formation of molecular aggregates. By studying
a system of unprecedented size, LBPD simulations were able to explore
a branched disordered fibril-like structure that had never been described
by computer simulations before \cite{LBMDbio3}. The results show
that hydrodynamics forces also steer the growth of the leading largest
cluster and impact the aggregation kinetics and the fluctuations of
the oligomer sizes, by favouring the fusion and exchange dynamics
of oligomers between aggregates. 

\subsection{Towards Computational Physiology and Medicine}

Physiological flows offer one of the most attractive applications
of the LB framework to real-life situations, with high potential social
impact in utilizing computer simulations to diagnose pathologies,
prognose a medical condition or even guiding clinical intervention
\cite{NOBLE1,NOBLE2,fenner2008europhysiome,patronis2018modelling}.

In the age of evidence-based medicine, the decision-making process
needs to be optimized by using evidence from well-designed and well-conducted
research. Although all medicine has some degree of empirical support,
the evidence-based approach requires that only the strongest data
coming from meta-analyses, systematic reviews, and randomized controlled
trials can be used to inform clinical recommendations. Incidentally,
this spawns major opportunities for the synergistic operation with
machine-learning techniques \cite{ML}, a delicate subject we shall,
return to at the end of this paper.

Physiological flows, for instance, are conditions where a biofluid
circulates in complex anatomical conduits and networks, examples ranging
from blood flow to lymphatic circulation, to airways, the urinary
system and so on. 

Blood flow has made the subject of intense research over the last
decades, the application of LB to blood flows experienced a major
burst of activity \cite{BUICK2003,DUPIN2003,SUN2003,SUN2006,DUPIN2006,SUI2007,BOYD2007,SUI2008,HUA2008,DUPIN2008,MACMECCAN2009,WU2010,XIONG2012,SHI2013,XU2013,KRUGER2013,KRUGER2014,HEMO1,REASOR2012},
with several applications to coronary, carotid and cerebral blood
flow. Of course, this is no surprise, since these are macroscale applications
for which the most conventional NS hydrodynamics appears perfectly
adequate. 

It is to be stressed that \emph{statistical averaging is of little
meaning in a physiological context, since each individual is a story
of her/his own}. Rather, the detailed access to the patient-specific
4D hemodynamic data across scales of motion, may offer a quantum leap
in the quality and accuracy of pre-emptive medicine \cite{ARTOLI2003,HIRABAYASHI2004,PELLICCIONI2007,CAIAZZO2009,HEMO4,HEMO5,GROEN2013,YUN2014,OMORI}.

\begin{figure}
\begin{centering}
\includegraphics[scale=1.0]{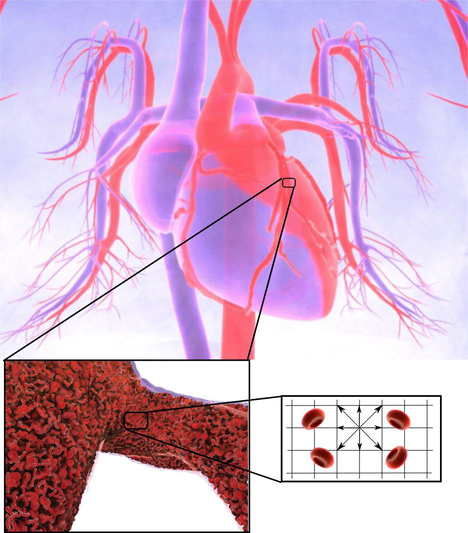} 
\par\end{centering}
\caption{Three-level representation of blood flow: upper panel, continuum macroscale
(cm); bottom left panel, the granular nature of red-blood cells starts
to be apparent (100 microns), and is fully revealed at the level of
the cell spacing (10 microns, bottom-right panel) \cite{MUPHY2}.\label{fig:FIGBLOOD}}
\end{figure}

This is why a fully 4D (three-space dimensions and time) real-time
numerical and visual access at the blood dynamic flow patterns from
microns all the way up to the full-scale geometry, can disclose unprecedented
opportunities for personalized and precision medicine (see Fig.\ref{fig:FIGBLOOD}). 

Again, a few numbers may help conveying a concrete sense of what is
meant here. Based on the estimate \ref{eq:CPU}, the computational
complexity of a 4D real-time LBPD simulation covering four spatial
decades ($L=10^{4}$) and just a single circulation time ($T\propto L$),
is of the order of 
\[
\mathcal{C}\sim10^{4}\times10^{16}Flops=10^{2}\mbox{ExaFlops}
\]
This is at least a factor ten short of the real timespan needed to
collect significant diagnostics, so let us take $1000$ Exaflops (1
Zettaflop) instead. On a current-time Petaflop computer, this makes
$10^{6}$ seconds wall-clock time, about two weeks, which is not exactly
what one would label as real-time. A prospective Exaflop computer,
though, would complete the job in some 20 minutes, thus bringing the
real-time task within direct clinical fruition. 

LB simulations of physiological conduits offer an exciting opportunity
due to its most practical asset: simplicity in handling complex geometries
and in automated mesh-generation. To the best of our knowledge, such
simplicity remains unparalleled as compared to grid methods for the
numerical solution of NS equations. Another strength of LB comes from
its local structure in space and time. As typical in the study of
unsteady flows, the flow patterns are particularly rich and accurate
once unsteadiness is taken into account, as for the study of gas flows.
Hemodynamics also sets a case, due to the large excursions of local
Reynolds number (going from virtually zero in microcapillaries to
nearly $10,000$ in the aorta), whereby unsteadiness promotes both
local and global patterns. Most importantly, blood is pumped into
the vessels by a pulsatile injection rate, a situation that requires
time-explicit boundary conditions. In biomechanics, the ratio of transient
inertial to viscous forces, the Womersely number, can range from $10^{-3}$
in capillaries to $10$ in the aorta, calling for the direct time-explicit
solution in the most general case. The applications are widespread,
but it is worth recalling the study of coronary and carotid arteries,
as critical vessels that subtend to the oxygenation of the heart muscle
or the brain. Any anomaly in the blood flow would cause major risks
of heart attack or stroke to the patient. In such networks of arteries,
geometric complexity is highly non-trivial; particularly challenging
is the handling of conditions where narrowings and plaques give rise
eccentric passages and tiny spaces, at times as large as a handful
of red blood cells. 

The way pressure is distributed in arteries has thus great physiological
relevance, since the supply and demand of oxygen in organs, primarily
the heart muscle, is regulated by the distribution of pressure in
vessels. It is well known that narrowings and blockages lead to strong
pressure losses, with consequent starvation of the tissues depending
on blood circulation. 

Here LB has a technical, yet very important, point. When looking at
the possible build-up of atherosclerotic plaques, the LB framework
offers facilitated access to the wall shear stress tensor. This can
be computed \emph{locally} based on the non-equilibrium populations,
via: 

\[
\overleftrightarrow{S}(\vec{r},t)=\frac{\nu\omega}{c^{2}}\sum_{i}\vec{c}_{i}\vec{c}_{i}(f_{i}-f_{i}^{eq})
\]
thereby dispensing with expensive and inaccurate finite-difference
operations.

As mentioned several times in this review, a truly major asset of
LB is parallel performance, with a large potential for the biofluidic
context. As in other applications, the LB mesh is plain cartesian
and, for most biofluidic applications, single-resolution is sufficient.
Due to the excellent scalability of the basic method, one may suppose
that parallel performance would persist also for intricate vessel
networks. Unfortunately, in this case the computational domain covers
a sparse region of space, typically filling space by less than $10$\%.
This is definitely a challenge, even for a highly scalable LB solver
and countermeasures have to be taken. Possible solutions to this issue
are discussed in the Appendix. 

LB is today widely used to study biofluidics and blood flow. The fast
turnaround makes it an excellent candidate for rapid screening in
clinical practice, with the prospective possibility to even predict
the outcome of clinical intervention. Stenting, angioplasty, flow
diverting, aneurism wiring, to mention but a few, are all possible
applications of the LBPD methodology. Clearly, this is not the end
of the story, as moving parts, as concerning valve placement or compliant
vessel deformations can be taken into account by using the various
schemes available today, starting from the IBM to handle the moving
walls \cite{FANG2002,HOEKSTRA2003,DESCOVICH2013,DEROSIS2014}.

Microcirculation is interesting in itself. Due to the red blood cells
constituents of blood that reaches up to $45$\% in volume in humans,
two main concurrent effects take place: the Farhaeus-Lindqvist effect,
whereby the average concentration of red blood cells decreases as
the diameter of the containing vessel is decreased; the second effect
is the viscosity change with the diameter of the vessel it travels
through. The two effects arise because red blood cells move preferentially
over the center of the vessel, leaving the plasma at the wall, thus
lowering the near-wall dissipation effects. The Farhaeus-Lindqvist
effect becomes visible in the range between $10$ and $300$ micrometers. 

In recent years, the study of microcirculation witnessed an upsurge
of interest in recent years by using mesoscale particle methods \cite{noguchi2005shape,HEMO1,HEMO2,mcwhirter2009flow,CLAUSEN2010,KARNIA,TOSCHI-RBC,HEMO3,PONTRELLI,MATYKA}.
In capillaries of lateral size of $100$ microns and below, the motion
of red blood cells reveals highly non-trivial signatures of granularity
and deformability. For capillaries with a diameter of a few microns,
erythrocytes undergo large deformations in order to squeeze into the
vessel and the globules are able to crawl into the micrometer-sized
space. On the other hand, when looking at larger-scale circulation,
in the $100-500$ microns range, it is generally sufficient to consider
blood cells as rigid bodies. The grand-challenge here is to \emph{reach
up to physiological scales (1-10 cm) while retaining essential micro-features,
the finite-size of red blood cells (8 micron) in the first place}.
This is of major interest for many reasons; the granular nature of
blood may have a significant impact on the recirculation patterns
in the proximity of natural geometrical irregularities, such as bifurcations,
stenoses, aneurysms, or man-made ones, like stents and other medical
devices. 

Micro-to-macro hemodynamics is particularly rich, showing a peculiar
distribution of oxygen-carrying cells at every bifurcation depending
on the local Reynolds number, and with far reaching consequences on
physiology. Erythrocytes exhibit both a tumbling motion and the tank-threading
effect, whereby the cell membrane can slide under a shear force \cite{KELLER1982},
two conditions that have deep impact on blood rheology. Plasma skimming
near the arterial walls has important consequences on the local and
global circulation in order to optimize the oxygen supply chain keeping,
at the same time, the flow speed high in the capillaries. Further
consequences relate to the most common of cardiovascular diseases
since atherosclerosis depends on the uptake of lipidic material by
the arterial wall and, ultimately on the near-wall shear stress. The
discussion is still open and its outcome is extremely important to
understand the causes of myocardial infarction for diagnostic or pre-emptive
medicine.

Cellular hemodynamics is an open branch of research. A direct extension
of the LBPD method can account for suspended bodies, for the explicit
presence of cells suspended in plasma. This is a typical case where
the hydrodynamic medium hosts particles with finite-size, anisotropic
shape, in fact oblate ellipsoids, that represent red blood cells to
first approximation. Diverse community software packages \cite{MAZZEO2008,MUPHY7,CLAUSEN2010,HEUVELINE2007},
such as OPENLB, MUPHY or HEMELB, are making their way to offer several
cell-type capabilities and performance so much so that studying flows
composed by red blood cells and leukocytes becomes extremely attractive.
Previous work showed the complex hydrodynamic interplay between cells
of different shapes, with the margination of leukocytes and their
rolling along the vessel wall \cite{MUNN2008}.

All LB assets show great value for deployment in hemodynamics. 

To appreciate the computational complexity of a typical blood flow
system, let us consider a coronary arterial tree and estimate the
number of mesh voxel needed to fill the sparse volume occupied by
the vessels to $10^{7}$. The problem is typically advective and the
number of timesteps to cover a pulsatile cycle is $\sim10^{6}$. When
red blood cells are also simulated, the computational effort on the
PD side can easily exceed the LB component by one or two orders of
magnitude. Consequently, ranging from LB to LBPD simulations of a
coronary tree, requires

\[
\mathcal{C}\sim10^{2}-10^{4}\mbox{PetaFlops}
\]

As discussed previously, with Exaflops computers at hand, this may
be brought close to the realm of real-time simulation. 

Besides the most appealing features, the downsides of LB owe to be
mentioned too. An important point concerns low-Mach flow circulation.
The virtually incompressible nature of blood flow (with Ma$<10^{-3}$)
can impact negatively on accuracy of simulations, in particular when
the pressure distribution is the goal of the investigation. The compressibility
error of LB scales as $Ma^{2}$, and making it negligible typically
requires lowering the simulated Mach number, $Ma^{*}$, or equivalently,
the numerical typical velocity, since $u/c=Ma^{*}$. However, the
price to pay is a corresponding reduction of the LB timestep, down
to values can be easily as small as microseconds, potentially undermining
its practical efficiency for simulating biofluidics. The answer to
the conundrum comes from practice and from the observation that, even
at such low timestep, computing efficiency is sufficient to resolve
highly complex fluid flows. It is worth noting that the optimal working
conditions can be problem-specific, and a study versus resolution
and simulated Mach is recommended.

Besides blood flow, LB applied to biofluidics is also witnessing medical
applications to airways, ranging from nasal to pulmonary flows \cite{LINTERMANN2011,KRAUSE2010,EITEL2010,FREITAS2008,MIKI2012,LI2011}.
This is yet another story, where compressibility effects become much
less important and the focus shifts towards very intricate geometries
in the presence of collapsible walls. Understanding how air flows
in these regions, the consequence of rather small but crucial imperfections,
the transport of small molecules or odorants, are all applications
that draw great profits from the possibility of solving fluid mechanics
in a multiphysics scenario in order to study peculiar conditions.
Again, LB and LBPD are particularly well suited to computationally
embrace to the presence of multiple agents in solution. 

\section{SIMULATIONS AT THE PHYSICS/CHEMISTRY/BIOLOGY INTERFACE: DREAMING
AHEAD}

\textcolor{black}{The growth in computing power that is expected to
be sustained for the next decades will spawn tremendous opportunities
to gain new insights into a series of fundamental problems dealing
with complex states of flowing matter in general, and in particular
those relevant to biology and medicine. }

\textcolor{black}{For the sake of reference, let us speculate what
can and hopefully will be possible once LBPD operates at Exascale
performance. Since this Review is mostly intended to discuss applications
at the physics/chemistry/biology interface, in the sequel we shall
focus our attention mostly on that paramount scenario.}

\textcolor{black}{To that purpose, we wish to reiterate that the LBPD
description of biological systems is based on a mesoscopic picture,
whereby molecular details are incorporated within suitable coarse-grained
terms in the effective kinetic equation for the solvent and coupled
to stochastic particle models of the biological molecules. The art,
as usual, is to incorporate the least amount of molecular details
required to describe the essential physical phenomena under scrutiny.}

\subsection{Future Challenges: Towards Extreme LBPD computing }

The inclination of LB for parallel processing comes from the fact
that within the LB formalism \emph{information always travels along
straight straight streamlines, regardless of the physical complexity
of the emergent flow structure. }This marks a major divide versus
macroscopic formulations, whereby information moves along material
lines defined by the space-time dependent flow field $\vec{u}(\vec{r};t)$.
The point is key to achieve outstanding parallel efficiency also in
the case of geometries having real-life complexity, like those that
often occur in biological problems where shape and function are tightly
correlated, as described in previous Sections. Albeit highly technical,
this point is absolutely crucial to achieve the levels of extreme
scalability (extreme LB = XLB), which are mandatory to access the
``disruptive'' applications described in the previous Sections.
In a way, we may compare these technological advances to the development
of a new experimental technique/device aimed at exploring new states
of matter: Tera electronvolts in high-energy physics, Teramolecules
in computer explorations at the physics-chemistry-biology interface.
Needless to say, scaling up to millions and soon billions of computing
cores, surely does not comes for free, especially when the geometry
is not regular; it must be won via very advanced and dedicated programming
strategies (see Appendix and the recent prospective paper \cite{AMATI}).

\begin{figure}
\begin{centering}
\includegraphics[scale=0.6]{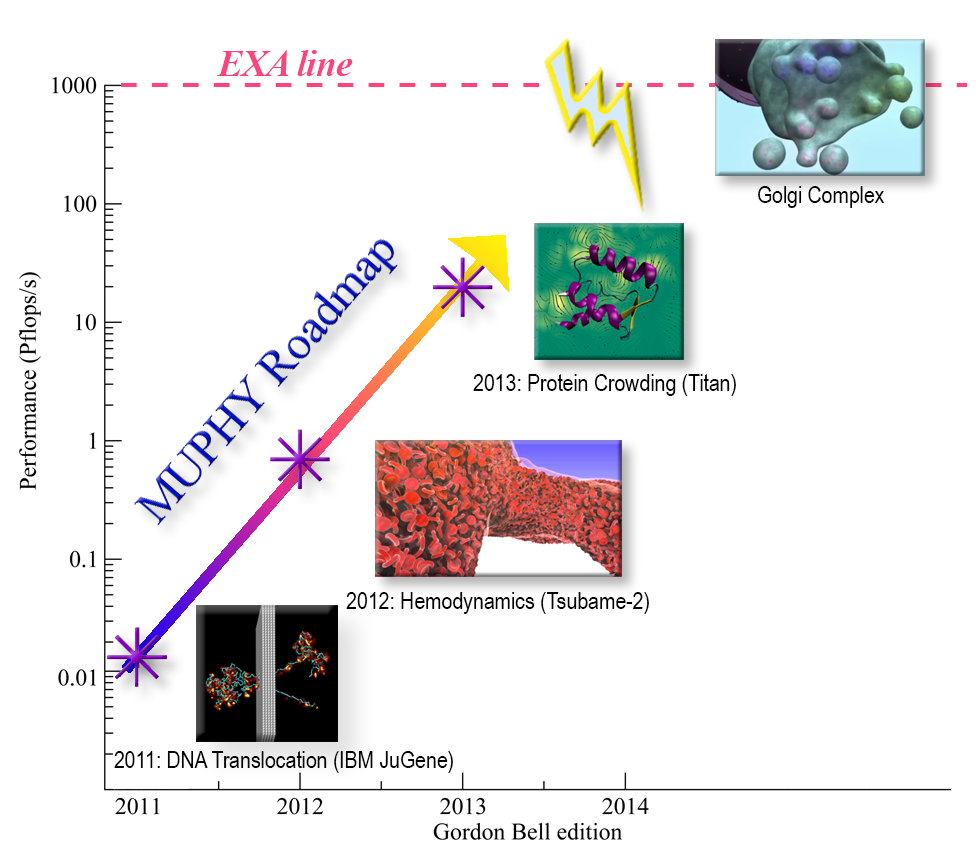} 
\par\end{centering}
\caption{The roadmap of the MUPHY code, based on the Gordon Bell performance
obtained over the years (see \cite{MUPHY2} and references therein).\label{fig:Roadmap}}
\end{figure}

However, once such efforts are put in place, the results are extremely
rewarding, on virtually any parallel platform. In the last decade
a few multiscale codes, coupling LB for the fluid motion with various
forms of (stochastic) particle methods for the dynamics of floating
bodies within the flow have been developed \cite{MUPHY7,WALBERLA,ROSSINELLI}.
Among these, MUPHY is a fully scalable LBPD code which has been successfully
used for the simulation of a variety of biofluidic applications \cite{MUPHY1,MUPHY4,MUPHY7}.
These include biopolymer translocation, multiscale hemodynamics and,
lately, proteins. MUPHY has attained fairly impressive parallel performance,
with an escalating progression from $11$ TeraFlops/s for biopolymer
translocation on the IBM Jugene (2011), to $0.7$ PetaFlops/s for
multiscale hemodynamics on Tsubame (2012), up to a world-record (to
the best of our knowledge) of $20$ PetaFlops/s (sustained performance)
for protein crowding on Titan (2013). Although such figures refer
to leading-edge supercomputing experiments rather than fully-fledged
biofluidic applications, they point to a tremendous potential for
prospective applications at the physics-chemistry-biology interface,
suggesting to proceed along the roadmap illustrated in Fig.\ref{fig:Roadmap}.
However, extracting such potential on upcoming Exascale architectures
faces with a number of challenging issues. Since a successful handling
of these issues is key to open up new and otherwise unconceivable
LBPD applications at the PCB interface, in the Appendix we shall provide
a relatively detailed coverage of the main technical topics, including
\textcolor{black}{\emph{i)}} how to best exploit modern CPU and accelerators
architectures for LBPD simulations; \textcolor{black}{\emph{ii)}}
how to exploit at their best modern CPU and accelerators architectures
for the simulation of LB methods; \textcolor{black}{\emph{iii}}\textcolor{black}{)}
how to overlap computation and communication, so as to hide the overheads
of the latter, \textcolor{black}{\emph{iv)}} how to secure a balanced
load among millions of computing cores in realistic geometries such
as the cell or blood vessels. 

Having laid down the two-legged engine for fluids and molecules, a
stringent requirement is to design accurate kernels to exchange forces
among fluids and particles of different kinds: hydrodynamic, solvation,
dispersion, electrostatic and so on. This program requires accessing
algorithmic methods from a vast array of options and organizing various
types of biological agents, representing proteins, nucleic acids,
lipids, sugars, and so on. At the core, one must devise new algorithms
that allow reproducing environments of increasing complexity, either
from the bottom-up molecular route, or from the top-down macro/mesoscopic
perspective. Most importantly, the plan requires not only the representation
of multiple agents (fluids, macro or small molecules, continuum-based
solutes) but also the architected orchestration of their evolution,
as they move across multiple scales. This presents an outstanding
challenge in both conceptual and technical respects.

At the time of this writing, a fully-fledged multiscale numerical
strategy is still lacking, thus making the subject of intense research.
Since multiple particle-based and field-based agents need to coexist
within a unified computational framework, the search for a robust
and universal strategy to achieve their seamless coexistence is still
pretty open.

A first step forward is to devise numerical methods whereby only the
fluid component is allowed to cross scales: generally speaking, these
fall in the class of multi-grid fluid-dynamics, where adaptive, multi-resolution
or body-fitted meshes are used to resolve fluid dynamic patterns via
static or automatic mesh generation. Although pretty laborious, these
methods have been developed for several decades and are presently
in the position of providing dramatic savings in both memory and number-crunching
requirements.

A second, much more challenging aspect, regards the evolution of multiple
Eulerian or Lagrangian \emph{transmuting} agents, meaning by this
that such agents are capable of changing their identity and representation
on-the-fly, depending on the local physics in point.

By its very mesoscopic nature, LB is conceptually at a vantage point
for multiscale/level coupling, both upwards, towards continuum representations
and downwards, towards atomistic models. For all its conceptual appeal,
such high-level program requires a concrete computational substantiation
in order to turn theoretical ideas into actual computational tools.

A most useful LB asset in this respect, is the cartesian mesh. Owing
to its simplicity, mesh construction, management and extensions offer
a particularly flexible approach to handle complex biological settings.
As two or more neighboring scales are juxtaposed in space/time, mesh
refinement corresponds to increasing the number of mesh points in
proximity of geometrical variations, where mass or momentum gradients
are most likely to attain peak values. 

To date, a number of multigrid LB schemes have been proposed, some
of which are customarily used in large-scale academic codes \cite{LAGRAVA2012,MUPHY7}
or in industrial applications \cite{CHEN2003}. The common approach
is to consider two juxtaposed meshes, a factor two ratio in spacing.
A possible strategy is to exchange information between two neighboring
meshes that overlap in some finite region of space and to exchange
populations by spatial and temporal interpolation schemes. This hand-shaking
procedure must make sure that not only the flow fields, mass and momentum
density, are continuous across the surface, but also their fluxes.
This leads to a specific map between the discrete populations in the
coarse and fine grids. Another approach is to consider the fluxes
across neighboring meshes by using a finite-volume description of
the LB method, and to exchange these components \cite{CHEN2006}.
Yet another, very recent but promising approach, is to merge standard
LB with unstructured finite-volume LB formulations in correspondence
with sharp features of the flow \cite{HYBLBE}.

Given the underlying mesh connectivity (e.g., by taking the D3Q19
mesh as a reference), the mesh nodes that are connected by a complete
($18$) set of mesh neighbors of the same mesh spacing are called
``saturated'' nodes, as opposed to the ``unsaturated'' ones, which
have only an incomplete ($<18$) set of mesh neighbors connecting
nodes from different meshes.

For the sake of clarity, let us define $\mathcal{S}^{1}$, $\mathcal{S}^{2}$,
$\mathcal{S}^{3}$, ..., as a sequence of \emph{scopes} that describe
the ownership of fluids and particles to a single scale and help to
coherently organize the computation across scales. 

By construction, each scope accommodates a single cartesian mesh $\mathcal{M}^{1}$,
$\mathcal{M}^{2}$, $\mathcal{M}^{3}$, .... 
\[
\mathcal{M}^{k}\in\mathcal{S}^{k}
\]

A given scope $\mathcal{S}^{k}$ contains a set of fluids $\mathcal{F}^{k,\alpha}$
and a set of Lagrangian particles $\mathcal{P}^{k,\alpha}$ 
\[
\sum_{\alpha}\{\mathcal{F}^{k,\alpha}\}\cup\{\mathcal{P}^{k,\alpha}\}\in\mathcal{S}^{k}
\]

By associating a fluid to a scope, by design, we posit that LB fluids
are forced to belong to a given cartesian mesh and cannot cross the
interface between different meshes. On the other hand, since particles
are, by definition, grid-free they are allowed to move anywhere in
space. As a result, their ownership to a given scope is purely a matter
of organization. 

However, particles can change their physical identity or even disappear
(change to class ``nihil'') as they cross scales. In fact, a particle
belonging to a physical domain can swap to another particle type (possibly
with a different representation) as it crosses scale boundaries, without
necessarily leading to a discontinuous trajectory. 

As a further development, one can transform Lagrangian agents into
Eulerian ones and vice versa, a procedure which involves not only
interpolation but also projection from particles to LB distributions
via hydrodynamic fields and direct sampling of particles coordinates
directly from the LB distribution.

This ``transmutational'' functionalities raise genuinely new issues,
both conceptual (compliance with the basic principles of statistical
physics) and computational, i.e the design of and efficient manipulation
of the corresponding data-structures. It is conceivable that they
might spawn the needed for dedicated ``transmutational'' multiscale/level
software as well.

Coming back to our conceptual scheme, meshes are best organized in
a hierarchical order, such that the $k$-th level mesh, $\mathcal{M}^{k}$,
has spacing $\Delta x^{k}$ ordered in an ascending sequence $\Delta x^{k}=2\Delta x^{k-1}=\cdots=2^{k-1}\Delta x^{1}$,
thus the mesh spacing doubles at each level increase.

Juxtaposed meshes feature a finite overlapping or compenetration between
neighboring meshes \cite{LAGRAVA2012}, and the extent of the compenetration
region is taken as $L(\mathcal{M}^{k}\cap\mathcal{M}^{k-1})=\Delta^{k}$.

In its simplest implementation, all boundary conditions operate on
the finest mesh, covering regions where the strongest gradients need
to be resolved, typically in the proximity of solid walls. In multigrid
LB, let us consider the propagation in time of the LB fluid across
a set of scales labelled $0$, ..., $K$ and illustrate the scheme
based on interpolation/extrapolation of fluid populations rather then
on the fluxes.

To that purpose, spatial averaging of a generic field $A$ living
on mesh $\mathcal{M}$ is given by 
\begin{equation}
\bar{A}(\vec{r}^{k},t)=\frac{1}{q}\sum_{p\in{\cal N}}A(\vec{r}^{k}+c_{p}^{k}\Delta^{min},t)\label{eq:filter}
\end{equation}
where ${\cal N}$ is the set of all the nodes of $M$ that are neighbors
of the nodes of $\mathcal{M}^{\prime}$ at $\vec{r}_{k}$, $q=Card({\cal N})$
and $\Delta^{min}$ is the smallest spacing between $\mathcal{M}$
and $\mathcal{M}^{\prime}$. When $\mathcal{M}$ is finer than $\mathcal{M}^{\prime}$,
averaging/interpolation corresponds to a low-pass filter. 

Another useful tool is time-averaging between the latest updated time
of $\mathcal{M}$ ($t_{i}^{\mathcal{M}}$ ) and the previous one ($t_{i-1}^{\mathcal{M}}$),
which is needed to estimate the field $A$ at time $t=t_{i}^{\mathcal{M}}=(t_{i-1}^{\mathcal{M}}+t_{i}^{\mathcal{M}})/2$: 

\[
\bar{A}(\vec{r}^{k},t)=\frac{1}{2}\left[\bar{A}(\vec{r}^{k},t_{i}^{\mathcal{M}})+\bar{A}(\vec{r}^{k},t_{i-1}^{\mathcal{M}})\right]
\]
In the above, the quantities $\bar{A}$ within the square brackets
can be evaluated from Eq. (\ref{eq:filter}).

Multigrid LB proceeds by a local timestepping frequency and by rescaling
the relaxation time according to the local mesh in action. Each scope
agent is evolved according to a specific propagator. For either LB
fluids or particles, the evolution timestep reflects the ownership
to a given scale, with its own specificity. For LB, mesh spacing and
timestep are strictly related, so that in a single streaming operation,
information hops between neighboring nodes of the same mesh. For particles,
the timestep is an independent quantity and, as long as the energies
involved are well sampled and the simulation is stable, one can tweak
the timestep to achieve optimal performance.

When information travels across neighboring meshes, the lattice speed
should not change between meshes and fluid velocity and pressure should
be continuous across the interface. Given a reference lengthscale
$\ell$ and timescale $\tau$, all other scales are such that $\ell^{k}=\Delta x^{k}\ell$
and $\tau^{k}=\Delta t^{k}\tau$. For the time marching, it is possible
to adopt the convective scaling, which leaves the flow speed invariant
across scales, i.e. $u^{k}=u^{k-1}=...=u^{1}$.

Then, the timestep $\Delta t^{k}$ is related to $\Delta x^{k}$ via
\[
\frac{\Delta x^{k}}{\Delta t^{k}}=\frac{\Delta x^{k-1}}{\Delta t^{k-1}}=\cdots=\frac{\Delta x^{1}}{\Delta t^{1}}
\]
By matching the Reynolds number across all scales to a single reference
value $\textrm{Re}=u\ell/\nu$, where $u$ is a typical flow speed,
$\frac{u^{k}\ell^{k}}{\nu^{k}}=\cdots=\frac{u^{1}\ell^{1}}{\nu^{1}}\equiv Re$,
we obtain the scaling relation for the fluid viscosity, namely: 
\[
\nu^{k}=2\nu^{k-1}=\cdots=2^{k-1}\nu
\]
Given the reference viscosity $\nu$, a scale-specific frequency $\omega^{k}$
is derived. In fact, in order to have a single global kinematic viscosity,
each species has the same viscosity $\nu$, related to the relaxation
frequency via eq. (\ref{eq:viscosity}).

It is important to appreciate that the local nature of the LB methodology
makes the construction of hydrodynamic moments and gradients, such
as the deviatoric stress, for multi-resolution as simple as in the
single-resolution case. This marks a neat plus as compared to NS schemes,
where gradients need to be computed through stencils or other numerical
templates. 

However, when considering embedded particles, a number of adaptations
have to be catered for. The single-mesh described in Section \ref{sec:GeneralizedLB}
requires interpolation/extrapolation schemes that are well-posed in
terms of smoothness and robustness. The scheme then adapts to the
situation where particles cross mesh (or scale) boundaries.

\begin{figure}
\begin{centering}
\includegraphics[scale=0.6]{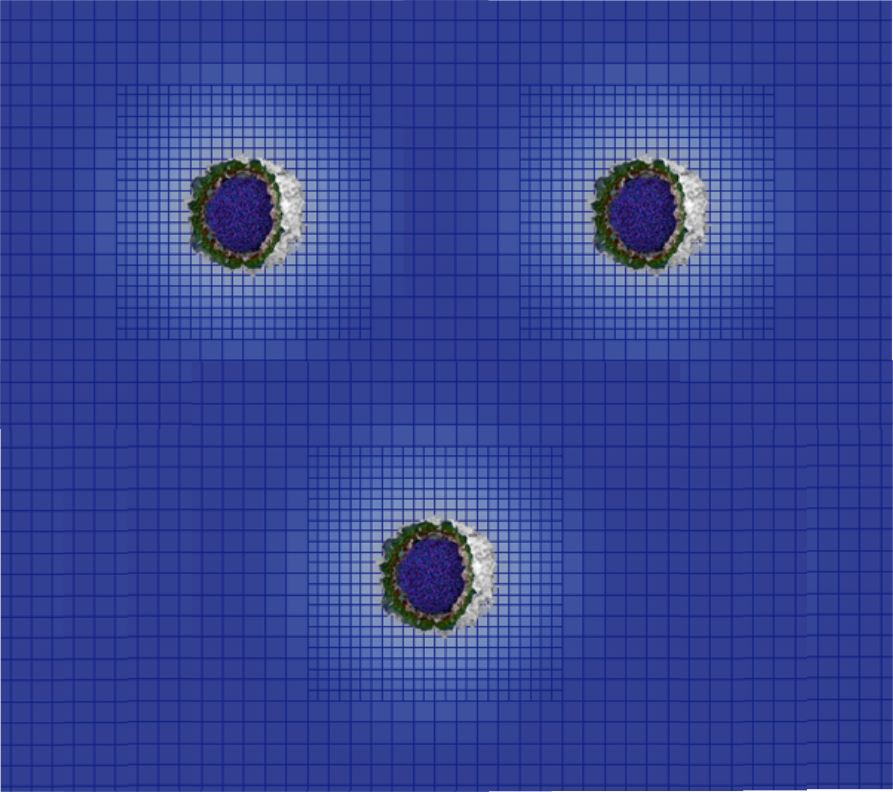} 
\par\end{centering}
\caption{The flow structure around a macromolecule resolved by a multi-resolution
mesh being finer in proximity and inside the macromolecule\label{fig:MGproteins}.}
\end{figure}

A particularly desirable extension to simulate biosystems concerns
the case when the mesh locally readapts to accommodate the presence
of moving macromolecules. This operation implies refining the LB mesh
on-the-fly by following the instantaneous position of the molecular
agents. As the macromolecules diffuse and encounter each other, assemble
or undergo structural transitions, a finer level of meshes must follow
the macromolecule like a shadow (see Fig.\ref{fig:MGproteins}).

Owing to the simplicity in constructing the cartesian mesh, the price
to pay for automatic mesh refinement is, in principle, negligible
although the associated software management can become nonetheless
pretty demanding.

The availability of multi-resolution techniques makes it increasingly
attractive to design fully-fledged, LBPD-based multiscale/level environments. 

The underlying multi-resolution mesh is the basic support to move
information between juxtaposed regions, whereas Eulerian and Lagrangian
agents cross the boundaries and unveil their essential nature locally
in space and time. Note that in this context, we refer to Eulerian
and Lagrangian, as to two distinct \emph{levels} of description, each
of which does embrace multiple scales. Here, we follow Noble's definition
\cite{NOBLE2}, according to which \emph{scale} pertains to the physical
extent over which information is distributed in space and time, whereas
\emph{level} touches at the mathematical/computational organization
of such information. In this terminology, \char`\"{}transmutation\char`\"{}
is basically a change of level. 

As a result, at each scale, the following four quadrants are involved
(see Fig. \ref{fig:EulerLagrange}): 
\begin{itemize}
\item \emph{EE (Eulerian-Eulerian)}: LB populations that coarsen and refine 
\item \emph{LE (Lagrangian-Eulerian)}: Particles that transform into LB
populations 
\item \emph{EL (Eulerian-Lagrangian)}: LB populations that transform into
particles 
\item \emph{LL (Lagrangian-Lagrangian)}: Particles that coarsen or refine 
\end{itemize}
\begin{center}
\begin{figure}
\begin{centering}
\includegraphics[scale=0.6]{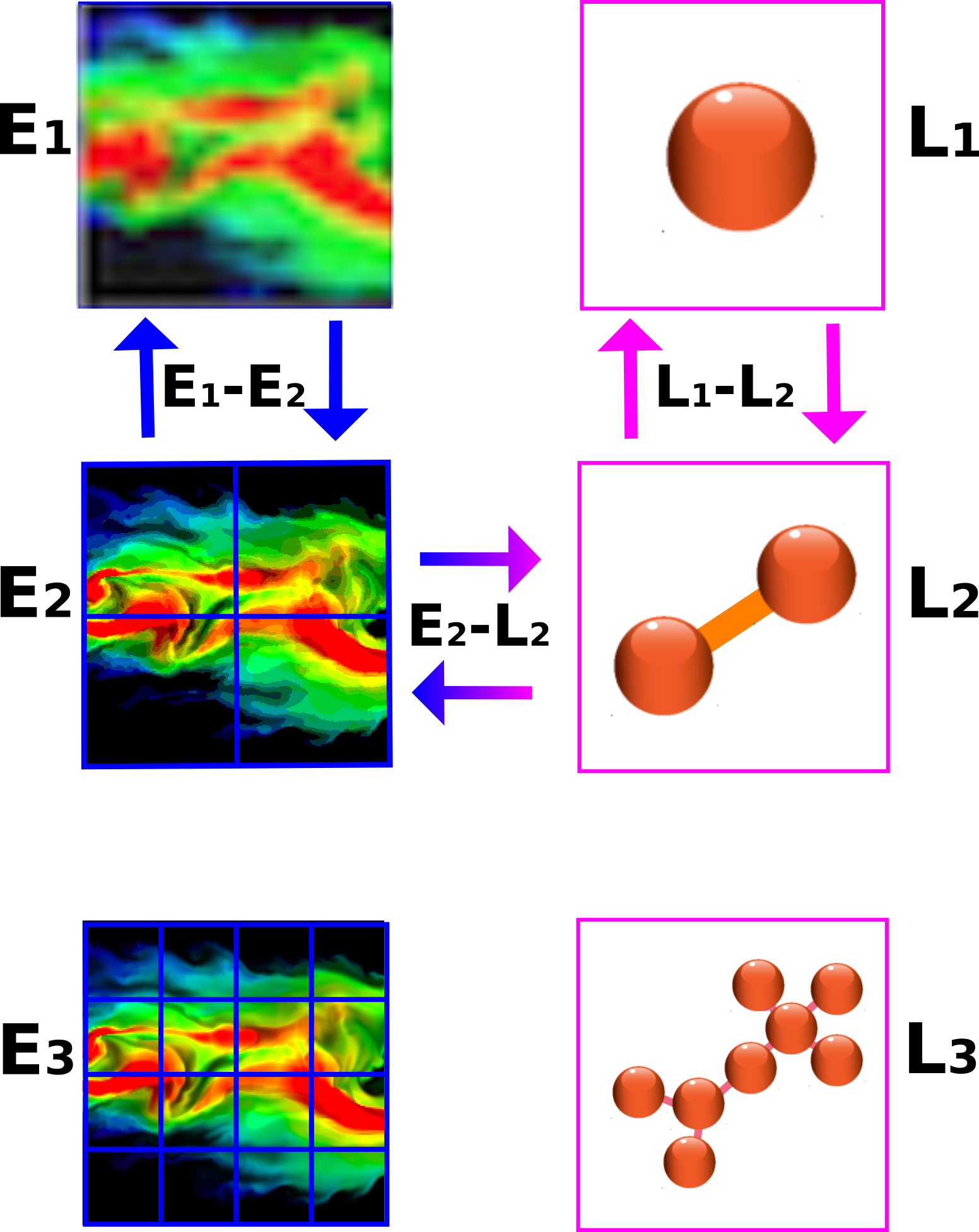} 
\par\end{centering}
\caption{Diagram for the multiscale approach where Eulerian (E) and Lagrangian
(L) agents can move across scales and across representations. The
boxes of different size on the left column indicate the coarse/fine
grain of the Eulerian representation, the ball-and-stick representations
on the right column indicate particle-based molecules at different
resolution. The arrows indicate the exchange of agents between E-E
scales, L-L scales or E-L representations. \label{fig:EulerLagrange}}
\end{figure}
\par\end{center}

As discussed above, the EE transform can be accomplished within LB-adapted
multigrid techniques. The LL transform also involves homogeneous quantities
and adjusting their interactions to the required scale appears to
be doable within currently existing methods \cite{PRAPROTNIK2008}.
A typical example of LL transform is between a detailed representation
of a polypeptide to an elastic network and vice versa. Coarsening
implies loss of information by projection and it is easier to handle
than the reverse case, namely reconstruction of the information lost
in the projection step, a inherently non-zero-error task.

The cross-level terms LE and EL are specific of LBPD, hence less consolidated.

We note that the mechanism that regulates the crossing is inherently
different for LE versus EL transitions, since the single-particle
identity is inevitably lost in LE. For the EL case, we need to guess
the identity of individual particles, and even more unwieldy, construct
the topology of extended molecules.

A fully-fledged multiscale/level LBPD framework still awaits for conceptual
validation from the point of view of fundamental statistical physics
as well as in terms of computational implementations.

Hence, making an educated guess on its future is far from being an
easy task. Yet, we can make a conservative prediction about the concurrent
spatio-temporal scales that can be accessed once Exascale computing
becomes available.

The LBPD code MUPHY was reported to deliver $20$ PetaFlops/s for
extreme simulations involving $20$ billions fluid sites and nearly
$70$ millions particles. By naive linear extrapolation, an Exaflop
computer would permit to scale these figures up by another factor
$50$, leading to $1$ trillion fluid sites and nearly $10$ billion
particles. This corresponds to four decades in space, the best one
can expect \emph{without} any of the multigrid/level sophistications
described above. Once such multi scale/level strategies are in place,
another two orders of magnitude can reasonably be envisaged (in a
plain multigrid scenario this corresponds to about seven levels of
refinement, which is well within the current capabilities of multigrid
LB solvers). The resulting LBPD tool would then be able to handle
\emph{six} decades in space and time.

With these mind-boggling advances in mind, several exciting scenarios
may open up to the LBPD strategy, which we next describe in some detail.

\subsection{Protein crowding}

Protein trafficking in cellular compartments is deeply connected to
the structural and diffusional behavior of proteins under crowding
conditions. Realizing that the interior of cells is characterized
by high concentrations of macromolecules and, depending on the organelle
or sub-cellular location, 5 to 40\% of the total volume is occupied
by macromolecules, those proteins conduct their activity in extremely
crowded environments. Therefore, organelle functioning depends on
the structural and dynamical response of proteins to dense packing
conditions, a critical yet elusive element of cellular organization.
While advances are made through in vitro studies, crowding effects
can force molecules in cells to behave in radically different ways
as compared to test-tube assays \cite{ellis2001,zhou2008}.

The way that cells utilize intracellular spatial features to optimize
their signaling characteristics is still not clearly understood. The
physical distance between the cell-surface receptor and the gene expression
machinery, fast reactions, and slow protein diffusion coefficients
are some of the properties that contribute to the intricacy.

Extracellular signals captured by receptor proteins on the cell surface
are transduced inward to control target proteins or gene expression.
Two interconnected underpinnings of this cellular response are molecular
mobility (e.g., diffusion and active transport) and the signal transduction
reactions. Despite their importance, limited attention has been paid
to the former biophysical properties of the cellular environment,
which can contribute to overall signaling characteristics of the system
by introducing non-linear signal delays. The Stokes\textendash Einstein
relation implies slow diffusion rate of protein macromolecules, which
are key players in the signaling. The significance of diffusion in
reaction\textendash diffusion systems becomes key whenever reactions
are comparatively faster than diffusion rates.

\begin{figure}
\begin{centering}
\includegraphics[scale=1.00]{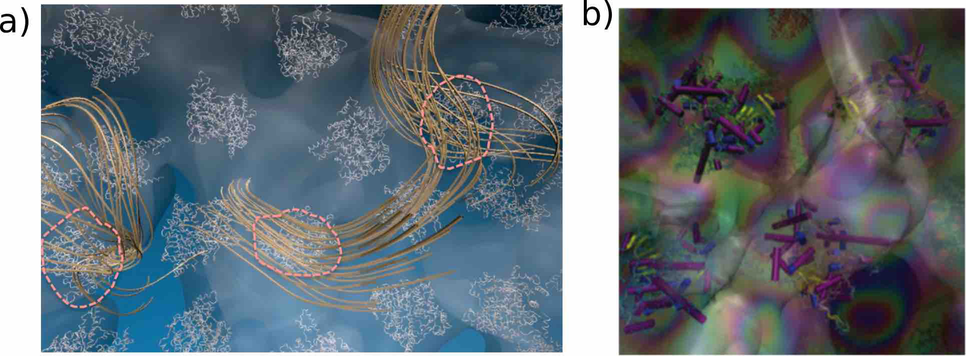} 
\par\end{centering}
\caption{Snapshot of globular proteins in solution, shown in different visual
representations. a) Proteins are shown with a wire-like bundles and
the surrouding hydrodynamic signal is represented via an isosurface
of its velocity field. Streamlines generated by selected proteins
in the crowded macromolecular environment showing how the hydrodynamic
disturbance propagates in the aqueous solution. b) Proteins represented
via conventional ribbon-sticks style and the constant velocity isosurface
that illustrates the complexity of the flow structure.\label{fig:crowding-streams}}
\end{figure}

Extremely high protein density in the intracellular space, commonly
called molecular crowding, can magnify the spatial effect. In a typical
cell, the total macromolecular density is $50-400$ mg/ml, far higher
than typical in vitro conditions (1\textendash 10 mg/ml). If a solution
contains $30\%$ by volume of identical globular molecules, less than
$1\%$ of the remaining space is available to an additional molecule
of the same size due to the excluded volume effect caused by steric
repulsion, resulting in a mutual impenetrability of macromolecular
solutes. In such environment, slow (5\textendash 20 times lower than
saline solutions) translational diffusion is still observed, which
exhibits the footprint of anomalous diffusion \cite{TAKAHASHI2005,sterpone2014,timr2019multi}.
Anomalous diffusion is defined as sub-linear scaling of mean-squared
displacement of the molecule over time, and is used as a measure for
cytoplasmic crowding. Molecular crowding, as exemplified by two snapshots
of simulation shown in Fig.\ref{fig:crowding-streams}, can also alter
protein activities and break down classical reaction kinetics.

Ideally, to reproduce crowding effects, with the ensuing anomalous
diffusion and protein encounters, simulation methods should be able
to track coarse-grained shapes and sizes of molecules and their positions
in three-dimensional space. Proteins stay localized in certain compartments
as a result of cell compartmentalization and non-covalent weak interactions
such as ionic, van der Waals, hydrogen bonds and hydrophobic-polar
interactions. Weak interactions, which can also influence the reaction
and diffusion rates of molecules should be considered during simulation.
As a first estimate, the problem being dominated by diffusive motion,
one should consider the simulation of a cubic system of side $L=10^{3}$
lattice units. For such systems, the LB and PD contribute almost equally
to the computational effort. With a unit diffusion coefficient in
lattice units, the computational complexity scaling as $L^{3}T=L^{5}$
is particularly challenging, resulting in an estimated complexity
\[
\mathcal{C}\sim10^{1}-10^{2}\mbox{ ExaFlops}
\]
Clearly, high scalability is mandatory to support the handling of
such large intracellular systems. 

Biological interactions are quite promiscuous and their occurrence
cannot be predicted based on conformational considerations. It is
essential to account for chemical specificity in order to capture
the local details of molecular recognition and signaling. LBPD stands
as a very appealing approach to address the plethora of questions
related to signaling in the protein crowd. 

Having reduced drastically the number of degrees of freedom for the
water solvent down to the stringent economics of LB, the PD engine
should be already sufficient to track enzymatic activity and related
pathways. This is sufficient for a subset of possible conditions in
the cell, but still not satisfactory for the general case. As described
earlier, an essential feature is to resolve the local hydrodynamic,
thermodynamic and chemical patterns in the surroundings of active
sites, with their peptidic components, hydrogen exchanging groups
or metal atoms. Local mesh refinement would entail dramatic benefits
in terms of reducing the cost of the far-field regions, and so would
the corresponding adaptive timestepping. Leveraging the minimalistic
LB mesh is precisely the option at stake, to reduce the number of
degrees of freedom (by a factor $2^{3}$ gain) together with the timestep
(by a factor $2$ gain) at every resolution change. The same advantage
applies for every partial differential equation that can be solved
on-the-fly via a concurrent LB scheme, say the ADR equation, for electrostatics,
or for other modeling purposes. If the same LB mesh is used for solving
concurrent PDE via conventional finite-difference approaches, one
should refer to the consolidated multigrid algorithms. 

\begin{figure}
\begin{centering}
\includegraphics[scale=0.8]{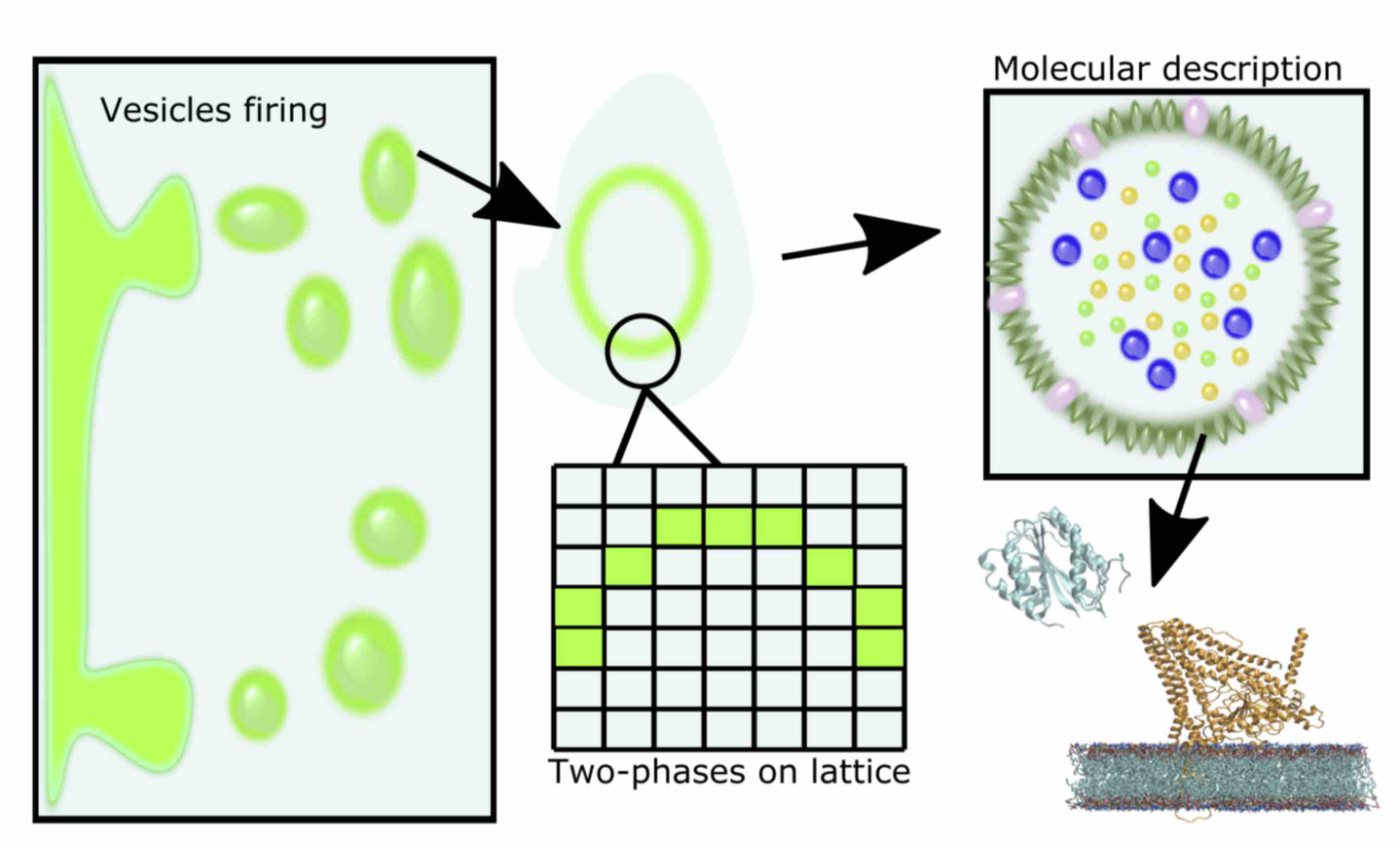} 
\par\end{centering}
\caption{Multiscale representation of vesicles transporting proteins. Vesicle
firing from a membrane where the vesicles are modeled as immiscible
fluid phase separating from the aqueous host, the central panel zooms
the vesicle bilayer with liquid water in the interior, the right panel
depicts a molecular representation of the vesicle containers with
an heterogeneous protein suspension in the interior and membrane proteins
embedded in the bilayer.\label{fig:vesicles}}
\end{figure}

\subsection{Direct simulation of full-scale cell compartments: Golgi and neuronal
firing}

Cellular organization relies upon the operation of several organelles,
from the nucleus to ribosomes, mitochondria, lysisomes, Golgi, etc.
Each organelle lives its own life separated by confining lipid bilayers.
These specialized subunits are in action and continuously interplay
one another. The scenario is extraordinarily rich and rich and no
numerical framework can hope to capture the whole picture. Yet, a
few characterizing elements make the LB framework particularly appealing.
With the Exascale capability at our doorsteps, LB can be taken to
the next level: from the study of basic biofluidic processes to the
direct simulation of full-scale cellular compartments, like protein
cargoes, vesicles and possibly even full-scale organelles. The endeavor
commands the integration of the LBPD paradigm within a broad scope
software infrastructures, including mechanical models of biological
structures, ranging from all-atom molecules to elastic networks for
membranes and so forth~\cite{LBMDbio1}.

In order to design the most appropriate strategy, it is instructive
to take a look to a few processes involving macromolecules that are
transported within cellular environment. During the cell life cycle,
proteins are continuously translated and delivered to specific cellular
locations by traversing different membrane structures. However, most
molecules, including proteins, are too large to pass directly through
membranes. Instead, large molecules are loaded into small membrane-wrapped
containers called vesicles. Vesicles are constantly forming - especially
at the plasma membrane, the Endoplasmic Reticulum, and the Golgi apparatus,
or simply the Golgi. Once formed by exocytosis, vesicles deliver their
contents to destinations within or outside the cell. On the different,
yet related, scenario of neurotransmission, signaling neurotransmitters
are released by a neuron and bind to and activate the receptors of
another neuron. Neurotransmission is the essential process of communication
between two neurons with synaptic transmission and firing relying
on the release of neurotransmitters. The latter are stored in vesicles
in the axon terminal. Different mechanisms involve partial opening
and then re-closing of vesicles, together with the fusion of vesicles
with the membrane.

The emerging picture is that the entire arsenal of the LBPD approach,
entailing single or multiphase variants, in presence or absence of
suspended macromolecules, provides a powerful, flexible and self-contained
framework to solve multiple levels of cellular biology. A closer look
at how proteins and vesicles interplay unveils the type of challenge
the numerical approach has to tackle. The way proteins are transferred
inside in cellular compartments is fascinating. Vesicles form when
the membrane bulges out and pinches off. Then it travels to its destination,
where it merges with another membrane to release its cargo. In this
way, proteins and other large molecular cargoes are transported without
ever having to cross a membrane. Even more, the mechanism underlying
the formation of vesicles is budding and is deeply assisted by proteins.
When vesicles bud, they wear ``coats\textquotedblright{} and when
coat proteins assemble at the member, they force the lipid bilayer
to begin to bend. As they gather at the membrane, coat proteins may
also select the specific cargo that is packaged into the forming vesicle.
As more coat proteins are aged, they shape the surrounding membrane
into a sphere. Finally, once a coated vesicle pinches off, the coat
falls off, and the cargo-filled vesicle is ready to travel to its
final destination. The plain fact shows that the proteins-fluid system
is inherently two-way such that chemical specificity as much as the
vesicular chemical composition has to be correctly included to convey
the required molecular realism.

The feasibility of reproducing full-size organelles can be analyzed
by considering the typical size of the Golgi. This organelle has a
lateral size of $2.5$ microns and is composed of $10^{12}$ atoms,
that is, the ``TeraSize\textquotedblright . The typical timespan
for morphogenesis spans between one microsecond and a minute. In operating
conditions, one femtosecond timestep is required for the accurate,
bottom-up description of molecular trajectories. Given the burgeoning
progress of computing power promises for the forthcoming years to
reach the ExaFlops/s capabilities. Sustained by the development of
modeling techniques and specialized algorithms, the available power
will soon allow facing astonishing assemblies of macromolecules above
the microsecond timescale and, at the next level, targeting entire
cellular compartments. Under such conditions, one can expect that,
on an Exascale computer, one could simulate the system evolution in
full at a cost of 
\[
\mathcal{C}\sim10^{4}-10^{6}\mbox{ ExaFlops}
\]
corresponding to about $1-10$ days of wall-clock time.

Clearly, the path to simulate full-scale compartments can be a multi-stage
process and a first step is to represent lipidic membranes at coarse
LB level, namely by neglecting the molecular character and chemical
specificity of the confining components, as sketched in Fig.\ref{fig:vesicles}.
Being a peculiar visco-elastic fluid, the membrane can be handled
by the multiphase LB, e.g. via Shan-Chen or free-energy approaches.
The description has to cope with the fact that a membrane has finite
thickness, therefore any multiphase approach should reproduce not
only the interfacial properties of lipidic bilayer, but also its phase
diagram, including the multiple shapes of micelles, vesicles, etc,
with their tendency to deform. The formation of lipidic chains and
their preferential orientation under external forces induce changes
in shapes from circular to elongated, as observed in experiments.
Complex fluid\textendash fluid interfaces featuring mesoscale structures
with adsorbed particles are key elements of biological components.
For such a schematic approach, the expectation is demanding but one
should keep in mind that confiners and carriers can provide a minimal,
yet satisfactory, level of realism.

\subsection{Biochemical reactivity and signaling pathways}

As we have discussed earlier on, LBPD can reproduce the interplay
between flow and macromolecules in a relatively large, cell-like,
environment. Dealing with biochemical reactivity is a different, huge
sector that calls for a full deployment of the simulation capabilities\cite{TAKAHASHI2005}.
Application of LB to reproduce advection-diffusion-reaction phenomena
is an option. However, biochemical reactions are not always classifiable
according to simple statistical rules, as for the Michaelis-Menten
or Logistic laws \cite{VOIT2013}. 

In biology details matter since a small minority of active sites immersed
in a jungle of organic groups can make the whole difference. In addition,
metabolic and synthetic reactions can occur in bulk conditions within
a single phase (homogeneous reactions) or in a multiphase environment,
typically at the interface between regions (heterogeneous reactions),
in no-flow or flow conditions, as for example the enzyme reactions
on the surface of the blood vessels.

Although there are many possible reactive events in action, they all
fall into two broad categories: oxidation and reduction, the motion
of functional groups within or between molecules, the addition and
removal of water and the bond-breaking reactions. Most reactions are
catalyzed by proteins, RNA or DNA. A different class of reactions
involves electrostatic, electrodynamic and hydrophobic interactions,
where electrons are not shared and covalent bonds are not modified.
Enzymes have the property to increase the rate of the reactions and
are specific to the reactant molecules, also known as the substrates,
interacting with high affinity. Finally, the activity of enzymes is
regulated in a number of ways, controlling the rate and amount of
products formed. Examples of regulation include cofactors binding
to the enzymes, or the presence of reaction products that inhibit
the reaction \cite{fersht2017structure}.

The range of biological activity does not result from many different
types of reactions, but rather from a few simple reactions, occurring
under many different situations. Thus, for example, water can be added
to a carbon-carbon double bond as a step in the breakdown of many
different compounds, including sugars, lipids, and amino acids. One
could model this compact set of reactions by utilizing the LBPD apparatus.
Within a classical description this is feasible indeed, except that
bond breaking and formation requires using electronic structure methodologies,
notably by using one of most successful theories to date, the (quantum)
Density Functional Theory. The good news is that, given the small
number of active sites present in macromolecules, this stands as a
perfect candidate to embed a numerical solver for electrons within
a classical solver for particles, and ultimately within the LB overarching
framework. The quantum-mechanical molecular mechanics (or QM-MM) approach
is an active avenue of research today and proceeds along similar ideas
as the LBPD scheme. Clearly, the numerical details of LB and QM methods
are different in nature, but basic similarities can be found. Indeed,
like LBPD, the QM-MM procedure is based on a combination of Lagrangian
(classical molecular dynamics) and Eulerian (quantum electronic structure)
components. One may push the similarities even further by utilizing
LB to solve the electronic structure too. Work in this direction \cite{MENDOZA2014}
has shown promises for the future.

To the point that, in the far-future, one may fancy of an unprecedented
four-level QM-MM-PD-LB unified multiscale structure, ranging from
electronic scales all the way up to the cellular ones, the overlap
link being MM-PD. Incidentally, three-level structures of this sort
have now been in place for two decades, although their routinely operation
seems to remain somewhat unwieldy \cite{KAX98}. 

In prospect, evolutions of the LBPD methodology to cope with reacting
systems should benefit from one of its major assets. Reactions control
species interconversion and involve the breaking and forming of covalent
bonds, often catalyzed by enzymes. Dealing with particles or molecules
that change identity on-the-fly, their molecular connectivity and
can undergo multiple reaction channels, is a tall order for the Lagrangian
treatment. Instead, electron transfer and chemical breaking and forming
can take advantage of the probabilistic nature of the kinetic representation.
In this respect, one can envisage interconversion between the Eulerian
and Lagrangian representations for molecules that are likely to react,
and choose between the optimal treatment, possibly on-the-fly. An
estimate of the computational resources must embrace the full scale
organelles, for which a region of $L=10^{4}$ sites in size must be
catered for. With diffusive-reactive scaling, $T\simeq L^{2}$ and
some $k\simeq10^{5}$ flops/site/step, we obtain
\[
\mathcal{C}\sim10^{7}\mbox{ ExaFlops}
\]
corresponding to about three months wall-clock time on a Exaflop computer. 

Assuming sufficient accuracy to model the chemical reactions is available,
the next objective is to examine the structure and dynamics of cellular
functioning at system level, rather than the characteristics of isolated
regions. Collective properties of networks, such as their efficiency
and robustness, emerge as a central characteristic in system biology.
Needless to say, the understanding of these properties may deeply
impact medicine, bolstering the emerging notion of network-medicine,
as opposed to the \char`\"{}magic-bullet\char`\"{} approach of genomics.
Advances in this direction can take full advantage of LBPD once sufficient
chemical specificity is incorporated via coarse-grained force fields.
Caution must be exercised to find the optimal balance between resolving
local details and the collective properties of the metabolic network.
As it stands, this is an ideal scenario to develop innovative multiscale/level
methodologies for the biological context. 

\subsection{Hemostasis}

Another grand challenge is the study of hemostasis, a crucial healing
mechanism in which molecular specificity and chemical reactivity contribute
on an equal footing. Hemostasis is the immediate response of the body
to stop bleeding from within a damaged blood vessel. It is the first
stage of wound healing and involves a blood change from the liquid
to the gel state (coagulation). When endothelial injury occurs, the
endothelial cells stop secreting coagulation and aggregation inhibitors
and secrete instead the globular glycoprotein von Willebrand factor
(vWF), which uncoils and initiates the maintenance of hemostasis after
injury. The overall process is governed by Virchow's triad which
comprises composition of blood, wall surface reactivity and material
flow. A multiscale simulation approach stands out as a prime route
to gain a better understanding of such complex and life-essential
mechanism. 

\begin{figure}
\begin{centering}
\includegraphics[scale=1.2]{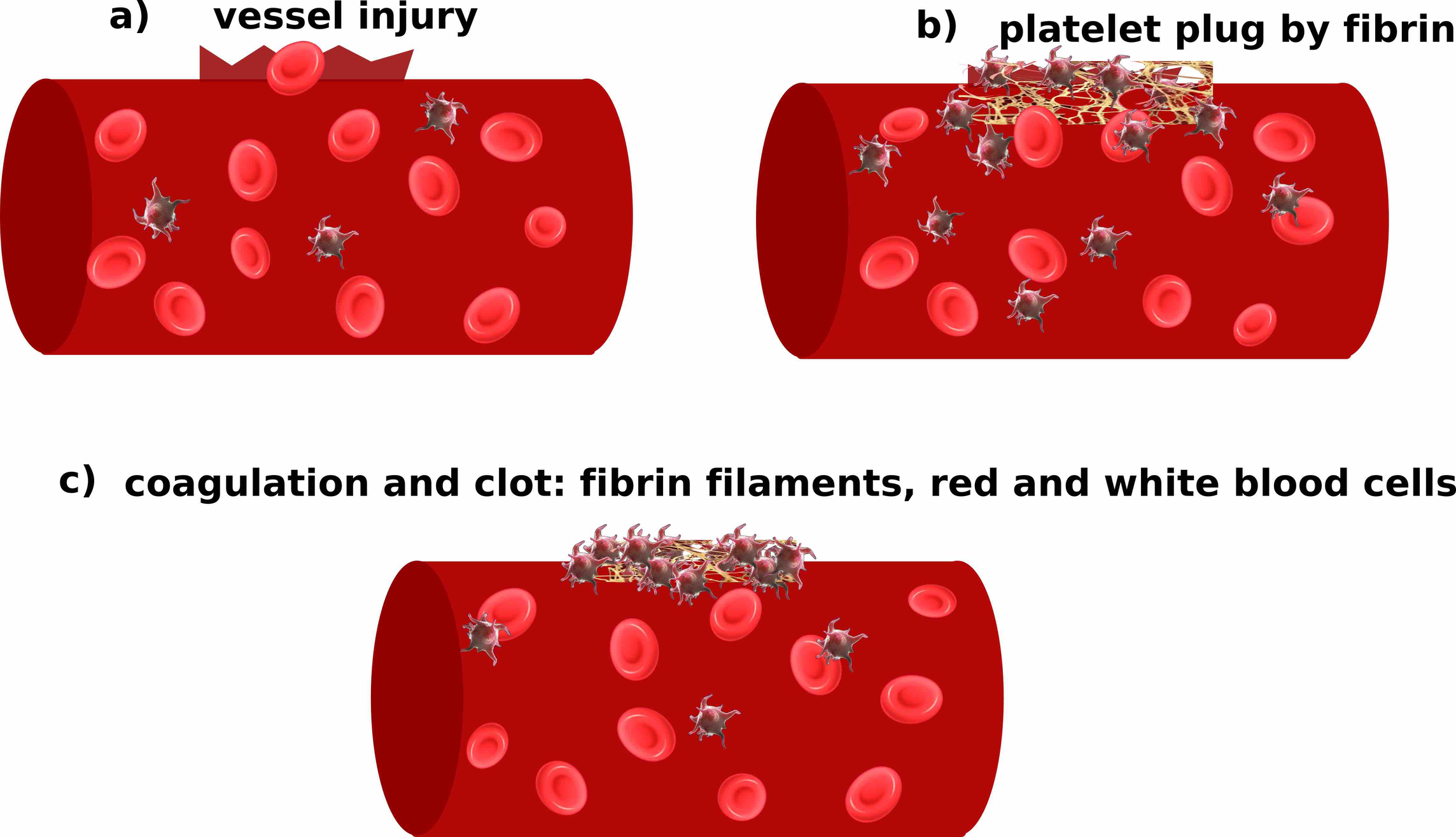} 
\par\end{centering}
\caption{The major phases of hemostasis following the vessel injury: a) a platelet
plug is formed to rapidly stop the initial bleeding; b) a mesh of
fibrin is made at the injury point to make the platelet plug stronger;
c) finally the clot is formed by the coagulation of fibrin, red and
white blood cells.\label{fig:hemostasis}}
\end{figure}

As we delve into the details, it becomes clear that hemostasis shows
articulated features, as sketched in Fig.\ref{fig:hemostasis}, highlighting
the multistep and multiscale elements in action. It proceeds along
three subsequent steps that seal the injury until tissues are repaired:
vasoconstriction, temporary blockage by a platelet plug and formation
of a blood clot. Vascular spasm is the first response to constrict
the blood vessels and reduce the blood loss. Second, platelets stick
together to form a temporary seal via the so-called primary hemostasis:
platelets adhere to damaged endothelium to form the plug and then
degranulate as activated by the vWF. Finally, coagulation takes place
and reinforces the platelet plug with fibrin threads that act as the
``molecular glue''. In this picture, platelets are key to the process:
the plug forms almost directly after the vessel has ruptured and within
seconds and disrupted platelets adhere to the sub-endothelium surface.
Within a minute the first fibrin strands begin to intersperse among
the wound and just in a few minutes later, the plug is completely
formed by fibrin. During the process, a dozen clotting proteins are
activated in a sequence known as the coagulation cascade to hold it
in place, the so-called secondary hemostasis. Here, red and white
blood cells are trapped in the mesh which causes the primary hemostasis
plug to become harder: the resultant plug is called as thrombus or
clot.

The role of blood flow is even more important than it might appear
at first sight. Since the 70s, experiments have demonstrated that
shear stress strongly affects the activation of platelets and their
adhesion to the injured tissue. However, there is still confusion
about the mechanism that regulates platelet arrangement and about
the reasons why sites of disturbed flow appear to be more prone to
platelet deposition. 

When the level of shear exceeds a certain threshold, platelet aggregate
even in the absence of any chemical agonist and without any modification
of vWf. The reason for this behavior remains unclear. Shear-enhanced
exposure or the alteration in the structure of receptors on the platelets
membrane increases the frequency of collisions where particle migration
is known to increase with shear up to three orders of magnitude above
the Brownian value, due to the enhanced collision frequency, primarily
between platelets and erythrocytes. Furthermore, erythrocytes are
known to enhance shear-induced platelet adherence not only mechanically,
but also chemically, through the release of the platelet agonist adenosine
diphosphate.

The uncoiling of bound globular vWf at elevated shear is responsible
for the increased platelet deposition and upon platelet activation,
the release reaction feedback amplifies the hemostatic system. At
high shear, adhesion requires the synergistic action of several receptors
on the platelets membrane and on other ligands. Upon activation, platelets
change their shape from discoid to spherical, release their granule
content and increase their stickiness among themselves, as a function
of the amount of local shear stress. Chemistry plays a major role,
as a multitude of reactions lead to the production of thrombin, which
is a key enzyme in the hemostatic process as well as a strong platelet
agonist. In the course of hemostasis, fibrin fibers which surround
the platelet aggregate and stabilize it against the shear forces in
the flowing blood.

Shear stress and saturation-dependent changes in surface reactivity
influence thrombus growth and the adhesion and aggregation of platelets
to reactive materials. Simplified models based on fluid dynamic and
species conservation equations can match in-vitro experimental data,
as regarding the initial phase of platelet deposition, when thrombus
growth can be neglected, while accounting for shear stress and changes
in surface reactivity. However, taking thrombus growth into account
results in a free-boundary problem, with fully coupled fluid dynamic
and species conservation equations, again a scenario that calls for
the LBPD apparatus, as witnessed by a few pioneering studies \cite{OUARED2005,CHOPARD2006,HARRISON2007,HARRISON2008,OUARED2008,TAMAGAWA2009}.
In these studies, activation of platelets in the bulk flow and subsequent
agonist production were not included as a part of the model. On the
other hand, flow could be explained by using a shear-independent adhesion
rate. By using such a model, predictions on the flow structure were
improved in some parts of the flow chamber, such as in stagnation
points, whereas notable discrepancies remained in some other parts.

In order to be effective, the LBPD approach should include the combined
effects of shear stress, changes in surface reactivity, and aggregate
growth in modeling both hemostasis and thrombosis. The practical outcome
would be paramount, i.e \emph{assist the minimization of thrombus
formation in vascular prostheses without the use of strong anticoagulants}.
This is all important in bioengineering, to design materials with
improved surface properties, also for shape optimization techniques
in flow conditions vs platelet deposition. Due to its fundamental
character, the influence of shear stress demands the inclusion of
the full coupling of flow and thrombus growth in models intended to
capture the long-term behavior of platelet deposition, with the potential
to enlighten the basic mechanisms taking place in further kinds of
adhesion processes. Given the extended multiscale nature of hemostasis,
Exascale computers may still be insufficient to solve the problem.
For an estimate of computational requirements, we take a cubic box
of side $L=10^{4}$, in order to acknowledge the need for micron-scale
resolution of regions of the order of the centimeters. With a diffusive-reactive
scaling $T\simeq L^{2}$ and a computational density of the order
of $k\simeq10^{6}$ flops, leading to: 

\[
\mathcal{C}\sim10^{8}\mbox{ ExaFlops}
\]
Such figure could be reduced by making systematic assumptions on the
process and analyzing the different phases of primary and secondary
hemostasis at different stages.

Notwithstanding the mentioned current limitations, it is of paramount
importance to consider the simulation of hemostasis for medical purposes.
Hemostasis is life-essential because it can go wrong in atherosclerotically
narrowed vessels. Ruptured plaques and elevated shear rates may induce
the formation of platelet-rich thrombi that may eventually become
life-threatening by occluding the vascular lumen. In fact, a major
number of deaths is due to thrombotic events provoked by disorders
of the hemostatic system. Severe consequences are triggered if the
thrombus detaches from the vessel wall and travels through the circulatory
system. If the clot reaches the brain, heart or lungs, it can lead
to stroke, heart attack or pulmonary embolism, respectively. Here
again, the full potential of LBPD for medical purposes cannot be understated,
and a full-scale deployment of the method in complex arterial networks
should be considered. Again, the large spread of scales, the presence
of multiple agents and the need for specialized solvers for chemical
reactivity, calls for similar, if not more sophisticated, high-performance
techniques that push the computational limits to their extreme.

\section{PCB Modeling versus Big Data Science}

All along this paper, we have advocated the mesoscale physics-inspired
modelling of complex phenomena at the interface between physics, chemistry
and biology, as a promising avenue towards the ultimate goal of benefitting
medical science and clinical practice. Before concluding, it is worth
to mention prospective connections of LBPD with the current burgeoning
trend towards the use of big data and machine learning techniques
in science. Leaving aside the most aggressive instances of big data
\cite{anderson2008end}, which can be readily commented away, \cite{coveney2016big,succi2018big,hosni2017forecasting},
it is undeniable that data science, and notably Physics-Aware Machine
Learning (PAML), bears major potential to enhance the LBPD scenario.
By PAML we refer to the machine-learning scenario whereby neural networks
are designed in such a way as to incorporate physical constraints
directly into their architecture, see \cite{karpatne2017physics,Karnia2018}
and references therein.

We note in fact that, due to its inherent mesoscale nature, LBPD will
necessarily be exposed to increasing parametrizations, as it proceeds
towards enhanced biological fidelity. For instance, PAML techniques
could prove of great value in automating the search of effective potentials
providing an optimal match to the desired biological properties, such
as the mechanical response of red-blood cells (size, shape, stiffness
...), so as to improve the description of their interaction with tissue
cells. More ambitiously, PAML could even help automating the choice
of the relevant degrees of freedom which characterise the mesoscale
formulation of the problem, thus helping to strike an optimal balance
between computational efficiency and biological fidelity. 

\section{SUMMARY AND PERSPECTIVE}

The Lattice Boltzmann method has undergone major progress over the
last decade, moving from an alternative technique for solving Navier-Stokes
hydrodynamics, to a versatile computational strategy to simulate complex
states of matter across many scales of motion, including micro and
nanoflows of relevance to biological processes. This quantum leap
has been fuelled by major advances of the LB ``technology'' alone
and by its successful coupling to particle methods, i.e. the Lattice
Boltzmann-Particle Dynamics paradigm illustrated in this Review.

The Lattice Boltzmann-Particle Dynamics paradigm has given access
to a new level of complexity in the description of phenomena occurring
at the physics-chemistry-biology interface. In this Review, we have
focussed our attention on the possibility of reaching up to scales
of direct relevance to clinical applications, thus portraying the
grand-dream of a mesoscale physics-based approach to precision-medicine.
This grand-dream has been illustrated through a series of actual examples
which, albeit not quite there yet, support the expectation that, once
Exascale computing is with us, the dream will come true. The task
is neither simple nor straightforward, but its scientific and societal
impact cannot be overstated.

\section{Acknowledgements}

The research leading to these results has received funding from the
European Research Council under the Horizon 2020 Programme Grant Agreement
n. 739964 (``COPMAT''). One of the authors (SS) wishes to acknowledge
enriching discussions with Profs. A. Cavalli, B. Chopard, P.V. Coveney,
E. Kaxiras, A. Hoekstra, M. Levitt, D. Noble, G. von Heijne and P.
Wolynes. One of the authors (SM) wishes to acknowledge fruitful exchanges
with U.M.B. Marconi, P. Derreumaux, H. Chen, E. Kaxiras, and F. Sterpone.

\section*{APPENDIX: HIGH-PERFORMANCE LBPD COMPUTING}

The LBPD paradigm described in the present Review connects two basic
computational pillars, a lattice-bound treatment of the fluid-field
with an off-lattice handling of the discrete particle dynamics. Given
the markedly distinct nature of the associated data structures, the
optimal merge of these two components must necessarily be realized
through a careful trade-off between the two. In this Appendix we provide
an overview of the main technical issues which concur to achieve such
compromise.

\subsection{High Performance Simulations of Particle Dynamics}

The computational requirements of PD simulations have always limited
their applicability to short time intervals, but advances in parallel
algorithms and special-purpose hardware (GPU, FPGA, ASIC, \emph{etc}.)
have recently extended the scope of such simulations to much longer
timescales. The state-of-the-art platform for high performance and
parallel execution of PD simulations is the Anton 2 system developed
by DE Shaw Research \cite{ANTON2}. Anton 2 performs the entire PD
computation within custom ASICs that are tightly interconnected by
a specialized high-performance network. A key component of the Anton-2
design is a set of new mechanisms devoted to efficient fine-grained
operations. The resulting architecture exploits at its best the parallelism
of PD simulations, which fundamentally consists of a large number
of fine-grained computations involving individual particles or small
groups of them. By providing direct hardware support for fine-grained
communication and synchronization, Anton 2 allows these computations
to be distributed across an increased number of functional units while
maintaining high utilization of the underlying hardware resources.
Fine-grained operation is exposed to software via distributed shared
memory and an event-driven programming model, with hardware support
for scheduling and dispatching small computational tasks. Anton 2
breaks the microsecond-per-day barrier on million-atom systems, allowing
larger biomolecules such as ribosomes to be simulated for much longer
timescales.

\subsection{Achieving High Performance for Lattice Boltzmann methods\label{SNperf}}

In general the performance of the LB on most platforms is memory-bandwidth
limited, that is, the rate by which the set of LB populations can
be read off and written to the memory is the main bottleneck. This
issue and its consequences can be understood looking at, for instance,
the widely used D3Q19 model in which there are 19 populations that
need to be read and write twice from the memory. Data are moved from/to
memory once for the collision phase and once for the streaming phase.
If the size of the memory word used to store each population is $WS$,
then $4\times19\times WS$ bytes are moved for each point of the lattice.
The number of floating point operations depends on the collision operator
only (there are no floating point operations during the streaming
phase) but it is safe to assume that it does not exceed 300 so, using
single-precision floating point format ($WS=4$), the arithmetic intensity\footnote{The arithmetic intensity of a numerical procedure is defined as the
ratio between the number of floating point operations and the number
of bytes moved from/to the memory.} of the LB update procedure is $~\frac{300}{4\times19\times4}\simeq1$.
On (virtually) any modern platform the number of bytes that can be
moved from/to the memory in a unit of time (e.g., in a nanosecond)
is smaller that the number of (floating) operations that the computing
cores can execute during the same unit of time (assuming that the
operands are available in the registers) so it is the memory that
limits the throughput also in the ideal situation in which the access
to the memory achieves its peak performance. The performance of LB
codes is typically given in terms of \textit{Millions of FLuid nodes
Updates Per Second (MFLUPS)}. On a platform with a memory-bandwidth
$B_{max}$, the peak performance in MFLUPS is: 
\[
P_{mflups}=\frac{B_{max}}{4\times19\times WS\times10^{6}}
\]
However, to alleviate the problem of the memory bandwidth several
variants of the LB method have been proposed. One of the most widely
used is the so-called \emph{fused} implementation, in which the \emph{collision}
and the \emph{streaming} phases of the LB update are combined in a
single procedure that either reads the populations from the source
locations and collides them (\emph{pre-stream}) or collides them and
stores the result directly to the target locations (\emph{post-stream}).
The advantage is that the populations are read from and written to
the memory just once for each time step.

\begin{figure}
\begin{centering}
\includegraphics[scale=0.5]{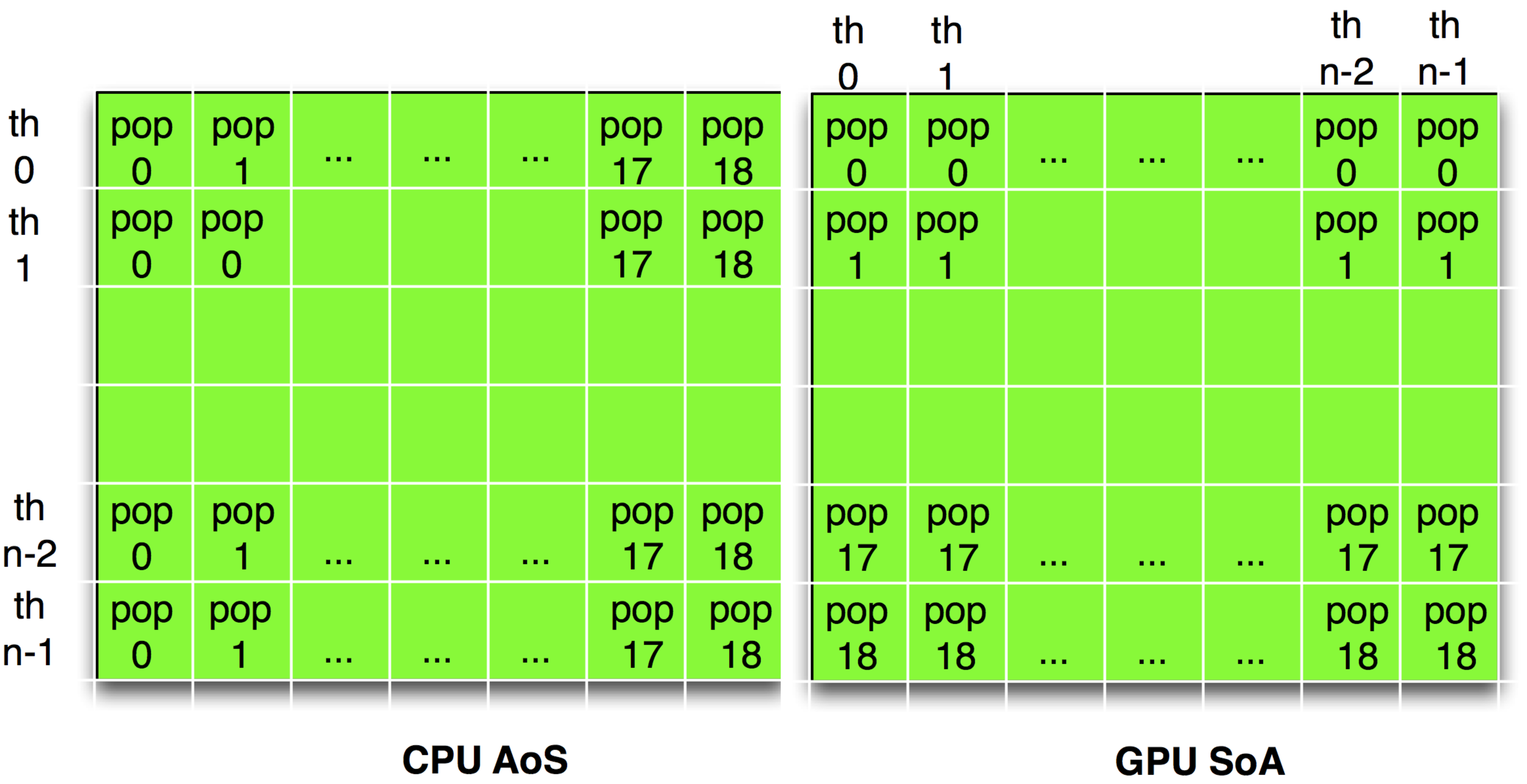} 
\par\end{centering}
\caption{CPU vs. GPU optimal data layout. Row major ordering is assumed for
storing multidimensional arrays in memory (typical of \emph{C} and
C++ languages). AoS stands for Array of Structures, SoA stands for
Structure of Arrays. \label{AoSvsSoA}}
\end{figure}

In this way the achievable peak performance doubles. A \emph{fused}
implementation increases the complexity of the code (and, most of
the times, the memory requirements). Unfortunately, the actual performance
may be significantly lower than $P_{mflups}$ since the memory-bandwidth
is hardly exploited at its best due to the memory access pattern of
the LB. The situation is the same for \emph{any} platform (i.e., a
general purpose CPU or an accelerator) but the techniques to improve
the situation, that is to reach a higher percentage of $P_{mflups}$
strongly depend on the features of the memory hierarchy and on the
programming model. For instance, there are several alternatives for
the data layout of the populations in memory. It is possible to store
all populations of a single lattice point close each other in a data
structure containing 19 floating point values. The populations of
all lattice points will form an \emph{Array of Structures} (figure
\ref{AoSvsSoA}) that can exploit the data-locality principle of cache-based
memory hierarchies like those found in a general purpose CPU. However
that layout prevents from exploiting vector instructions available,
for instance, on Intel CPU, because populations of different mesh
nodes are stored in non-contiguous memory locations. For this reason,
specially on accelerators like the Nvidia GPUs, a different layout,
in which each population is stored in a single array and the whole
data set of populations forms a \emph{Structure of Arrays}, must be
used so that neighboring threads access contiguous memory locations
according to a principle of thread-locality. Many studies have been
carried out, some of them in the recent past, proposing variants in
the usage of these two data-layouts with special attention to the
Nvidia GPUs which expose a complex memory hierarchy to the explicit
control of the programmer. Although their main purpose is the optimization
of the LB update, those works provide useful indications to enhance
the performance of other procedures, like those for the solution of
PDE, that need to use 3D \emph{stencils} to access data.\\

Very recently, new technologies promise a significant boost in performance
for any memory bandwidth-limited application, including the LB method.
In particular, multi-dimensional memory-processor interfaces provide
a much higher bandwidth. Preliminary tests with a recent generation
Nvidia GPU card (featuring the so-called \emph{Pascal} architecture)
that offers a memory bandwidth up to 720 GBytes/sec., show an improvement
of the LB performance of a factor 3 with no change in the source code.

As mentioned several times in the present Review, the populations
update procedure of the Lattice Boltzmann method is very suitable
to parallel processing and usually achieves a very good efficiency
on shared memory systems. The only drawback is that a parallel \emph{fused}
implementation requires a double memory buffer for storing the populations:
at each iteration, one of the buffers is used as source of the populations
and the other as target; at the end of the iteration the role of the
two buffers swaps. Actually, following a tricky ordering, the collision
and streaming phases could be carried out without requiring a double
buffer but only with a serial update procedure. As a consequence,
a parallel implementation of the \emph{fused} procedure for the D3Q19
model requires, at least, $19\times2\times WS$ bytes of memory for
each lattice site (actually more, because memory is required also
for the hydrodynamic variables, i.e., density, velocities, etc.) so
that large scale simulations may not fit in the memory of a shared
memory system. In those cases or simply for reducing the simulation
times by exploiting many more computing resources, it is necessary
to resort to a distributed system with multiple computing nodes. We
discuss the main issue of that approach in the next Section \ref{sec:overlap}.

\subsection{Overlap between computation and communication \label{sec:overlap}}

Exascale computing platforms will be very likely based on super clusters
of powerful computing nodes, possibly equipped with accelerators like
GPUs. A general and detailed discussion of the challenges posed by
the efficient exploitation of such platforms is beyond the scope of
the present work, however, at least one specific issue, related to
the scalability of large scale LBPD simulations, deserves to be mentioned,
namely the need of overlapping the computation and the communication
stages of the LB algorithm, so as to ``hide'' the overhead of the
latter behind the former.

\begin{figure}
\centering{}\includegraphics[scale=0.8]{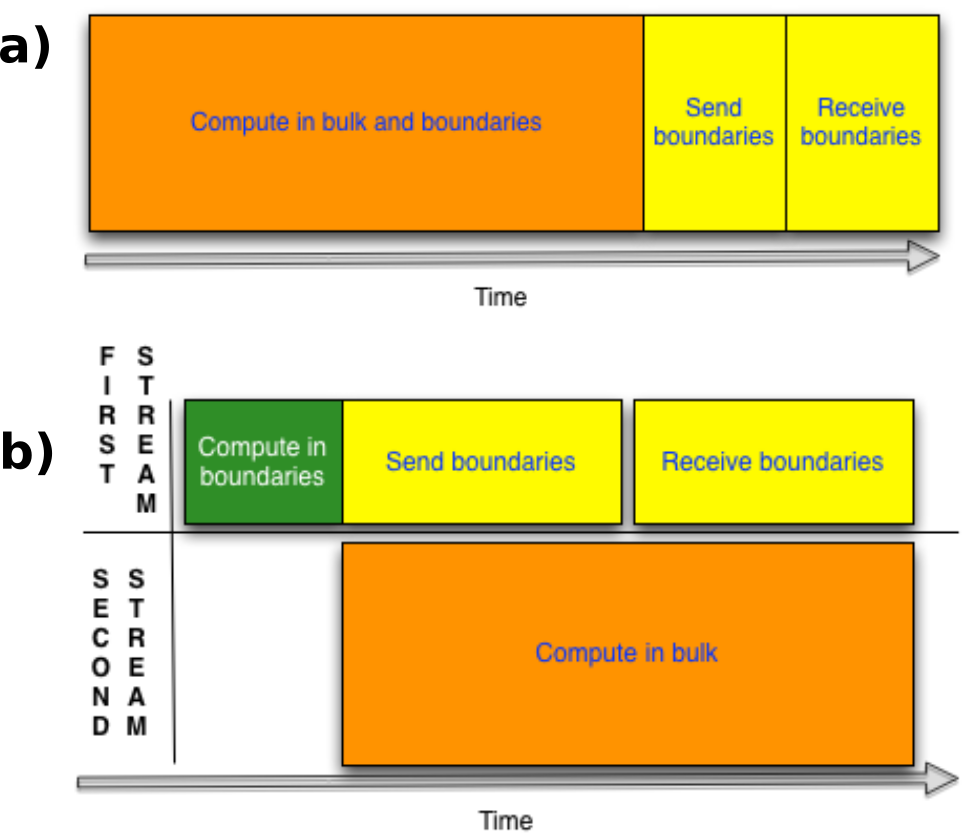}\caption{Communication schemes: a) with no overlap between communication and
computation and b) multi-GPU scheme using two streams.}
\label{fig:simplempi} 
\end{figure}

When more than one Computing Node (CN) is available for the simulation
of a system, it is quite natural to apply a \emph{domain} decomposition.
With this approach, each computing node is responsible for a subset
of the whole mesh. Here we assume that the domain has a regular geometry
postponing the discussion of irregular and/or sparse geometries to
the next subsection (\ref{sec:irdo}). When $N_{CN}$ are available,
each CN $i$ needs to know, for the update of the nodes in its own
boundaries, the value of variables belonging to the nodes in the boundaries
of its neighbors, so, at each iteration, it must (assuming a simple
one-dimensional domain decomposition and periodic boundary conditions): 
\begin{enumerate}
\item \begin{trivlist} 
\item Send data belonging to the points of its bottom boundary to CN $(i-1)\%N_{CN}$; 
\item Send data belonging to the points of its top boundary to CN $(i+1)\%N_{CN}$.
\end{trivlist} 
\item \begin{trivlist} 
\item Receive data sent by CN $(i-1)\%N_{CN}$; 
\item Receive data sent by CN $(i+1)\%N_{CN}$. \end{trivlist} 
\item Update data belonging to the nodes of its subdomain $i$ (both bulk
and boundaries); 
\end{enumerate}
That ``naive'' scheme is represented in Figure \ref{fig:simplempi}
(panel a). We define it as naive because computation and communication
are carried out one after the other whereas they can be overlapped
to a large extent when accelerators are used for the computation.
We now briefly describe how the overlap works for one of the most
common accelerators in use at the present time.

\subsubsection{Effective Multi-GPU CUDA programming\label{sec:multigpu}}

CUDA, the programming environment of the Nvidia GPU, supports concurrency
within an application through \emph{streams} \cite{CUDAProg}. A stream
is a sequence of commands that execute in order. Different streams,
on the other hand, may execute their commands out of order with respect
to each other or concurrently. By using two streams on each GPU it
is possible to implement the following scheme that assigns one stream
to the bulk and one to the boundaries of the LB domain. 
\begin{enumerate}
\item starts to update the boundaries by using the first stream; 
\item first stream: 
\begin{itemize}
\item copy data in the boundaries from the GPU to the CPU; 
\item exchange data between nodes by using MPI; 
\item copy data in the boundaries from the CPU to the GPU; 
\end{itemize}
\item second stream: 
\begin{itemize}
\item updates the bulk; 
\end{itemize}
\item starts a new iteration. 
\end{enumerate}
The overlap with this scheme, also shown in Figure \ref{fig:simplempi}
(panel b), is between the exchange of data within the $boundaries$
(carried out by the first stream and the CPU) and the update of the
bulk (carried out by the second stream). The CPU acts as a data-exchange-coprocessor
of the GPU. Non-blocking MPI primitives should be used if multiple
CPUs are involved in the data exchange.

Recently, Nvidia announced NVLink, a new high-speed interconnect technology
for GPU-accelerated computing. Supported on SXM-2 based Tesla V100
accelerator boards, NVLink significantly increases performance for
both GPU-to-GPU communications, and for GPU access to system memory.
Programs running on NVLink-connected GPUs can execute directly on
data in the memory of another GPU as well as on local memory. That
feature should further improve the scalability of LBPD simulations
running on large clusters of GPUs.

\subsection{Sparse and irregular geometries \label{sec:irdo}}

A number of LB applications may use regular and dense geometries for
which domain decomposition is, most of the times, straightforward
(e.g., a uniform decomposition along one, two or three directions).
However, in bio-fluidics, soft matter or porous media simulations
the geometry is often neither regular nor dense. In those situations
it is not possible or, at least, it is a waste of memory to store
LB populations in a simple, regular multi-dimensional data structure
whose size would be proportional to the \emph{bounding box} of the
domain. It is much more convenient to follow other approaches for
storing only the minimal set of populations required for the simulation
of non-solid nodes of the mesh. In the present Section we describe
two possible alternatives.

\subsubsection{Indirect Addressing\label{indadd}}

The first solution relies on a linearized indirect addressing scheme
\cite{Chopard}\cite{Schultz}. Each node of the LB lattice is labeled
with a \emph{tag} that identifies it as belonging to a specific subregion
of the computational domain (i.e., fluid, wall, inlet, outlet, solid).
Mesh nodes may be grouped according to their features in several one-dimensional
arrays, so that there is an array of fluid nodes, an array of \emph{wall}
nodes, an array of \emph{inlet} nodes, etc., with the exception of
\emph{solid} nodes that do not need to be stored at all since they
refer to inactive regions of the domain.

\begin{figure}[h]
\centering{}\includegraphics[scale=0.7]{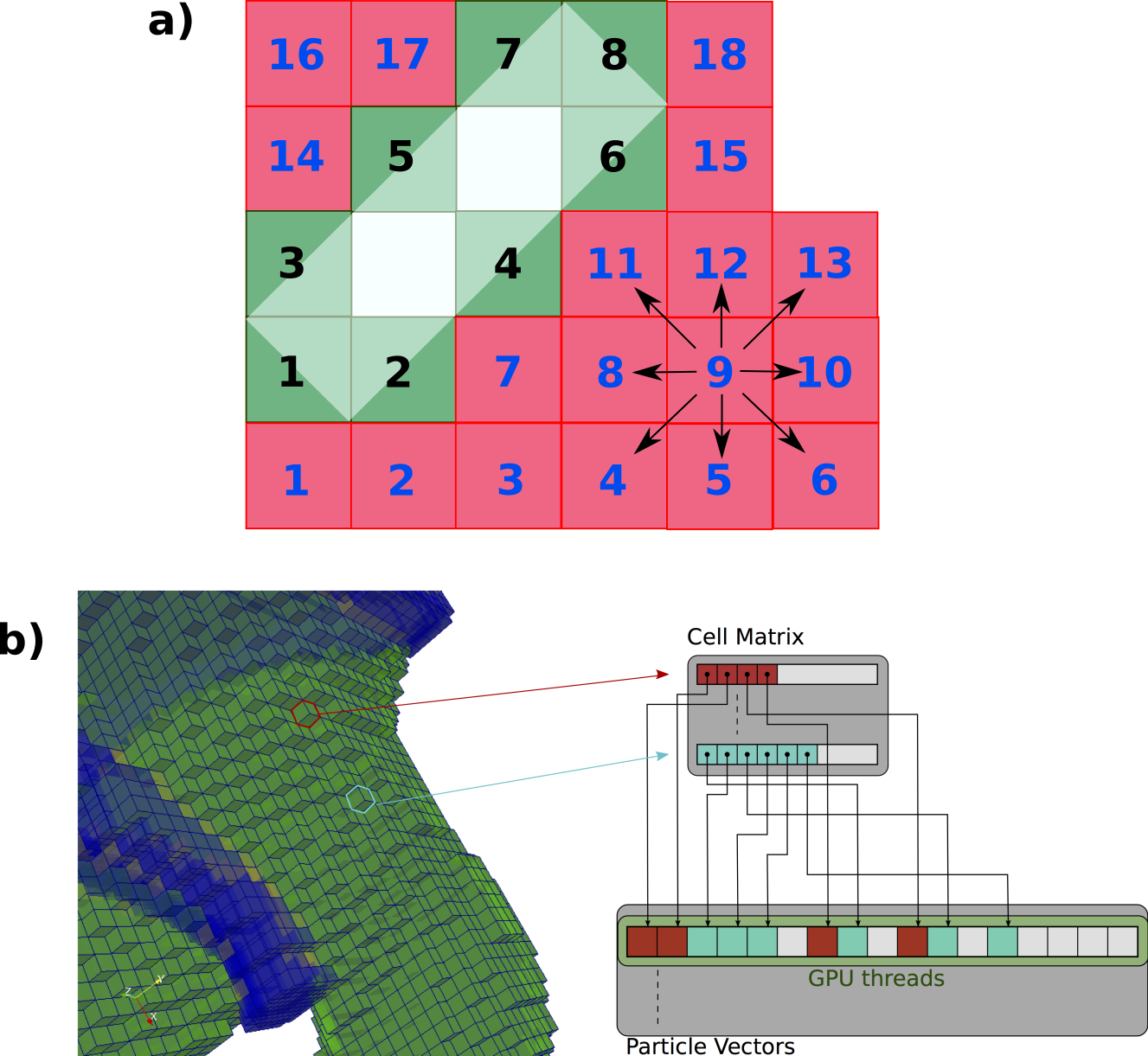}
\caption{Organization of mesh nodes according to the indirect addressing scheme
for an irregular domain. a) In the D2Q9 scheme, red (dark grey) squares
correspond to fluid regions and green (light grey) squares to wall
regions. Inactive nodes are in white and are not numbered since they
are not stored in memory. Entries in the connectivity matrix are shown
for fluid node $9$. b) Indirect addressing on the GPU for a 3D case.
Each GPU thread handles a subset of fluid nodes and a subset of populations.}
\label{fig:indadd} 
\end{figure}

As a consequence, \emph{homogeneous} nodes (i.e., all fluids nodes,
all wall nodes, etc.) are contiguous in memory regardless of their
geometrical distance. This type of data organization requires, for
each node, an additional data structure (\emph{connectivity matrix})
that contains the list of all positions, within the above mentioned
one-dimensional arrays, of its neighboring nodes (see Fig. \ref{fig:indadd}
for a simple 2DQ9 case).

With this approach only the nodes playing an active role in the Lattice
Boltzmann dynamics need to be accessed and stored in memory, resulting
in huge savings in storage requirements, despite of the additional
data structure \cite{Hoekstra}, for most (non-trivial) geometries.
An indirect addressing scheme allows to support very flexible domain
decomposition strategies, a fundamental requirement for a good load
balancing among computational tasks. For instance, the MUPHY code
\cite{CAP} supports all possible Cartesian decompositions (along
$X,Y,Z,XY,XZ,YZ,XYZ$) and \emph{custom} decompositions, e.g., those
produced by graph/mesh partitioning tools like METIS\cite{METIS}
or SCOTCH\cite{SCOTCH}, which are necessary for distributing the
computational load in an even manner in case of very irregular domains.

Those graph-based procedures utilize, most of the times, a graph bisection
algorithm that is completely unaware of the geometry of the computational
domain. However the lack of geometrical information degrades the quality
of the partitioning as the number of partitions increases, in which
case the subdomains reduce to highly irregular shapes with large contact
areas between subdomains and large communication loads. A possible
solution is to combine the graph-based partitioning with a \emph{flooding}-based
approach (also known as graph-growing method) according to the following
procedure: the mesh is first partitioned, in a given number of subdomains
(\emph{e.g., 256}). If the mesh needs to be partitioned in a finer
number of parts, say $256*P$, with $P$ an integer $\ge2$, then
each of the 256 domains is further divided according to the following
\emph{flooding} scheme: starting from a seed mesh point, a region
is iteratively grown in a isotropic way until the number of mesh points
equals $N_{i}/P$ (with $N_{i}$ being the number of mesh point in
the $i$-th original partition). As the condition is met, the visited
mesh points are assigned to a computational resource. Subsequently,
a new growth procedure starts from a new seed until all points in
the subdomain are assigned to a new computational resource. In \cite{MUPHY4}
it is shown how the distribution of tasks versus the number of neighbor
tasks with which they exchange data tends, for a large number of tasks,
to stabilize instead of increasing up to much higher values as it
would happen with a pure graph-based partitioning approach.

\subsubsection{Tiling and Blocking\label{tiling}}

Another possible solution is to ``tile'' the sparse geometry using
much smaller (with respect to the original domain size) regular, square
or cubic (depending on the dimension of the original domain), tiles
\cite{Tiling}. One of the advantages of the tiling is that during
a single LB iteration, the tiles can be processed independently and
in any order provided that values at the tile edges are correctly
propagated. Moreover the update of each square or cubic tile can be
carried out according to the simple addressing scheme used for the
case of regular geometries. However also a tiling procedure introduces
some overheads due to the presence of solid nodes inside tiles and
additional memory requirements for saving information about tiles
placement. An interesting variant of this approach is described in
\cite{WALBERLA} where a given geometry is divided using a hierarchical
structure of ``patches'' composed of ``blocks''. For sparse geometries,
empty blocks can be removed reducing memory usage and computational
complexity. Blocks correspond to leaves in a distributed forest of
octrees \footnote{An octree is a tree data structure in which each internal node has
exactly eight children. Octrees are most often used to partition a
three-dimensional space by recursively subdividing it into eight octants} and are quite sophisticated data structures designed rather for efficient
multi-processor implementations, where load balancing and communication
may affect performance. This approach supports quite naturally grid
refinement procedures, however the load balancing may suffer from
the granularity of the blocks.

\subsubsection{Parallel Particle Dynamics in Irregular Domains\label{PDID} }

Multiple techniques for Parallel Particle Dynamics have been put forward
over the years. In particular, PD has now reached a pretty good degree
of efficiency when dealing with regular geometries. However, in presence
of highly-irregular domains, like those found in bio-fluidic devices
or in physiological conduits, several critical issues arise related
to the calculation of forces and migration of particles among subdomains.
For instance, irregular subdomains imply irregular contact surfaces
and, in principle, irregular communication patterns. The geometrical
tests for particle ownerships and exchange of particles among domains
require strategic decisions that affect the efficiency of stand-alone
PD as much as the LBPD multiphysics applications.

The first general solution to those problem was presented in \cite{ourCiCP}.
The proposed method relies on two basic notions, proximity and membership
tests. Those tests are used to discriminate particles according to
their position relative to the geometry of the domains. Proximity
tests are used to select the particles that have out-of-domain interactions
and are used to perform inter-domain forces computation. The membership
tests regard the assignment of particles to domains and exploit a
tracking method to associate particles position to the domains morphology.
Moreover, the critical regions around the contact surfaces of the
subdomains are approximated so that is computationally simple to find
a superset of the particles located inside those regions and to apply
the tests only to those particles. This is possible by covering each
subdomain with identical box-shaped cells.

There are several other issues that would deserve attention in the
design and implementation of large-scale LBPD based simulations. However,
for the sake of brevity, here we just mention them: \emph{i)} whether
it is better to use only accelerators to run the simulation leaving
the CPU as a sort of communication and I/O co-processor of the accelerators
or if it makes sense to develop hybrid codes running the LBPD code
on both the CPU and the accelerators; \emph{ii)} whether it is actually
possible to develop portable LB high-performance codes by using directive-based
software and finally, \emph{iii)} how to implement some form of fault-tolerance
within the LBPD, so as to secure prompt and error-free recovery from
hardware/software failures in multi-million and possibly billion-core
computing environments.

\settowidth{\nomlabelwidth}{QM-MM}
\printnomenclature{}

\nomenclature{ADR}{Advection-Diffusion-Reaction}

\nomenclature{ASIC}{Application Specific Integrated Circuits}

\nomenclature{ATP}{Adenosine triphosphate}

\nomenclature{BE}{Boltzmann equation\\}

\nomenclature{BGK}{Bhatnagar-Gross-Krook\\}

\nomenclature{CGFF}{coarse grained force fields\\}

\nomenclature{DSMC}{Direct Simulation Monte Carlo} 

\nomenclature{DFT}{Density Functional Theory}

\nomenclature{DPM}{diffused particle model} 

\nomenclature{ELB}{Entropic Lattice Boltzmann} 

\nomenclature{EPM}{Extended Particle Model} 

\nomenclature{EE}{Eulerian-Eulerian}

\nomenclature{EL}{Eulerian-Lagrangian}

\nomenclature{FLBE}{Fluctuating Lattice Boltzmann} 

\nomenclature{FPGA}{Field Programmable Gate Array} 

\nomenclature{GI}{Galilean invariance}

\nomenclature{GPU}{Graphics Processing Unit} 

\nomenclature{HI}{hydrodynamic interactions}

\nomenclature{IBM}{Immersed Boundary Method} 

\nomenclature{LB}{Lattice Boltzmann} 

\nomenclature{LBPD}{Lattice Boltzmann-Particle Dynamics} 

\nomenclature{LE}{Lagrangian-Eulerian} 

\nomenclature{LGCA}{Lattice Gas Cellular Automata} 

\nomenclature{LBM}{Lattice Boltzmann Method} 

\nomenclature{LBE}{Lattice Boltzmann Equation} 

\nomenclature{LL}{Lagrangian-Lagrangian} 

\nomenclature{MB}{Maxwell-Boltzmann} 

\nomenclature{OPEP}{Optimized Potential for Efficient Protein}

\nomenclature{PCB}{Physics Chemistry Biology} 

\nomenclature{PD}{Particle Dynamics} 

\nomenclature{PDE}{Partial Differential Equations} 

\nomenclature{PPM}{point particle model} 

\nomenclature{PAML}{Physics Aware Machine Learning} 

\nomenclature{QM-MM}{quantum-mechanical molecular mechanics} 

\nomenclature{NESS}{non equilibrium steady states} 

\nomenclature{NS}{Navier-Stokes} 

\nomenclature{RKT}{Reverse Kinetic Theory} 

\nomenclature{SC}{Shan-Chen}

\nomenclature{vWF}{von Willebrand Factor}

\bibliographystyle{apsrev}
\bibliography{./lbmd}

\end{document}